\documentclass[12pt]{article}
\usepackage{amsmath}
\usepackage{amssymb}
\usepackage{enumitem}
\begin{document}
\vskip 2cm
\begin{center}
{\sf {\Large Abelian   $p$-form ($p =1, 2, 3$) gauge theories as the field theoretic models for the Hodge theory}}

\vskip 1.5cm

{\sf R. Kumar$^{(a)}$\footnote {Present address: S. N. Bose National Centre for Basic Sciences, 
Kolkata-700 098, W. B., India }, S. Krishna$^{(a)}$, A. Shukla$^{(a)}$ and R. P. Malik$^{(a,b)}$}\\
$^{(a)}$ {\it Physics Department, Centre of Advanced Studies,}\\
{\it Banaras Hindu University, Varanasi - 221 005, (U.P.), India}\\
$^{(b)}$ {\it DST Centre for Interdisciplinary Mathematical Sciences,}\\
{\it Faculty of Science, Banaras Hindu University, Varanasi - 221 005, India}\\
{\small {\sf {e-mails: raviphynuc@gmail.com; skrishna.bhu@gmail.com; ashukla038@gmail.com;  rpmalik1995@gmail.com}}}

\end{center}

\vskip 1cm

\noindent
{\bf Abstract:} 
Taking the simple examples of an Abelian 1-form gauge theory in two $(1+1)$-dimensions, 
a 2-form gauge theory in four (3+1)-dimensions and a 3-form gauge theory in six 
$(5+1)$-dimensions of spacetime, we establish that such gauge theories respect, in  
addition to the gauge symmetry transformations that are generated by the first-class 
constraints of the theory, additional continuous symmetry transformations. We christen 
the latter symmetry transformations as the dual-gauge  transformations. We generalize 
the above gauge and dual-gauge  transformations to obtain the proper (anti-)BRST and 
(anti-)dual-BRST transformations for the Abelian 3-form gauge theory within the framework of 
BRST formalism. We concisely mention such symmetries for the 2D free Abelian 1-form and 4D
free Abelian 2-form gauge theories and briefly discuss their topological aspects in our present endeavor.
We conjecture that any arbitrary Abelian $p$-form gauge theory would respect 
the above cited  additional symmetry in $D = 2p$ ($ p = 1, 2, 3,...)$ dimensions of spacetime. 
By exploiting the above inputs, we establish that the Abelian 3-form gauge theory, in six 
($5 + 1$)-dimensions of spacetime, is a perfect model for the Hodge theory whose discrete and 
continuous symmetry transformations provide the physical realizations of all aspects of the 
de Rham cohomological operators of differential geometry. As far as the physical utility of 
the above nilpotent symmetries is concerned, we demonstrate that the 2D Abelian 1-form  gauge theory is a perfect 
model of a new class of topological theory and 4D Abelian 2-form as well as 6D Abelian 3-form gauge  theories 
are  the field 
theoretic models for the quasi-topological field theory.



\newpage

\noindent

\section{Introduction}
\label{intro}
It has been well-established that the four $(3 + 1)$-dimensional (4D) non-Abelian 1-form
$\big(A^{(1)} = dx^\mu A_\mu \big)$ gauge theories are at the heart of standard model of high
energy physics where there is a stunning degree of agreement between theory and experiment.
Two of the central shortcomings of the standard model of particle physics are the experimental
observation of the mass of the neutrinos and, so far, no conclusive experimental detection of the
Higgs particles\footnote{In recent experiments at LHC, it has been claimed that
the Higgs particle has been experimentally observed. However, this claim is debatable
at the moment and there is no hundred percent certainty about its identification (as Higgs particle).}
which provide masses to the gauge bosons and fermions of the
standard model of particle physics. It has been possible to construct models that provide masses to
the (non-)Abelian 1-form gauge bosons without taking any recourse to the  Higgs mechanism.
These models  are based on the merging of 1-form and 2-form [$B^{(2)} 
= (1/2!) (dx^\mu \wedge dx^\nu) B_{\mu\nu}$] gauge fields through the 
topological coupling [1--4].
In an exactly similar fashion, the 2-form gauge boson has been shown to acquire a mass through the topological
coupling with the 3-form gauge field (see, e.g., [5]).
Thus, there has been a renewed interest in the study of the
higher $p$-form ($ p = 2, 3, 4,...$) gauge theories. One of the central purposes of our
present investigation is to discuss some {\it novel} continuous and
discrete symmetry transformations that are associated with the 
6D Abelian 3-form gauge theory and briefly mention about  such symmetries in the context of our earlier works on 
the 2D Abelian 1-form and 4D Abelian 2-form gauge theories.

In recent years, it has become fashionable to study  the higher $D$-dimensional $[(D -1) + 1]$
(with $D = 5, 6, 7,...$) $p$-form tensor gauge fields because these fields appear in the quantum excitations
of the (super)string theories and related extended objects (see, e.g., [6--8]). In fact, as is well-known,
the quantum versions of the (super)strings themselves live in dimensions of spacetime that are higher than the 
 observable {\it physical} four (3 + 1)-dimensions of spacetime (at present level
of available energy scale). Thus, from the point of view of  modern developments in
(super)string theories, it is important to study higher $p$-form gauge theories in higher dimensions
($D > 4$) of spacetime.  There is yet another motivation to study, particularly, 
higher Abelian $p$-form ($p \geq 2$) gauge theories in higher dimensions $(D > 4)$ of spacetime. 
In our very recent paper on the existence of the (anti-)dual Becchi-Rouet-Stora-Tyutin 
(BRST) symmetries [9], we have claimed that the dual-gauge- and (anti-)dual BRST symmetries would
always exist for any arbitrary Abelian $p$-form gauge theory in the specific $D$-dimensions of spacetime (when
$D = 2 p$).  In other words, we have proposed that, besides the usual gauge- and corresponding nilpotent
(anti-) BRST symmetries, any arbitrary Abelian $p$-form gauge theory would be {\it also} endowed with the dual-gauge- 
and corresponding nilpotent (anti-)dual BRST symmetry transformations  in the spacetime dimensions $D = 2p$.

To follow the current trends and to
corroborate the above assertions, we have chosen the 6D Abelian 3-form gauge theory to 
demonstrate that it respects the dual-gauge- and nilpotent (anti-)dual BRST [or (anti-)co-BRST] 
symmetries along with the {\it usual} gauge- and corresponding nilpotent (anti-)BRST symmetries. 
As a consequence, the present theory provides a tractable field theoretic model for the Hodge
theory in six (5 +1)-dimensions of spacetime as do the Abelian 1-form gauge theory in two 
(1 + 1)-dimensions [10--12] and Abelian 2-form gauge theory in four (3 + 1)-dimensions of spacetime [13--16].
In all the above theories, we have obtained the physical realizations of the de Rham cohomological operators
of differential geometry [17--20] in the language of discrete and continuous symmetry transformations.
Furthermore, we have deduced the analogue of the celebrated Hodge decomposition theorem in the quantum
Hilbert space of states for the above theories.

In our present investigation, we have taken the (anti-)BRST invariant coupled (but equivalent) Lagrangian 
densities from our earlier works [21,22] on the superfield approach to BRST formalism for the Abelian
3-form gauge theory in any arbitrary $D$-dimensions of spacetime where we have established the connection
of the Curci-Ferrari (CF) type restrictions with the geometrical objects called gerbes (see, e.g., [22] for
details). It is the specific property of the six (5 +1) dimensional (6D) spacetime that the kinetic
term $\big((1/24)\, H^{\mu\nu\eta\kappa} H_{\mu\nu\eta\kappa}\big)$ of the above Lagrangian densities can
be linearized [cf. (53), (54)] by exploiting the 6D Levi-Civita tensor 
($\varepsilon_{\mu\nu\eta\kappa\rho\sigma}$).  
This linearization enables us to derive the dual-gauge- and off-shell nilpotent 
(anti-)dual BRST transformations besides the usual gauge- and off-shell nilpotent 
(anti-)BRST  transformations. We have deduced  a bosonic symmetry in the theory which
is obtained from the {\it suitable} anticommutators between (anti-)BRST and (anti-)co-BRST 
symmetry transformations. We show, in our present endeavor, that there are, in totality, six {\it useful} 
continuous symmetries in the theory that include the usual ghost-scale symmetry transformations 
as well. There also exists a set of discrete symmetry transformations
in the theory which plays a very crucial role in our present discussions on the proof of 
our present 6D free Abelian gauge theory to be a tractable field theoretic model for the Hodge theory.

The physical relevance of all these studies, it may be pointed out, is the observation that the 2D
free Abelian 1-form gauge theory provides a new model [23] for the topological field theory (TFT) which
captures  a part of the key features of Witten-type TFT and some salient features of Schwarz-type TFT.
In an exactly  similar fashion, we have been able to prove that the 4D Abelian 2-form gauge theory is a 
model for the quasi-topological field theory (q-TFT) [24]. In our present endeavor, we briefly review
topological aspects of these 2D and 4D Abelian theories.
We demonstrate, in our present endeavor, that 
present 6D Abelian 3-form gauge theory is {\it also} a model for q-TFT apart from being
a cute and precise model for the Hodge theory. To study the physical constraints on the theory,  
we exploit the Hodge decomposition theorem in
the quantum Hilbert space of states  and choose the physical state to be the {\it harmonic} state of the theory
[which is (anti-)BRST as well as (anti-)co-BRST invariant and, hence, is the most symmetric state in the theory].
We have already chosen such a physical state (i.e. the harmonic state) 
in our earlier works on the proof of the {\it exact} topological
nature of  free 2D Abelian 1-form gauge theory [25,10].

The following factors have spurred our curiosity and interest in pursuing our present investigation.
First, to put our claim [9] on a firmer-footing, it is essential for us to show that the 6D Abelian
3-form gauge theory is {\it also} endowed with the dual-gauge- and off-shell 
nilpotent (anti-)co-BRST symmetry transformations
as we have been able to show in the cases of 2D Abelian 1-form and 4D Abelian 2-form gauge theories. Second, our
present exercise  helps us to establish that the 6D Abelian 3-form gauge theory is a perfect example of
Hodge theory and a  model for the q-TFT [which is similar in contents and textures as 
the 4D Abelian 2-form gauge theory (see, e.g., [24])].
Finally, it is always challenging to explore some new features that turn up in the study of the
higher $D$-dimensional ($D> 4$) and higher $p$-form ($p \geq 2$) gauge theories that are, in some
sense,  generalizations of the usual 4D gauge theories based on the 1-form potentials 
(that provide the basis for the standard model of particle physics).

The material of our present investigation is organized as follows. In Sect. 2, we discuss the 
continuous (dual-)gauge transformations and discrete symmetry transformations for the Abelian 
1-form, 2-form and 3-form gauge theories in 2D, 4D and 6D spacetime, respectively. We {\it also} 
make brief comments on the nilpotent (anti-)BRST and (anti-)co-BRST symmetry transformations associated 
with the   4D  Abelian  2-form and 2D Abelian 1-form gauge   theories and their relevance in the proof of 
(quasi-)topological nature of these theories which have been discussed in our earlier works (see, e.g. [10,23,24]
for details).
Our Sect. 3  is devoted to the discussion of (anti-)BRST symmetries and corresponding conserved charges for 
the 6D Abelian 3-form gauge theory. In Sect. 4, we discuss about the (anti-)co-BRST symmetries 
and corresponding charges for the above 6D Abelian 3-form gauge theory. We derive the anticommutators 
of (anti-)BRST and (anti-)co-BRST symmetries in Sect. 5 and deduce a single bosonic symmetry and 
its corresponding charge. We take up the ghost-scale symmetry in Sect. 6 and derive its 
conserved charge. In Sect. 7, we discuss the algebraic structures of {\it all} the conserved 
charges and devote time on the analysis of cohomological aspects of the above algebraic structures. Our
Sect. 8 is devoted to the proof that our present theory is {\it also} a model for the q-TFT.
Finally, we summarize our central results, make some concluding remarks and point out a few future directions 
for further investigations in Sect. 9.

In our Appendices  {\bf A}, {\bf B} and {\bf C}, we discuss some explicit computations that have been used in 
the main body of our text (i.e. the derivation of coupled Lagrangian densities (53), (54) 
and the extended BRST algebra [cf. (102)]). Further, in our Appendix {\bf D}, we discuss very 
briefly the self-duality condition for the general $D = 2p$ dimensional Abelian $p$-form 
gauge theory. Our Appendix {\bf E} is devoted to the discussion of {\it extra} bosonic continuous symmetries
that are also present in our Abelian 3-form gauge theory.

{\it Notations and conventions:}
We adopt here the conventions and notations such that the background flat $D$-dimensional 
Minkowskian spacetime manifold is endowed with a metric that has signatures ($+1, -1, -1,...$) so that the dot 
product between two non-null vectors $P_\mu$ and $Q_\mu$ is:  $P \cdot Q = P_\mu\, Q^\mu = P_0\, Q_0 - P_i \,Q_i$, 
where the Greek indices $\mu, \nu, \rho,... = 0, 1, 2,..., D-1$ correspond to the spacetime
directions and the Latin indices $i, j, k,... = 1, 2, 3,..., D-1 $ stand for the space directions 
only. We also follow the convention:
$(\delta B_{\mu\nu}/\delta B_{\rho\sigma}) = (1/2!)\, (\delta^\rho_\mu \,\delta^\sigma_\nu 
- \delta^\rho_\nu \,\delta^\sigma_\mu)$, $\;(\delta A_{\mu\nu\eta}/\delta A_{\rho\sigma\kappa}) 
= (1/3!)\, [\delta^\rho_\mu\, (\delta^\sigma_\nu \,\delta^\kappa_\eta$ $- \delta^\kappa_\nu\, \delta^\sigma_\eta) + \delta^\sigma_\mu\,(\delta^\kappa_\nu\, \delta^\rho_\eta - \delta^\rho_\nu\, \delta^\kappa_\eta) 
+ \delta^\kappa_\mu\,(\delta^\rho_\nu\, \delta^\sigma_\eta - \delta^\sigma_\nu\, \delta^\rho_\eta) ]$, etc., 
where $B_{\mu\nu}$ and $A_{\mu\nu\eta}$ are the totally antisymmetric tensor gauge fields. 
We denote on-shell as well as off-shell nilpotent (anti-) BRST and (ant-)co-BRST symmetry
transformations by $s_{(a)b}$ and $s_{(a)d}$, respectively. The 
Minkowskian spacetime remains in the background and it
does not {\it directly} play any role in our discussions because we focus on only the internal symmetries
of our theories.

\section{Preliminaries: Dual-gauge symmetries}

In our subsection  {\bf 2.1}, we briefly mention the key points connected with the (dual-) gauge 
transformations for the Abelian 1-form  and 2-form gauge theories in 2D and 4D flat Minkowskian 
spacetime, respectively [10--16]. Our subsections {\bf 2.2} and {\bf 2.3} deal with the topological 
features of the 2D Abelian 1-form and 4D Abelian 2-form gauge theories. 
In subsection {\bf 2.4},  we discuss about the continuous 
(dual-)gauge- and discrete symmetries for the 6D Abelian 3-form gauge theory.

\subsection{Abelian 1-form and 2-form gauge theories: Continuous and discrete symmetries}

We begin with the two $(1+1)$-dimensional (2D) gauge-fixed Lagrangian density for a free 
Abelian 1-form gauge theory in the Feynman gauge (see, e.g., [10--12])
\begin{eqnarray}
{\cal L}_{(1)} &=& - \, \frac{1}{4}\; F_{\mu\nu}\,F^{\mu\nu} - \frac {1}{2} \;(\partial \cdot A)^{2}
\equiv \frac {1}{2}\; E^{2} - \frac {1}{2} \;(\partial \cdot A)^{2},
\end{eqnarray}
where $F_{\mu\nu} = \partial_\mu A_\nu - \partial_\nu A_\mu$ is derived from the 2-form 
$[F^{(2)} = (1/2!) (dx^\mu \wedge dx^\nu)\; F_{\mu\nu}]$ which owes its origin to the 
application of an exterior derivative $d = dx^\mu \partial_\mu$ (with $d^2 = 0$) on a 1-form 
($A^{(1)} = dx^\mu A_\mu $) defined in terms of the gauge potential $A_\mu$. The gauge-fixing 
term ($\partial \cdot A$) is obtained from the application of a co-exterior derivative 
$\delta = -\; * d *$ (with $\delta^2 = 0$) on the 1-form gauge connection $A^{(1)}$. 
Here $(*)$ is the Hodge duality operation on the 2D spacetime flat Minkowskian manifold. 
In 2D spacetime, the only existing component of the curvature tensor $F_{\mu\nu}$ ($\mu, \nu,... = 0, 1$) is the 
electric field (i.e. $F_{01} = - F_{10} = E$) only.

The gauge- and the dual-gauge symmetry  transformations ($\delta_g,\, \delta_{dg}$) for the 
above gauge-fixed Lagrangian density (1) are given by (see, e.g., [10--12] for details)
\begin{eqnarray}
\delta_{g} A_\mu (x) = \partial_\mu\, \Omega(x),\qquad\qquad 
\delta_{dg} A_\mu (x)= - \,\varepsilon_{\mu\nu} \,\partial^{\nu} \,\Sigma(x),
\end{eqnarray}
where $\Omega(x)$ and $\Sigma(x)$ are the infinitesimal local gauge- and dual-gauge parameters, respectively, 
and $\varepsilon_{\mu\nu}$ is the 2D Levi-Civita tensor with $\varepsilon_{01} = +1 = - \varepsilon^{01}$. 
The latter satisfies $\varepsilon_{\mu\nu}\,\varepsilon^{\mu\nu} = - 2!,\;
\varepsilon_{\mu\nu}\,\varepsilon^{\mu\lambda} = - \delta^\lambda_\nu,$ etc. 
It should be noted that, under the infinitesimal gauge ($\delta_g$) and  dual-gauge ($\delta_{dg}$) 
symmetry transformations, the curvature tensor $F_{\mu\nu}$ and the gauge-fixing term ($\partial \cdot A$)
remain invariant, respectively. One can check that, under the above infinitesimal transformations (2), the 
Lagrangian density transforms as follows:
\begin{eqnarray}
\delta_{g} {\cal L}_{(1)} = - \,(\partial \cdot A)\,(\Box\, \Omega), 
\qquad\qquad  \delta_{dg} {\cal L}_{(1)} 
= E\,(\Box\,\Sigma).
\end{eqnarray}
We note that the Lagrangian density remains invariant under the above symmetry transformations 
(2) if we impose the conditions $\Box \;\Omega(x) = 0$ and $\Box\; \Sigma(x)  = 0$ from outside.
However, we obtain a {\it perfect} symmetry invariance of the modified version of the above
Lagrangian density within the framework of BRST formalism where the (dual-)gauge 
symmetry transformations are generalized  to their counterparts (anti-)co-BRST and (anti-)BRST symmetries.
The latter symmetries are nothing but the supersymmetric type (on-shell as well as off-shell nilpotent) symmetry transformations (see, e.g., [10--12]).

In two dimensions of spacetime, the self-duality condition on the Abelian 1-form gauge 
connection is defined in terms of the Hodge duality ($*$) operation as 
\begin{eqnarray}
* \;A^{(1)}\; = dx^\mu \;(- \;\varepsilon_{\mu\nu} \;A^\nu ) \;\equiv \;dx^\mu \;\tilde A_\mu.
\end{eqnarray}
Thus, we observe that $\tilde A_\mu = -\; \varepsilon_{\mu\nu} A^\nu$. Under the transformation 
$A_\mu \to \tilde A_\mu = -\;\varepsilon_{\mu\nu} A^\nu$, it can be checked that 
${\cal L}_{(1)} \to - {\cal L}_{(1)} $
 because of the fact that $(\partial \cdot A) \leftrightarrow E.$  
As a consequence, the Lagrangian density ${\cal L}_{(1)}$ is {\it not} self-duality invariant. 
However, it is obvious that, under the discrete symmetry transformations
$A_\mu \to \mp\;i\; \varepsilon_{\mu\nu} A^\nu$, we have 
${\cal L}_{(1)} \to {\cal L}_{(1)}$. These discrete symmetry transformations are at 
the heart of the existence of dual-gauge symmetry transformations in the theory. The former   provides
a physical realization of the Hodge duality ($*$) operation of differential geometry [10--12]. 
Furthermore, these symmetries are the reasons
behind the existence of exactly similar kind of restrictions on the (dual-)gauge parameters $\Sigma(x)$ 
and $\Omega(x)$ for the (dual-)gauge invariance of ${\cal L}_{(1)}$ [cf. (3)]. The above 
continuous and discrete symmetries have been exploited in the case of (non-)interacting 2D 
Abelian theories  within the framework of BRST formalism and these theories have been shown 
to be the models for the Hodge theory (see, e.g., [10--12]) as, the interplay  of these symmetries,
provide the physical realizations of all aspects of the de Rham cohomological operators of 
differential geometry [17--20] in terms of the above physical symmetry transformations.
It will be noted that the discrete symmetries exist {\it only} in the Feynman gauge. 
Thus, the most symmetric theory (respecting the discrete as well as continuous symmetries) 
picks up the Feynman  gauge in a very clear fashion.

Against the above backdrop,  let us look at the 4D free Abelian 2-form gauge theory which is 
described by the gauge-fixed Lagrangian density (in the Feynman gauge) as 
\begin{eqnarray}
{\cal L}_{(2)} = \frac{1}{12}\; H_{\mu\nu\eta}\,H^{\mu\nu\eta} 
 + \; \frac{1}{2}\;(\partial_\mu B^{\mu\nu})\,(\partial^{\eta} B_{\eta\nu}),
\end{eqnarray}   
where $H_{\mu\nu\eta} = \partial_\mu B_{\nu\eta} + \partial_\nu B_{\eta\mu} 
+ \partial_\eta B_{\mu\nu}$ is the totally antisymmetric curvature tensor 
derived from the 3-form [$H^{(3)} = d B^{(2)} 
= (1/3!)\; (dx^\mu \wedge dx^\nu \wedge dx^\eta) \; H_{\mu\nu\eta}$]. The latter is obtained from the 
application of the exterior derivative $d$ on the 2-form [$B^{(2)} = \frac{1}{2!}(dx^\mu \wedge dx^\nu) \; B_{\mu\nu}$] 
antisymmetric ($B_{\mu\nu} = - B_{\nu\mu}$) tensor gauge field $B_{\mu\nu}$.
The gauge-fixing term (in the Feynman gauge) can be obtained by the action of the 
co-exterior derivative $\delta$ on the 2-form gauge field [$\delta B^{(2)} = (\partial^\nu B_{\nu\mu})dx^\mu $]. 
The above Lagrangian density (5) transforms as
\begin{eqnarray}
\delta_{g} {\cal L}_{(2)} &=& (\partial_\mu B^{\mu\nu})\;\big[\Box\; \Omega_\nu 
- \partial_\nu (\partial \cdot \Omega)\big],\nonumber\\
\delta_{dg} {\cal L}_{(2)} 
&=& \frac {1}{3!}\; \varepsilon_{\mu\nu\eta\kappa}\; H^{\mu\nu\eta}\;\big[\Box\; \Sigma^{\kappa} - \partial^{\kappa} 
(\partial \cdot \Sigma)\big],
\end{eqnarray} 
under the following gauge- and dual-gauge transformations:
\begin{eqnarray}
\delta_g B_{\mu\nu} = (\partial_\mu \Omega_\nu - \partial_\nu \Omega_\mu ), 
\quad\qquad \delta_g H_{\mu\nu\eta} = 0, \nonumber\\
\delta_{dg} B_{\mu\nu} =  \varepsilon_{\mu\nu\eta\kappa} \partial^{\eta} \Sigma^{\kappa}, 
\quad\qquad \delta_{dg} (\partial^{\mu} B_{\mu\nu}) = 0,
\end{eqnarray}
where $\Omega_\mu (x)$ and $\Sigma_\mu (x)$ are the infinitesimal local gauge- and dual-gauge  
parameters and $\varepsilon_{\mu\nu\eta\kappa}$ is the 4D Levi-Civita tensor with 
$\varepsilon_{0123} = +\,1 \equiv  -\, \varepsilon^{0123}$. Furthermore, we note that 
$\varepsilon_{\mu\nu\eta\kappa}\;\varepsilon^{\mu\nu\eta\kappa} = -\, 4!, \; 
\varepsilon_{\mu\nu\eta\kappa}\;\varepsilon^{\mu\nu\eta\lambda} = -\, 3! \;\delta^\lambda_\kappa,\;
\varepsilon_{\mu\nu\eta\kappa}\;\varepsilon^{\mu\nu\lambda\rho} = - \,2!\;\big(\delta^\lambda_\eta \;\delta^\rho_\kappa 
- \delta^\lambda_\kappa \; \delta^\rho_\eta \big)$, etc.
The key features of the above continuous (dual-)gauge transformations are 
the invariance of the gauge-fixing and kinetic terms, respectively. We note that, 
for the (dual-)gauge invariance in the theory, we have to impose $\Box \;\Sigma_\mu 
- \partial_\mu (\partial \cdot \Sigma) = 0, \; \Box\, \Omega_\mu 
- \partial_\mu (\partial \cdot \Omega) = 0$ in equation
(6) from outside. However, these restrictions are {\it not} required within the 
framework of BRST formalism and there exists a perfect symmetry in the theory  (see, e.g., [13--16]).

The Hodge duality ($*$) operation can be defined on the Abelian 2-form $B^{(2)}$ 
(on the 4D flat spacetime manifold) as:
\begin{eqnarray}
*\,B^{(2)} &=& \frac{1}{2!} \;(dx^\mu \wedge dx^\nu) \,\left(\frac{1}{2!}\, \varepsilon_{\mu\nu\eta\kappa}\, 
B^{\eta\kappa}\right)
\equiv \frac {1}{2!}\,(dx^\mu \wedge dx^\nu) \, { \tilde B_{\mu\nu}},\nonumber\\
\tilde B_{\mu\nu} &=& \frac{1}{2!}
\;\varepsilon_{\mu\nu\eta\kappa} \, B^{\eta\kappa}.
\end{eqnarray}
It turns out that the gauge-fixed Lagrangian density (5) respects the discrete symmetry 
transformations $B_{\mu\nu} \rightarrow \pm \,\frac{i}{2}\,\varepsilon_{\mu\nu\eta\kappa} B^{\eta\kappa}$ 
$\big($i.e. ${\cal L}_{(2)} \rightarrow {\cal L}_{(2)} \big)$ because the kinetic and gauge-fixing terms exchange 
with each-other $\big[$i.e. $\frac{1}{12} H^{\mu\nu\eta} H_{\mu\nu\eta} \longleftrightarrow 
\frac{1}{2} (\partial_\mu B^{\mu\nu}) (\partial^\eta B_{\eta\nu})\big].$
It is clear from (8) that the existence of this discrete symmetry transformations owes its 
origin to the self-duality (Hodge duality) condition. We further point out that the above 
(dual-) gauge symmetry 
transformations as well as discrete symmetry transformations have been exploited within the 
framework of BRST formalism and, in our earlier works [13--16], it has been established that
the Abelian 2-form gauge theory is a field theoretic model for the Hodge theory in the 4D 
Minkowskian flat spacetime (as it provides the physical realizations of the de Rham cohomological 
operators of differential geometry).

We wrap of this subsection with the remark that the
existence  of the discrete symmetry transformations for the 2D Abelian 1-form and 4D Abelian 2-form theories
owes its origin to the self-duality conditions on the 1-form and 2-form gauge fields
as illustrated in this subsection. The existence  of these discrete symmetry transformations are
{\it also} the root-cause of the presence of off-shell as well as on-shell
nilpotent (dual-)BRST as well as (anti-) dual BRST symmetries in the theory.

\subsection{2D Abelian 1-form gauge theory: A perfect model for a TFT}
\label{sec:3}
It is well-known that 2D Abelian 1-form ($A^{(1)} = dx^\mu\,A_\mu$) gauge field $A_\mu$ has, to begin 
with, only two degrees of freedom (d.o.f.). However, these d.o.f. can be gauged away 
due to the masslessness condition and the presence of gauge symmetry in the theory
 [25,26,27]. Thus, it becomes a topological field theory (TFT)
where there are no propagating  degrees of freedom. In our earlier  works [23],
it has been established that the 2D Abelian 1-form theory provides a tractable 
field theoretical model of a {\it new} class of TFT that captures a few aspects of Witten 
type TFT as well as some salient features  of Schwarz type TFT.
In this proof, the existence  of 
the nilpotent and absolutely anticommuting (anti-)co-BRST and (anti-)BRST symmetries play a very important role.

It is elementary to check that the following (anti-)BRST symmetries ($s_{(a)b}$) and (anti-) co-BRST 
[or (anti-)dual-BRST] symmetries 
($s_{(a)d}$):
\begin{eqnarray}
&&s_b A_\mu = \partial_\mu C, \quad\qquad s_b C = 0, \quad\qquad s_b \bar C = - i\, (\partial \cdot A),
\quad \qquad s_b E =0,\nonumber\\
&& s_{ab} A_\mu = \partial_\mu \bar C, \;\;\qquad s_{ab} \bar C = 0, \;\qquad s_{ab} C 
= + i\, (\partial \cdot A),\; \qquad s_{ab} E =0,\nonumber\\
&&s_d A_\mu = -\varepsilon_{\mu\nu}\partial^\nu \bar C, \,\qquad s_d \bar C = 0, \,\qquad s_d  C =  -i\,E, 
\,\qquad s_d (\partial \cdot A) = 0,\nonumber\\
&&s_{ad} A_\mu = -\varepsilon_{\mu\nu}\partial^\nu C, \;\;\quad s_{ad}  C = 0, \;\;\quad s_{ad} \bar C =  +i\,E, 
\;\;\,\quad s_{ad} (\partial \cdot A) = 0,
\end{eqnarray}
leave the following Lagrangian density [10]: 
\begin{eqnarray}
{\cal L}_{(b)} = \frac{1}{2}\, E^2 - \frac{1}{2}\, (\partial \cdot A)^2 - i\partial_\mu \bar C\, \partial^\mu C,
\end{eqnarray} 
quasi-invariant (i.e. $s_b {\cal L}_{(b)} = - \partial_\mu [(\partial \cdot A) \partial^\mu C],\; 
s_{ab} {\cal L}_{(b)} = - \partial_\mu [(\partial \cdot A) \partial^\mu \bar C],\;
s_d {\cal L}_{(b)} = \partial_\mu [E\, \partial^\mu \bar C],\; 
s_{ad} {\cal L}_{(b)} = \partial_\mu [E\, \partial^\mu C]$) because it transforms to 
the total spacetime derivatives. Hence, the action integral $S =  \int d^2x\; {\cal L}_{(b)}$ 
remains invariant for physically well-defined fields that vanish off at infinity. In the above,
the symbol ${\cal L}_{(b)}$ stands for the BRST-invariant Lagrangian density  corresponding to 
the gauge-fixed Lagrangian density in (1) and $(\bar C)C$ are the fermionic (i.e. $\bar C^2 = C^2 = 0, \; 
 C \, \bar C + \bar C\,  C = 0$) (anti-)ghost fields with ghost numbers $(-1)+1$, respectively.

The salient features of the above symmetries are 
\begin{enumerate}[label = $(\roman{*})$]
\item They are on-shell ($\Box C = \Box \bar C = 0$) nilpotent of order two ($s^2_{(a)b} = 0,\; s^2_{(a)d} = 0$)
and absolutely anticommuting ($s_b\, s_{ab} + s_{ab}\, s_b = 0, \; s_d\, s_{ad} + s_{ad}\, s_d = 0 $) 
in nature where the equations of motion are imposed  for its proof.
\item The kinetic term $[(1/2)\, E^2$], for the free 2D Abelian theory, 
remains invariant under the (anti-)BRST symmetry transformations. 
On the contrary, it is the gauge-fixing term $[-(1/2)(\partial \cdot A)^2]$ that remains unchanged under the 
(anti-)co-BRST transformations.
\item The kinetic  and gauge-fixing terms owe their origin to the exterior ($d = dx^\mu\, \partial_\mu$) and
co-exterior ($\delta = - * d*$) derivative because: 
\begin{eqnarray}
&&d\, A^{(1)} = \frac{1}{2!}\, (dx^\mu \wedge dx^\nu)\, F_{\mu\nu}, 
\qquad\qquad F_{\mu\nu} = \partial_\mu A_\nu - \partial_\nu A_\mu,\nonumber\\
&&\delta\, A^{(1)} = - * d* A^{(1)} = (\partial \cdot A),
\end{eqnarray}
where $F_{\mu\nu}$ has only one non-vanishing component $(F_{01} = E)$ in 2D (which is an electric field and 
it is a pseudo-scalar).
\item The nomenclature of (anti-)BRST and (anti-)co-BRST symmetries have  their genesis in the exterior
($d = dx^\mu\, \partial_\mu$) and co-exterior ($\delta = -*d*$) derivatives of differential geometry which are
nilpotent of order two. 
\item  Under the (anti-)BRST symmetries, it is the combination of the gauge-fixing  and Faddeev-Popov (FP)
ghost terms that become total spacetime derivatives. On the other hand,  is the kinetic and  FP ghost terms 
that lead to the emergence  of a total spacetime derivative in the case of on-shell nilpotent
(anti-)co-BRST symmetries.
\item The nilpotency of the (anti-)BRST and (anti-)co-BRST symmetries owe their origin to the nilpotency of the (co-)exterior derivatives.
\item Within the framework of the BRST formalism, there is {\it no} restriction from outside as (anti-)BRST
and (anti-)co-BRST symmetries are {\it perfect} symmetries of the theory. 
\item The self-duality condition (4) and ensuing discrete symmetry of the Lagrangian density (1)
can be generalized to:
\begin{eqnarray}
A_\mu\longrightarrow \pm \,i\,\varepsilon_{\mu\nu}\, A^\nu \qquad C\longrightarrow \pm \,i\, \bar C,
\qquad \bar C\longrightarrow \pm \,i\, C,
\end{eqnarray}
which turn out to be the symmetry of the Lagrangian density (10) and it can checked that
\begin{eqnarray}
s_{(a)d} \,\phi = \pm\, * s_{(a)b}\,*\phi, \qquad\quad \phi = A_\mu, \, C,\, \bar C,
\end{eqnarray}
where ($*$) is the discrete symmetry transformations (12) and $s_{(a)d}$ as well as  $s_{(a)b}$
are given in (9). The relation (13) provides  the realization of the relationships 
$\delta =  \pm \, *\, d\,*$. The ($\pm$) signs in (13) are decided  by a couple of successive discrete 
symmetry transformations (12) on the generic field $\phi$, namely;
\begin{eqnarray}
*\,* \,(\phi) = \pm\,\phi, \qquad\quad \phi = A_\mu,\, C,\, \bar C.
\end{eqnarray}
This should be contrasted with ($\pm$) signs present in $\delta = \pm\, *\, d\, *$ which are dictated  by 
the dimensionality of the spacetime manifold and degree of the form involved 
in the inner product on this very manifold. 
\end{enumerate}

The Lagrangian density ${\cal L}_{(b)}$ can be expressed   in the following two different ways 
\begin{eqnarray}
{\cal L}_{(b)} &=& s_b\left(i\, T_1\right) + s_d\left(i\, T_2\right) + \partial_\mu Y^\mu\nonumber\\
&=& s_{ab}\left(i\, P_1\right) + s_{ad}\left(i\, P_2\right) + \partial_\mu Y^\mu,
\end{eqnarray}
where $T_1 = (1/2)\, E\,C, \; T_2 = -(1/2)\,(\partial \cdot A)\,\bar C, \; P_1 = (i/2)\, E\, \bar C,\; 
P_2 = -(i/2)\,(\partial \cdot A)\,C$ and the expression for $Y^\mu$ is: $Y^\mu = (i/2)\, \partial^\mu[\bar C\,C]
\equiv (i/2)\,[(\partial^\mu \bar C)C + \bar C(\partial^\mu C)]$. Thus, in its appearance, the Lagrangian density
(15) is similar to the case of Witten-type TFT because it is able to be expressed, modulo a total spacetime derivative,
 as the sum of the (anti-)BRST and (anti-)co-BRST {\it exact} symmetries. However, we note that there is {\it no} topological shift symmetry in our theory which is the benchmark of a Witten-type TFT. Furthermore,  we point out that the appearance (15) is {\it not} like the Schwarz-type TFT because there is 
no piece in the Lagrangian density which
 can {\it not} be expressed in terms of the (anti-)BRST and (anti-)co-BRST symmetry transformations. 
However, the symmetries of the theory (where there is {\it no} topological shift symmetry) is just like the Schwarz-type TFT. Thus, our present 2D free Abelian 1-form theory belongs to a {\it new} class of TFT which captures  some aspects of Witten-type TFT and a few properties of the Schwarz-type TFT.

To re-confirm the {\it perfect} topological nature of our present theory, we note that 
 the following two topological invariants:
\begin{eqnarray}
I_k = \oint_{C_k} V_k,\qquad\quad J_k = \oint_{C_k} W_k,
\end{eqnarray}  
where $C_k$ are the $k$-dimensional homology cycles in the 2D spacetime manifold and $V_k$ and $W_k$ are the
$k$-forms that  are defined in terms of the fields of the theory. The on-shell BRST
invariant $k$-forms ($k = 0, 1, 2$) on the 2D spacetime manifold are
\begin{eqnarray}
&& V_0 =  - (\partial\cdot A)\, C, \nonumber\\
&& V_1 = (dx^\mu)\, \Big[i\, C\,\partial_\mu \bar C - (\partial\cdot A)\, A_\mu\Big], \nonumber\\
&& V_2 = \Big(\frac{dx^\mu\wedge dx^\nu}{2!}\Big)\; \Big[i\,A_\mu \,\partial_\nu \bar C 
- i A_\nu \,\partial_\mu \bar C  - i\, \bar C \, F_{\mu\nu} \Big].
\end{eqnarray}   
The counterparts of (14), w.r.t. the on-shell co-BRST invariant $k$-forms ($k= 0, 1, 2$)
on the above 2D spacetime manifold, are  
\begin{eqnarray}
&& W_0 = E \,\bar C, \nonumber\\
&& W_1 = (dx^\mu)\,\Big[ \varepsilon_{\mu\nu}\,\bar C\, \partial^\nu  C - i E\, A_\mu\Big], \nonumber\\
&& W_2 = \Big(\frac{dx^\mu\wedge dx^\nu}{2!}\Big)\; \Big[i\,\varepsilon_{\mu\nu} (\partial\cdot A)\, C
+ i\, \varepsilon_{\mu\rho}\, \partial^\rho C\, A_\nu 
- i \, \varepsilon_{\nu\rho}\, \partial^\rho C \, A_\mu  \Big].
\end{eqnarray}   
The above topological invariants obey the following recursion relations:
\begin{eqnarray}
s_b\, V_k = d V_{k-1},\qquad\quad s_d\, W_k = \delta W_{k-1},
\end{eqnarray}  
where $k= 1, 2$ for the $k$-forms. It is obvious that we shall also have  relations like (19)
for the topological invariants w.r.t. the on-shell nilpotent anti-BRST and
anti-co-BRST charges (see, e.g. [23,10] for details).

We wrap up this section with the remarks that our present 2D free Abelian  
1-form gauge theory is a tractable field theoretic model for a {\it new} 
type of TFT where the form of the Lagrangian density is like Witten-type TFT but
the nilpotent symmetries of the theory are just like the Schwarz-type TFT.
In our earlier works [10], it has been demonstrated that the 2D free Abelian 1-form gauge theory
is a perfect model for the Hodge theory because {\it all} the de Rham cohomological operators 
of differential  geometry find their   physical realizations in the language of  symmetry properties 
of the theory (and their corresponding charges). In our present endeavor, we have concisely mentioned only 
the topological aspects of this theory and clarified the mathematical origin of the discrete 
 symmetry in this theory. The cohomological aspects of the 2D Abelian 1-form theory can be found in 
 earlier work [10].

\subsection{ Abelian 2-form gauge theory: A model for the q-TFT}
\label{sec:4}

It is well-known that 4D free Abelian 2-form gauge field ($B_{\mu\nu}$) has one degree of freedom. Thus, it can not be a TFT.
However, it was speculated that, perhaps, this d.o.f. can {\it also} be gauged away if there are several symmetries 
in the theory. In our previous  works [14], it has been well-established that this 4D field theory is a perfect model
for the Hodge  theory. We have also discussed a few points connected with the topological nature of this theory.
In this section, however, we shall concisely focus on  the salient features of the topological aspects of 4D free Abelian 2-form theory and show that it is a possible candidate for the field theoretic model for a quasi-topological field theory (q-TFT) in
the physical four dimensions of spacetime.

We begin with the following  Lagrangian density for the free 4D Abelian 2-form theory
which incorporates   the gauge-fixing and FP ghost terms [28]:
\begin{eqnarray}
{\cal L}^{(0)}_b &=& \frac{1}{2}\, (\partial^\sigma B_{\sigma\mu} - \partial_\mu \, \phi_1)\,
(\partial_\rho B^{\rho\mu} - \partial^\mu \phi_1) -\partial_\mu \bar\beta \,\partial^\mu\beta \nonumber\\ 
&-& \frac{1}{2}\, \Big(\frac{1}{2}\, \varepsilon_{\mu\nu\eta\kappa}\partial^\nu B^{\eta\kappa} 
- \partial_\mu \phi_2 \Big)\,\Big(\frac{1}{2}\, \varepsilon^{\mu\zeta \rho\sigma}\partial_\zeta B_{\rho\sigma} 
- \partial^\mu \phi_2 \Big) \nonumber\\   &+& (\partial_\mu {\bar C}_\nu 
- \partial_\nu {\bar C}_\mu)\, (\partial^\mu { C}^\nu ) - \frac{1}{2} \,(\partial\cdot \bar C)\,
(\partial\cdot  C),
\end{eqnarray}  
where $({\bar C}_\mu)C_\mu$ fields are the Lorentz-vector and fermionic 
($C_\mu\, {\bar C}_\nu + {\bar C}_\nu\, C_\mu = 0, \; C_\mu^2 = {\bar C}^2_\mu = 0$, etc.) (anti-)ghost
fields and $(\bar\beta)\beta$ are the bosonic (anti-)ghost fields. Here the massless ($\Box \,\phi_1 = 0$)
field $\phi_1$ is a scalar field that has been used for the linearization   of the gauge-fixing term  and the
massless ($\Box \,\phi_2 = 0$) pseudo-scalar field $\phi_2$ has been introduced to linearize the kinetic term
of the free Abelian 2-form gauge field. The fields $\phi_1$ and $\phi_2$ are required for the 
most general form of linearization of the gauge-fixing and kinetic terms. We have discussed about these aspects of  linearization in our earlier works  [14,28]. 
The ghost numbers for the fields $({\bar C}_\mu)C_\mu$ are $(-1)+1$ and,  for $(\bar\beta)\beta$,  we have 
$(-2)+2$ because the latter are the ghost-for-ghost fields which  
are required  because of the stage-one 
reducibility in the Abelian 2-form theory. The theory becomes unitary 
because of the presence of all these (anti-)ghost fields. The Lagrangian density (20) 
is the generalized form of the starting Lagrangian density (5) for the 4D free Abelian 2-form 
gauge theory where appropriate linearizations have been performed  (see, e.g. [28,14] for details)

 The above Lagrangian density (20) is endowed with the following 
(anti-)BRST symmetry transformations [28,24]:
\begin{eqnarray}
&& s_{b} B_{\mu\nu} = (\partial_\mu C_\nu - \partial_\nu C_\mu), \;\;\qquad
 s_{b} C_\mu = \partial_\mu \beta, \;\;\qquad
 s_{b} \bar C_\mu = (\partial^\nu B_{\nu\mu} - \partial_\mu \phi_{1}), \nonumber\\
 && s_{b} \phi_{1} =  \frac{1}{2}\, (\partial \cdot C), \qquad\qquad
 s_{b} \bar \beta = - \frac{1}{2} \,(\partial \cdot \bar C), \qquad\qquad
 s_{b} (\beta, \,\phi_{2},\, H_{\mu\nu\eta}) = 0,\nonumber\\
&& s_{ab} B_{\mu\nu} = (\partial_\mu {\bar C}_\nu - \partial_\nu {\bar C}_\mu), \quad
 s_{ab}\bar C_\mu = - \partial_\mu \bar\beta, \quad
 s_{ab}  C_\mu = - (\partial^\nu B_{\nu\mu} - \partial_\mu \phi_{1}), \nonumber\\
 && s_{ab} \phi_{1} =  \frac{1}{2}\, (\partial \cdot\bar C), \qquad
 s_{ab} \beta = - \frac{1}{2} \,(\partial \cdot  C), \qquad
 s_{ab} (\bar\beta, \,\phi_{2},\, H_{\mu\nu\eta}) = 0,
\end{eqnarray}
because the Lagrangian density (20) transforms to the total spacetime derivatives:
\begin{eqnarray}
s_b {\cal L}_{(b)}^{(0)}&=& \partial_\mu\, \Big[\frac{1}{2}\, (\partial\cdot\bar C)(\partial^\mu\beta) +
\frac{1}{2} (\partial^\mu\phi_1)(\partial\cdot C) - (\partial^\mu\,C^\nu
- \partial^\nu\,C^\mu)(\partial_\nu\phi_1)  \nonumber\\
&-& \frac{1}{2}\,(\partial_\kappa B^{\kappa\mu})(\partial\cdot C) 
+ (\partial^\kappa\,B_{\kappa\nu}) (\partial^\mu\,C^\nu
- \partial^\nu\,C^\mu)\Big],\nonumber\\
s_{ab} {\cal L}_{(b)}^{(0)}&=& \partial_\mu\, \Big[\frac{1}{2}\, (\partial\cdot C)(\partial^\mu \bar\beta) +
\frac{1}{2} (\partial^\mu\phi_1)(\partial\cdot\bar C) - (\partial^\mu\,\bar C^\nu
- \partial^\nu\,\bar C^\mu)(\partial_\nu\phi_1) \nonumber\\
&-& \frac{1}{2}\,(\partial_\kappa B^{\kappa\mu})(\partial\cdot \bar C) 
- (\partial^\kappa\,B_{\kappa\nu}) (\partial^\mu\,\bar C^\nu
- \partial^\nu\,\bar C^\mu)\Big].
\end{eqnarray}
There is yet another set of nilpotent symmetries  in the theory which have been christened 
as the (anti-)co-BRST [or (anti-)dual-BRST] symmetries. These are:
\begin{eqnarray}
&&  s_{d} B_{\mu\nu} = \varepsilon_{\mu\nu\eta\kappa}
\partial^\eta \bar C^\kappa, \qquad  s_{d} \bar C_\mu = - \partial_\mu
\bar \beta, \qquad  s_{d} C_\mu =
\Big(\frac{1}{2}\, \varepsilon_{\mu\nu\eta\kappa} \partial^\nu B^{\eta\kappa}
- \partial_\mu \phi_{2}\Big), \nonumber\\
&& s_{d} \phi_{2} =  + \frac{1}{2} (\partial \cdot \bar C),\qquad \qquad
 s_{d} \beta = - \frac{1}{2} \,(\partial \cdot C), \qquad\qquad
 s_{d} (\bar \beta, \phi_{1},\, \partial^\mu B_{\mu\kappa}) = 0,\nonumber\\
&&  s_{ad} B_{\mu\nu} = \varepsilon_{\mu\nu\eta\kappa}
\partial^\eta  C^\kappa, \quad  s_{ad}  C_\mu =  \partial_\mu \beta, 
\quad  s_{ad} {\bar C}_\mu = - \Big(\frac{1}{2}\,
 \varepsilon_{\mu\nu\eta\kappa} \partial^\nu B^{\eta\kappa}
- \partial_\mu \phi_{2}\Big), \nonumber\\
&&  s_{ad} \phi_{2} =  + \frac{1}{2} (\partial \cdot  C),
\qquad
 s_{ad}\bar\beta = + \frac{1}{2} \,(\partial \cdot C), \qquad
 s_{ad} ( \beta, \phi_{1},\, \partial^\mu B_{\mu\kappa}) = 0,
\end{eqnarray}
under which the Lagrangian density (20) transforms to the total spacetime derivatives as listed below:
\begin{eqnarray}
s_d {\cal L}_{(b)}^{(0)}=&=& \partial_\mu\, \Big[\frac{1}{2}\,\left(\frac{1}{2}\varepsilon^{\mu\nu\eta\kappa}\,\partial_\nu\,B_{\eta\kappa} - \partial^\mu \phi_2 \right)(\partial\cdot \bar C) 
+ \frac{1}{2}\,(\partial\cdot  C)(\partial^\mu \bar\beta) \nonumber\\ 
&-&  \left( \frac{1}{2}\;\varepsilon_{\nu\eta\kappa\rho}\,\partial^\eta\,B^{\kappa\rho} 
- \partial_\nu \phi_2 \right)(\partial^\mu\,\bar C^\nu- \partial^\nu\,\bar C^\mu)\Big],\nonumber\\
s_{ad} {\cal L}_{(b)}^{(0)}=&=& \partial_\mu\, \Big[\frac{1}{2}\,\left(\frac{1}{2}\varepsilon^{\mu\nu\eta\kappa}\,\partial_\nu\,B_{\eta\kappa} - \partial^\mu \phi_2 \right)(\partial\cdot  C) 
+ \frac{1}{2}\,(\partial\cdot \bar C)(\partial^\mu\beta) \nonumber\\ 
&-&  \left(\frac{1}{2}\; \varepsilon_{\nu\eta\kappa\rho}\,\partial^\eta\,B^{\kappa\rho} - \partial_\nu \phi_2 \right)(\partial^\mu\, C^\nu- \partial^\nu\,C^\mu)\Big],
\end{eqnarray}
which demonstrate that the action integral $S = \int d^4x {\cal L}_{(b)}^{(0)}$ remains invariant
for the physical fields that vanish off at infinity. The key features of the (anti-)co-BRST continuous
 symmetry transformations are 

\begin{enumerate}[label = $(\roman{*})$]
\item The symmetries are on-shell nilpotent ($s_{(a)d}^2 = 0$) of order two
and they are absolutely anticommuting in nature (i.e. $s_d\,s_{ad} + s_{ad}\,s_d = 0$) where we have 
to use the equations of motion [28,24].
 \item The kinetic term $H_{\mu\nu\eta}$ [or ($\frac{1}{2} \,\varepsilon_{\mu\nu\eta\kappa}\, \partial^\nu B^{\eta\kappa} - \partial_\mu \phi_2$)] remains invariant under the (anti-)BRST symmetry transformations 
$s_{(a)b}$. On the other hand, it is the gauge-fixing term ($\partial^\mu  B_{\mu\nu} - \partial_\nu\phi_1$)
that remains invariant under the (anti-)co-BRST symmetry transformations.
\item The curvature   term $H_{\mu\nu\eta}$, in the kinetic term,  owes its origin to the exterior derivative.
On the contrary, the gauge-fixing term $\partial^\mu B_{\mu\nu}$ has its genesis in the co-exterior derivative 
of differential geometry.  
\item The nomenclatures of the (anti-)BRST and (anti)co-BRST symmetry transformations come 
from the invariances of the kinetic and gauge-fixing terms (that owe their origin to the 
exterior and co-exterior derivatives) under the (anti-)BRST and 
(anti-)co-BRST symmetry transformations, respectively. 
 \item The nilpotency ($d^2 =  \delta^2 = 0$) of the (co-)exterior derivative ($\delta)d$ plays an important role 
 in the reasons behind the nilpotency of (anti-)co-BRST and (anti-)BRST symmetry transformations. 
 \item Within the framework of BRST formalism, there are {\it no} restrictions like (6) on any fields
 of the theory because (anti-)BRST and (anti-)co-BRST symmetries are 
{\it perfect}  symmetries of the action integral.
The self-duality condition (8) ensuing the discrete symmetry transformations $B_{\mu\nu}\rightarrow 
\pm\, \frac{i}{2}\, \varepsilon_{\mu\nu\eta\kappa}\, B^{\eta\kappa}$ can be generalized  to
\begin{eqnarray}
&& B_{\mu\nu} =  \pm\, \frac{i}{2}\, \varepsilon_{\mu\nu\eta\kappa} \, B^{\eta\kappa}, 
 \quad\qquad \phi_{1} \rightarrow \pm \,i\, \phi_{2}, \quad\qquad
\phi_2 \rightarrow \mp\, \,i \,\phi_1, \nonumber\\
&& \beta \rightarrow \mp \,i \,\bar\beta, \qquad \bar\beta \rightarrow \pm \,i \,\beta, \qquad
C_\mu \rightarrow \pm\, i\, {\bar C}_\mu, \qquad {\bar C}_\mu \rightarrow \,\pm\, i \,C_\mu.
\end{eqnarray}
Under the above transformations, the Lagrangian density (20) remains invariant. It can be checked 
that we have the following relationship
\begin{eqnarray}
s_{(a)d} \,\Phi = \pm\, * s_{(a)b}\,*\Phi, \qquad\quad \Phi = B_{\mu\nu},
\, C_\mu,\, \bar C_\mu,\,\beta,\, \bar\beta, \, \phi_1,\, \phi_2,
\end{eqnarray}
where ($*$) is the transformations (25) and $s_{(a)b}$ as well as $s_{(a)d}$ are given in (21) and (22).
This relation (26) provides the realization of $\delta = \pm\, *\, d\, *$.
\end{enumerate}

To demonstrate that our present theory is a quasi-TFT (q-TFT), first of all, we observe that the 
following $k$-forms ($k = 0, 1, 2, 3, 4$) w.r.t. the on-shell nilpotent BRST symmetry transformations 
(18) are [24]
\begin{eqnarray} 
V_0(+2) &=& (\partial \cdot B)\beta,\nonumber\\
V_1(+1) &=& dx^\mu\,  \Big[\partial_\mu(\partial \cdot \bar C) \,\beta + (\partial \cdot B)\, C_\mu \Big],\nonumber\\
V_2(0) &=& \frac{1}{2!}\,(dx^\mu \wedge dx^\nu)\, \Big[\partial_\mu(\partial \cdot \bar C)\,C_\nu
 -\partial_\nu(\partial \cdot \bar C)\, C_\mu + B_{\mu\nu} \,(\partial \cdot B) \Big],\nonumber\\
V_3(-1) &=& \frac{1}{3!}\,(dx^\mu \wedge dx^\nu \wedge dx^\eta)\, \Big[\partial_\mu(\partial \cdot \bar C)\,B_{\nu\eta}
+ \partial_\nu(\partial \cdot \bar C)\, B_{\eta \mu} \nonumber\\
&+& \partial_\eta(\partial \cdot \bar C)\,B_{\mu\nu} + (\partial \cdot \bar C)\,H_{\mu\nu\eta}\Big],\nonumber\\
&\equiv & d\,\Big[\frac{1}{2!}(dx^\mu \wedge dx^\nu)\, (\partial \cdot \bar C)\, B_{\mu\nu}\Big], \nonumber\\
V_4(-2) &=& 0,
\end{eqnarray}
where the numbers in the round brackets, on the l.h.s.,  denote the ghost numbers of the invariants and 
$B_\mu = (\partial^\nu B_{\nu\mu} - \partial_\mu \phi_1)$. The above forms obey the following
 recursion relations  that are characteristic  features of a TFT: 
\begin{eqnarray}
s_b\, I_k = d\, I_{k-1}, \qquad\quad k = 1, 2, 3, 4. 
\end{eqnarray}
Here $I_k = \oint_{C_k} V_k$ are the topological invariants.

It is clear that the 0-form $V_0$ is a BRST-invariant quantity because $s_b V_0 = s_b [\partial^\mu(\partial^\nu B_{\nu\mu} - \partial_\mu \phi_1)\beta] = 0$ and 3-form is an exact  form because it is a 
total derivative. Hence, there is no 4-form in the theory 
even though this theory is defined on a 4D spacetime manifold. One can derive the invariants 
w.r.t. the anti-BRST symmetry transformations by exchanging the ghost and 
anti-ghost fields of the theory with suitable constant factors.

One can also enumerate the invariants $J_k = \oint_{C_k} W_k$ w.r.t. the on-shell nilpotent 
anti-co-BRST symmetries. These forms $W_k$ (with $k = 0, 1, 2, 3, 4$) are [24]
\begin{eqnarray} 
W_0(-2) &=& (\partial \cdot \cal B)\bar\beta,\nonumber\\
W_1(-1) &=& dx^\mu \, \Big[\partial_\mu(\partial \cdot  C)\,\bar \beta 
- (\partial \cdot {\cal B}) \,{\bar C}_\mu \Big],\nonumber\\
W_2(0) &=& \frac{1}{2!}\,(dx^\mu \wedge dx^\nu)\,\Big[\frac{1}{2}\, \varepsilon_{\mu\nu\eta\kappa}\,B^{\eta\kappa}\,
(\partial \cdot {\cal B}) - \partial_\mu (\partial \cdot  C) \,{\bar C}_\nu 
+ \partial_\nu\,(\partial \cdot  C) \,{\bar C}_\mu \Big],\nonumber\\
W_{3} (+1) &=& \frac{1}{4} \,(dx^\mu \wedge dx^\nu \wedge dx^\eta) \; \varepsilon_{\nu\eta\kappa\sigma}\,
\Big[ \partial_\mu (\partial \cdot C)\,
B^{\kappa\sigma} + (\partial \cdot  C)\, \partial_\mu  B^{\kappa\sigma}
\Big], \nonumber\\
&\equiv& d\; \Big[ \frac{1}{4}\; (dx^\mu \wedge dx^\nu) \,(\partial \cdot  C) \,
\varepsilon_{\mu\nu\kappa\sigma} \,B^{\kappa\sigma} \Big], \nonumber\\
W_{4}(+2)  &=& 0\; \qquad ( d^2 = 0 ),
\end{eqnarray}
 where ${\cal B}_\mu = \frac{1}{2}\, \varepsilon_{\mu\nu\eta\kappa}\, \partial^\nu B^{\eta\kappa} - \partial_\mu \,\phi_2$ and the numbers in the round brackets (on the l.h.s.) denote the ghost numbers of the corresponding
invariants of the theory. We note that 4-form, w.r.t. the co-BRST symmetries, {\it also} turns out to be zero 
because  its predecessor 3-form  $W_3 (+1)$ is an exact form. These invariants also obey the appropriate recursion
relations (i.e. $s_d W_k  = \delta W_{k-1}$).

Finally, we concentrate on the structures of the Lagrangian density (20)  in terms of the (anti-)BRST
and (anti-)co-BRST symmetry transformations and try to see whether this could be expressed 
as the sum of (anti-)BRST and (anti-)co-BRST {\it exact} form modulo some total derivative. In this connection, 
we note that the following is true [24] 
\begin{eqnarray}
&&s_d \,\Big[ \frac {1}{2}\, {\bar C}^\mu (\partial^\nu B_{\nu\mu}) + 
\frac {1}{2}\, (\partial \cdot C)\, \bar\beta + \frac{1}{2} \,(\partial \cdot \bar C)\, \phi_1\Big] \nonumber\\
&& + s_b \Big[- \frac {1}{2}\, (\partial \cdot \bar C)\,\beta  - \frac{1}{4}\, C^\mu \, (\varepsilon_{\mu\nu\eta\kappa}\, \partial^\nu B^{\eta\kappa}) - \frac{1}{2}\, (\partial \cdot C)\, \phi_2 \Big] \nonumber\\
&&= \frac{1}{2}\,(\partial^\nu B_{\nu\mu} - \partial_\mu \phi_1)
(\partial_\rho B^{\rho\mu} - \partial^\mu \phi_1) - \partial_\mu \bar\beta\, \partial^\mu \beta \nonumber\\
&&- \frac{1}{2} \,\Big(\frac{1}{2} \varepsilon_{\mu\nu\eta\kappa} \partial^\nu B^{\eta\kappa} 
- \partial_\mu \phi_2 \Big) \Big(\frac{1}{2} \,\varepsilon^{\mu\xi\rho\sigma} \partial_\xi B_{\rho\sigma} 
- \partial^\mu \phi_2 \Big) \nonumber\\ && + \Big( \partial_\mu  {\bar C}_\nu 
- \partial_\nu {\bar C}_\mu\Big)\, (\partial^\mu\, C^\nu)
- \Big( {\bf \frac{1}{2} + \frac{1}{2}} \Big) (\partial \cdot \bar C)(\partial \cdot C) + \partial_\mu X^\mu,
\end{eqnarray}
where the explicit expansion for $X^\mu$ is
\begin{eqnarray}
&&X^\mu =  \frac{1}{2}\,\Big[\Big(\partial_\nu B^{\nu\mu} - \partial^\mu \phi_1\Big )\phi_1 +
\partial^\mu \Big(\bar\beta\,\beta \Big)+  
\,\Big(\frac{1}{2} \,\varepsilon^{\mu\nu\eta\kappa}\, \partial_\nu B_{\eta\kappa}
- \partial^\mu\,\phi_2 \Big)\, \phi_2 \nonumber\\
&& -{\bar C}_\nu \Big( \partial^\mu  { C}^\nu - \partial^\nu { C}^\mu\Big) -
\Big( \partial^\mu  {\bar C}^\nu - \partial^\nu {\bar C}^\mu\Big)\, C_\nu \Big].
\end{eqnarray}
A close look at (30) and its comparison with (20) shows that {\it only} the {\it last} term
in (30) is different from the Lagrangian density (20). Thus, we conclude that the Lagrangian density
(20) {\it can not} be expressed precisely as the sum of the BRST and co-BRST {\it exact} terms.
It is also clear now that the {\it symmetric} energy momentum tensor for this theory
 can not be expressed as the sum of BRST and co-BRST exact terms. Hence, 
we infer that the 4D Abelian 2-form gauge theory is a {\it perfect} model for the Hodge theory
but it can be treated {\it only} as a model for the q-TFT because its properties are similar to
a perfect TFT but the Lagrangian density of this theory {\it can not} be expressed as the exact sum of
underlying BRST and co-BRST symmetries (cf. Sec. 8 also).

\subsection{Abelian 3-form gauge theories: Symmetries}
\label{sec:2}
Let us begin with the following gauge-fixed Lagrangian density for the 6D Abelian 3-form 
gauge theory in the Feynman gauge (see, e.g., [21,22] for details)
\begin{eqnarray}
{\cal L}_{(3)} = \frac{1}{24}\;H^{\mu\nu\eta\kappa} H_{\mu\nu\eta\kappa} 
+ \frac{1}{2}\; (\partial_\mu A^{\mu\nu\eta})\;(\partial^\rho A_{\rho\nu\eta}),
\end{eqnarray}
where the curvature tensor  $H_{\mu\nu\eta\kappa} = \partial_\mu A_{\nu\eta\kappa} 
-\partial_\nu A_{\eta\kappa\mu}+
\partial_\eta A_{\kappa\mu\nu}-\partial_\kappa A_{\mu\nu\eta}$ (of the kinetic term for the 
gauge field $A_{\mu\nu\eta}$) is derived from the following 4-form 
\begin{eqnarray}
H^{(4)} = d A^{(3)} = \frac {1}{4!}\left(dx^\mu \wedge dx^\nu \wedge dx^\eta \wedge dx^\kappa \right)
H_{\mu\nu\eta\kappa}.
\end{eqnarray}
Here $d = dx^\mu \partial_\mu $ is the exterior derivative and 3-form 
$ A^{(3)} = \frac{1}{3!}\, (dx^\mu \wedge dx^\nu \wedge dx^\eta)\,A_{\mu\nu\eta}$ 
defines the totally antisymmetric tensor gauge field $A_{\mu\nu\eta}$. The gauge-fixing term of (32) 
has its origin in the co-exterior derivative $\delta = - \,*\, d \,*\,$ as 
$\delta A^{(3)} = \frac {1}{2!}\, (dx^\mu \wedge dx^\nu) \, (\partial^\eta A_{\eta\mu\nu})$ 
where $*$ is the Hodge duality operation defined on the 6D Minkowskian flat 
spacetime manifold. We define the following 
local, infinitesimal  and continuous (dual-)gauge transformations ($\delta_{dg},\; \delta_g$):
\begin{eqnarray}
\delta_{dg} A_{\mu\nu\eta} &=& \frac {1}{2!}\; \varepsilon_{\mu\nu\eta\kappa\rho\sigma}\; 
\partial^\kappa \Sigma^{\rho\sigma}, \quad\qquad  \delta_{dg} (\partial^\eta A_{\eta\mu\nu}) = 0, \nonumber\\
\delta_{g} A_{\mu\nu\eta} &=& \partial_\mu \Omega_{\nu\eta} + \partial_\nu \Omega_{\eta\mu}
+\partial_\eta \Omega_{\mu\nu}, \qquad \delta_g H_{\mu\nu\eta\kappa} = 0,
\end{eqnarray}
where $\Sigma_{\mu\nu} (x)$ and $\Omega_{\mu\nu} (x)$ are the (dual-)gauge parameters in the theory.
Totally antisymmetric 6D Levi-Civita tensor satisfies 
$\,\varepsilon_{\mu\nu\eta\kappa\lambda\rho}\;\varepsilon^{\mu\nu\eta\kappa\lambda\rho} = -\, 6!,\;
\varepsilon_{\mu\nu\eta\kappa\lambda\rho}$ $\varepsilon^{\mu\nu\eta\kappa\lambda\sigma} 
= - \,5!\, \delta^\sigma_\rho$, etc., and we have chosen $\varepsilon_{012345} = +\, 1
 = -\, \varepsilon^{012345}$. We note that the gauge-fixing and kinetic terms, owing their origin to the 
(co-)exterior derivatives, remain invariant under the (dual-) gauge transformations, respectively.
Furthermore, we obtain the following transformations for the Lagrangian density ${\cal L}_{(3)}$ 
under the (dual-)gauge transformations (34), namely; 
\begin{eqnarray}
\delta_{dg} {\cal L}_{(3)} &=& - \frac{1}{4!}\; \varepsilon_{\mu\nu\eta\kappa\rho\sigma} \;H^{\mu\nu\eta\kappa}\;
\big[\Box \,\Sigma^{\rho\sigma} + \partial^{\rho} (\partial_{\lambda} \Sigma^{\sigma\lambda})
- \partial^{\sigma} (\partial_{\lambda} \Sigma^{\rho\lambda})\big], \nonumber\\
\delta_g {\cal L}_{(3)} &=& (\partial_\mu A^{\mu\nu\eta})\; \big[\Box\, \Omega_{\nu\eta}
+ \partial_\nu (\partial^{\rho} \Omega_{\eta\rho})
- \partial_\eta (\partial^{\rho} \Omega_{\nu\rho}) \big].
\end{eqnarray}
Thus, we observe that, for the (dual-)gauge invariance, exactly similar kind of restrictions must
be imposed on the (dual-)gauge parameters $\Sigma_{\mu\nu} (x)$ and $\Omega_{\mu\nu} (x)$: 
\begin{eqnarray}
&&\Box \,\Sigma_{\mu\nu} + \partial_{\mu} \;(\partial^{\lambda} \Sigma_{\nu\lambda}) 
- \partial_{\nu} \;(\partial^{\lambda} \Sigma_{\mu\lambda}) = 0, \nonumber\\
&&\Box \,\Omega_{\mu\nu} + \partial_\mu \;(\partial^{\lambda} \Omega_{\nu\lambda}) 
- \partial_\nu \;(\partial^{\lambda} \Omega_{\mu\lambda}) = 0.
\end{eqnarray}
The reason behind {\it these} restrictions is the existence of a discrete symmetry  invariance 
in the theory which we elaborate below in an explicit manner.

Let us consider the self-duality condition for the Abelian 3-form connection $A^{(3)}$ in the language 
of the Hodge duality $(*)$ operation (defined on a 6D Minkowskian manifold) as follows:
\begin{eqnarray}
*\,A^{(3)} = \frac {1}{3!}\, (dx^\mu \wedge dx^\nu \wedge dx^\eta)\, \tilde A_{\mu\nu\eta},\qquad 
\tilde A_{\mu\nu\eta} =  - \,\frac {1}{3!}\, \varepsilon_{\mu\nu\eta\kappa\rho\sigma}\, A^{\kappa\rho\sigma}.
\end{eqnarray}
As we have seen the importance of self-duality transformations in the context of 2D Abelian 1-form and 
4D 2-form gauge theories, under the following discrete transformations:
\begin{eqnarray}
A_{\mu\nu\eta} \;\longrightarrow \; \pm \;\frac {i}{3!}\; \varepsilon_{\mu\nu\eta\kappa\rho\sigma}\; 
A^{\kappa\rho\sigma},
\end{eqnarray}
the Lagrangian density ${\cal L}_{(3)}$ remains invariant.
We note, once again, that the symmetry transformations (38) owe their mathematical 
origin to the self-duality condition (37). In fact, the self-duality condition (37) 
is the root-cause for the existence of  dual-gauge symmetry transformations in the theory {\it and} the 
derivation of similar kind of restrictions in (36). We would like to lay emphasis on the
fact that the discrete symmetry transformations (38) would provide the physical realizations of the 
Hodge duality $(*)$ operation of differential geometry as we shall see later in the context of 
BRST formalism (see, Sect. 7 below). It is to be re-emphasized that the discrete symmetries are true only 
in the Feynman gauge.

We can linearize the kinetic and gauge-fixing terms by invoking the Nakanishi-Lautrup type auxiliary 
fields $K_{\mu\nu}$ and ${\cal K}_{\mu\nu}$ as given below:
\begin{eqnarray}
{\cal L}_{(K)} &=& \frac {1}{2}\, {\cal K}^{\mu\nu}{\cal K}_{\mu\nu} - {\cal K}^{\mu\nu} 
\left( \frac {1}{4!}\, \varepsilon_{\mu\nu\eta\kappa\rho\sigma}\; H^{\eta\kappa\rho\sigma} \right)\nonumber\\
&+& K^{\mu\nu}\left(\partial^\eta A_{\eta\mu\nu}\right) - \frac{1}{2}\, K^{\mu\nu}K_{\mu\nu}. 
\end{eqnarray}
The gauge- and dual-gauge transformations $(\delta_g, \delta_{dg}),$ for the fields of this
linearized version of the gauge-fixed Lagrangian density, are  
\begin{eqnarray}
&&\delta_{g} A_{\mu\nu\eta} = \partial_\mu \Omega_{\nu\eta} + \partial_\nu \Omega_{\eta\mu}
+\partial_\eta \Omega_{\mu\nu}, \quad\qquad \delta_g H_{\mu\nu\eta\kappa} = 0, \nonumber\\
&&\delta_g K_{\mu\nu} = 0,\qquad \quad \delta_g {\cal K}_{\mu\nu} = 0,  \nonumber\\
&&\nonumber\\
&& \delta_{dg} A_{\mu\nu\eta} = \frac {1}{2!}\; \varepsilon_{\mu\nu\eta\kappa\rho\sigma}\; 
\partial^\kappa \Sigma^{\rho\sigma}, \quad\qquad \delta_{dg} (\partial^\eta A_{\eta\mu\nu}) = 0,\nonumber\\
&&\delta_{dg} K_{\mu\nu} = 0,  \qquad \quad \delta_{dg} {\cal K}_{\mu\nu} = 0. 
\end{eqnarray}
One can check that, under the above infinitesimal (dual-)gauge symmetry transformations, the Lagrangian density 
${\cal L}_{(K)}$ transforms as:
\begin{eqnarray}
\delta_{dg}{\cal L}_{(K)} &=& - {\cal K}^{\mu\nu} \Bigl[\Box\Sigma_{\mu\nu} 
+ \partial_\mu (\partial^\eta \Sigma_{\nu\eta}) 
- \partial_\nu (\partial^\eta \Sigma_{\mu\eta}) \Bigr],\nonumber\\
\delta_g{\cal L}_{(K)} &=& K^{\mu\nu}  \Bigl[\Box \Omega_{\mu\nu} 
+ \partial_\mu (\partial^\eta \Omega_{\nu\eta}) - \partial_\nu (\partial^\eta \Omega_{\mu\eta}) \Bigr]. 
\end{eqnarray}
Thus,  once again, the restrictions (36) have to be imposed for the (dual-)gauge 
invariance of ${\cal L}_{(K)}$. This is due to the self-duality invariance in the theory.

The generalization of the discrete symmetry transformations (38) can be written for the
Lagrangian density ${\cal L}_{(K)}$, in terms of its fields, as
\begin{eqnarray} 
K_{\mu\nu}\; \rightarrow \;\pm \,i\,{\cal K}_{\mu\nu}, \quad 
{\cal K}_{\mu\nu}\; \rightarrow \;\pm \,i\,K_{\mu\nu}, \quad
A_{\mu\nu\eta} \;\rightarrow \; \pm \,\frac {i}{3!}\; \varepsilon_{\mu\nu\eta\kappa\rho\sigma}\, 
A^{\kappa\rho\sigma}.
\end{eqnarray}
It is interesting to point out that the Lagrangian density (39) can be further 
generalized by incorporating the Lorentz vector fields 
$\phi_\mu^{(1)}$ and $\phi_\mu^{(2)}$ as given below:
\begin{eqnarray}
{\cal L}_{(\phi, K)} &=& \frac {1}{2}\; {\cal K}^{\mu\nu}{\cal K}_{\mu\nu}
 - {\cal K}^{\mu\nu} \bigg( \frac {1}{4!}\, \varepsilon_{\mu\nu\eta\kappa\rho\sigma}\, 
H^{\eta\kappa\rho\sigma} 
+ \frac{1}{2} \left[\partial_\mu \phi_\nu^{(2)} - \partial_\nu \phi_\mu^{(2)}\right]\bigg) \nonumber\\
&-&\frac{1}{2}\; K^{\mu\nu}K_{\mu\nu}
+ K^{\mu\nu} \left(\partial^\eta A_{\eta\mu\nu} + \frac{1}{2} \left[\partial_\mu \phi_\nu^{(1)} 
- \partial_\nu \phi_\mu^{(1)}\right] \right),
\end{eqnarray}
\begin{eqnarray}
{\cal L}_{(\phi, \bar K)} &=& \frac {1}{2}\; \bar {\cal K}^{\mu\nu} \bar {\cal K}_{\mu\nu} + \bar {\cal K}^{\mu\nu} 
\bigg( \frac {1}{4!}\; \varepsilon_{\mu\nu\eta\kappa\rho\sigma}\; H^{\eta\kappa\rho\sigma} 
- \frac{1}{2} \left[\partial_\mu \phi_\nu^{(2)} - \partial_\nu \phi_\mu^{(2)}\right]\bigg) \nonumber\\
&-& \frac{1}{2}\; \bar K^{\mu\nu} \bar K_{\mu\nu}
- \bar K^{\mu\nu} \left(\partial^\eta A_{\eta\mu\nu} - \frac{1}{2} \left[\partial_\mu \phi_\nu^{(1)} 
- \partial_\nu \phi_\mu^{(1)}\right] \right),
\end{eqnarray}
where $\bar {\cal K}_{\mu\nu}$ and $\bar K_{\mu\nu}$ are the additional Nakanishi-Lautrup 
auxiliary fields that have been invoked for the most general form of the gauge-fixed Lagrangian densities.
It should be mentioned here that we have  freedom to add/subtract the 2-forms: 
$F^{(2)} = d\,\Phi^{(1)} = \frac{1}{2}\, \big(dx^\mu \wedge dx^\nu \big)\,\Big[\partial_\mu \phi_\nu^{(1)} 
\,-\, \partial_\nu \phi_\mu^{(1)}\Big],\;\; 
{\cal F}^{(2)} = d\tilde \Phi^{(1)} = \frac{1}{2} \,\big(dx^\mu \wedge dx^\nu\big)\,$ 
$\Big[\partial_\mu \phi_\nu^{(2)} 
\,-\, \partial_\nu \phi_\mu^{(2)}\Big]$ 
to the 2-forms $*\; H^{(4)} = \frac {1}{4!}\, \big(dx^\mu \wedge dx^\nu\big)\, 
\varepsilon_{\mu\nu\eta\kappa\rho\sigma} \;H^{\eta\kappa\rho\sigma}$
and $\delta A^{(3)} = \frac{1}{2!}\, \big(dx^\mu \wedge dx^\nu\big)(\partial^\eta A_{\eta\mu\nu})$ 
that are present in the 6D Lagrangian density (39). The above coupled set of Lagrangian densities will be
further generalized for the BRST analysis of the present theory in the forthcoming sections.

The equations of motion that emerge from the Lagrangian density ${\cal L}_{(\phi, K)}$ are:
\begin {eqnarray}
&&{\cal K}_{\mu\nu} = \frac{1}{4!}\, \varepsilon_{\mu\nu\eta\kappa\rho\sigma} H^{\eta\kappa\rho\sigma} 
+ \frac{1}{2}\, \big[\partial_\mu \phi_\nu^{(2)} 
- \partial_\nu \phi_\mu^{(2)}\big], \quad
\Box \phi^{(2)}_\mu 
- \partial_\mu \Big(\partial \cdot \phi^{(2)}\Big) = 0, \nonumber\\ 
&& K_{\mu\nu} = \partial^{\eta} A_{\eta\mu\nu}
+ \frac{1}{2}\, \big[\partial_\mu \phi^{(1)}_\nu - \partial_\nu\phi^{(1)}_\mu \big], \qquad \Box \phi^{(1)}_\mu 
- \partial_\mu \Big(\partial \cdot \phi^{(1)}\Big) = 0, \nonumber\\
&&\partial_\mu K_{\nu\eta} + \partial_\nu K_{\eta\mu} + \partial_\eta K_{\mu\nu} 
+ \frac{1}{2}\,\varepsilon_{\mu\nu\eta\kappa\rho\sigma}\, \partial^{\kappa} {\cal K}^{\rho\sigma} = 0, \qquad
\partial_{\mu}  K^{\mu\nu} = 0,\nonumber\\ 
&&\partial_\mu {\cal K}_{\nu\eta} + \partial_\nu {\cal K}_{\eta\mu} + \partial_\eta {\cal K}_{\mu\nu} 
+ \frac{1}{2} \,\varepsilon_{\mu\nu\eta\kappa\rho\sigma}\, \partial^\kappa K^{\rho\sigma} = 0,
\qquad \partial_\mu {\cal K}^{\mu\nu} = 0, \nonumber\\
&&  \Box A_{\mu\nu\eta} = 0,\qquad \Box {\cal K}_{\mu\nu} = 0, \qquad \Box K_{\mu\nu} = 0,
\end{eqnarray}
Furthermore, the equations of motion, that are derived from the coupled Lagrangian density 
${\cal L}_{(\phi, \bar K)}$, are same as (45) except the following equations:
\begin {eqnarray}
&&\bar {\cal K}_{\mu\nu} = - \frac{1}{4!} \; \varepsilon_{\mu\nu\eta\kappa\rho\sigma} H^{\eta\kappa\rho\sigma} 
+ \frac{1}{2} \big[\partial_\mu \phi_\nu^{(2)} - \partial_\nu \phi_\mu^{(2)}\big], \qquad
\partial_\mu \bar {\cal K}^{\mu\nu} = 0 , \nonumber\\
&&\bar K_{\mu\nu} = - \partial^{\eta} A_{\eta\mu\nu}
+ \frac{1}{2} \big[\partial_\mu \phi^{(1)}_\nu - \partial_\nu\phi^{(1)}_\mu\big], \qquad\qquad \Box \bar {\cal K}_{\mu\nu} = 0,\nonumber\\
&&\partial_\mu \bar K_{\nu\eta} + \partial_\nu \bar K_{\eta\mu} + \partial_\eta \bar K_{\mu\nu} 
+ \frac{1}{2}\; \varepsilon_{\mu\nu\eta\kappa\rho\sigma} 
\,\partial^{\kappa} \bar {\cal K}^{\rho\sigma} = 0,  \qquad \partial_\mu \bar K^{\mu\nu} = 0,\nonumber\\ 
&&\partial_\mu \bar {\cal K}_{\nu\eta} + \partial_\nu \bar {\cal K}_{\eta\mu}
+ \partial_\eta \bar {\cal K}_{\mu\nu} 
+ \frac{1}{2} \;\varepsilon_{\mu\nu\eta\kappa\rho\sigma} \;\partial^{\kappa}  \bar K^{\rho\sigma} = 0, 
\quad \Box \bar K_{\mu\nu} = 0.
\end{eqnarray}
We infer from the above equations that we have the following CF-type of restrictions:
\begin{eqnarray}
K_{\mu\nu} + \bar K_{\mu\nu} = \partial_\mu \phi^{(1)}_\nu - \partial_\nu \phi^{(1)}_\mu, \qquad
{\cal K}_{\mu\nu} + \bar {\cal K}_{\mu\nu} = \partial_\mu \phi^{(2)}_\nu - \partial_\nu \phi^{(2)}_\mu.
\end{eqnarray}
The above CF-type of conditions are responsible for the equivalence of the Lagrangian density 
(43) and (44) which can be {\it checked explicitly} (modulo some total spacetime derivatives). 
This is the reason that we call these Lagrangian densities as the coupled and equivalent 
Lagrangian densities for our present 6D Abelian 3-form gauge theory [as they are equivalent  due to (47)].

The discrete symmetry transformations (42) can be further generalized for the coupled 
Lagrangian densities (43) and (44) as given below:
\begin{eqnarray} 
&&A_{\mu\nu\eta} \;\longrightarrow \; \pm \;\frac {i}{3!}\; \varepsilon_{\mu\nu\eta\kappa\rho\sigma}\; 
A^{\kappa\rho\sigma}, \qquad\qquad
K_{\mu\nu}\; \longrightarrow \;\pm \;i\;{\cal K}_{\mu\nu}, \nonumber\\ 
&& {\cal K}_{\mu\nu}\; \longrightarrow \;\pm \;i\;K_{\mu\nu}, \quad
\phi_\mu^{(1)}\; \longrightarrow \; \pm\;i\; \phi_\mu^{(2)}, 
\;\quad \phi_\mu^{(2)} \;\longrightarrow\; \pm\;i\; \phi_\mu^{(1)},
\end{eqnarray}
\begin{eqnarray} 
&&A_{\mu\nu\eta} \;\longrightarrow \; \pm \;\frac {i}{3!}\; \varepsilon_{\mu\nu\eta\kappa\rho\sigma}\; 
A^{\kappa\rho\sigma}, \qquad\qquad \bar K_{\mu\nu}\; \longrightarrow  \;\pm \;i\;\bar {\cal K}_{\mu\nu}, \nonumber\\ 
&&\bar {\cal K}_{\mu\nu}\; \longrightarrow \;\pm \;i\;\bar K_{\mu\nu}, 
\quad \phi_\mu^{(1)} \;\longrightarrow\; \pm\;i\; \phi_\mu^{(2)}, 
\;\quad \phi_\mu^{(2)} \;\longrightarrow\; \pm\;i\; \phi_\mu^{(1)}.
\end{eqnarray}
The above transformations are the {\it symmetry} transformations for the Lagrangian
densities (43) and (44).
Under the following continuous, local and infinitesimal  (dual-)gauge transformations
\begin{eqnarray}
&&\delta_{dg} A_{\mu\nu\eta} = \frac {1}{2!}\; \varepsilon_{\mu\nu\eta\kappa\rho\sigma}\; 
\partial^\kappa \Sigma^{\rho\sigma}, \qquad 
\delta_{dg} \phi_\mu^{(2)} = \partial^\eta \Sigma_{\eta\mu} + \partial_\mu \chi, \nonumber\\ 
&& \delta_{dg} \big[\phi_\mu^{(1)},\; \bar {\cal K}_{\mu\nu},\;
 {\cal K}_{\mu\nu},\; K_{\mu\nu}, \; \bar K_{\mu\nu},\; (\partial^\eta A_{\eta\mu\nu})\big] = 0,  \nonumber\\
&&\nonumber\\ 
&&\delta_{g} A_{\mu\nu\eta} = \partial_\mu \Omega_{\nu\eta} + \partial_\nu \Omega_{\eta\mu}
+\partial_\eta \Omega_{\mu\nu}, \qquad \delta_{g} \phi_\mu^{(1)} \;
= \;\partial^\eta \Omega_{\eta\mu} \;+\;\partial_\mu  \zeta, \nonumber\\
&& \delta_g \big[\phi_\mu^{(2)},\; \bar K_{\mu\nu},\; K_{\mu\nu},\; {\cal K}_{\mu\nu},\;  \bar {\cal K}_{\mu\nu},
\; H_{\mu\nu\eta\kappa},\big] = 0,
\end{eqnarray}
the coupled Lagrangian densities (43) and (44) transform as:
\begin{eqnarray}
\delta_{dg} {\cal L}_{(\phi, K)} &=& - {\cal K}^{\mu\nu}\Big[\Box \Sigma_{\mu\nu} 
+ \frac {1}{2}\,\partial_\mu (\partial^\eta \Sigma_{\nu\eta}) 
- \frac {1}{2}\,\partial_\nu (\partial^\eta \Sigma_{\mu\eta}) \Big], \nonumber\\
\delta_g {\cal L}_{(\phi, K)} &=& K^{\mu\nu}\Big[\Box \Omega_{\mu\nu} 
+ \frac {1}{2}\, \partial_\mu (\partial^\eta \Omega_{\nu\eta})
- \frac {1}{2}\, \partial_\nu (\partial^\eta \Omega_{\mu\eta}) \Big],\nonumber\\
\delta_{dg} {\cal L}_{(\phi, \bar K)} &=&  \bar{\cal K}^{\mu\nu}\Big[\Box \Sigma_{\mu\nu} 
+ \frac {3}{2}\,\partial_\mu (\partial^\eta \Sigma_{\nu\eta})
- \frac {3}{2}\, \partial_\nu (\partial^\eta \Sigma_{\mu\eta}) \Big], \nonumber\\
\delta_g {\cal L}_{(\phi, \bar K)} &=& - \bar K^{\mu\nu}\Big[\Box \Omega_{\mu\nu} 
+ \frac {3}{2}\,\partial_\mu (\partial^\eta \Omega_{\nu\eta})
- \frac {3}{2}\,\partial_\nu (\partial^\eta \Omega_{\mu\eta}) \Big].
\end{eqnarray}
We note that the 2-forms $F^{(2)} = d \,\Phi^{(1)}$ and ${\cal F}^{(2)} = d \,\tilde \Phi^{(1)}$,
present in the Lagrangian densities (43) and (44), permit us to have the vector $U(1)$ gauge transformations 
$\Phi^{(1)} \longrightarrow  \Phi^{'(1)} = \Phi^{(1)} + d \,\zeta^{(0)}$ and 
$\tilde\Phi^{(1)} \longrightarrow  \tilde \Phi^{'(1)} = \tilde\Phi^{(1)} + d \,\chi ^{(0)}$. 
Thus, in the (dual-)gauge transformations (50), we have included the (dual-)gauge parameters $\chi$ and 
$\zeta$ corresponding to the zero-forms $\chi^{(0)}$ and $\zeta^{(0)}$. 
It can be checked that, under the following conditions:
\begin{eqnarray}
&&\Box \Sigma_{\mu\nu} 
+ \frac {1}{2}\,\partial_\mu (\partial^\eta \Sigma_{\nu\eta}) - \frac {1}{2}\,
\partial_\nu (\partial^\eta \Sigma_{\mu\eta})  = 0,\nonumber\\
&&\Box \Omega_{\mu\nu} 
+ \frac {1}{2}\, \partial_\mu (\partial^\eta \Omega_{\nu\eta}) 
- \frac {1}{2}\, \partial_\nu (\partial^\eta \Omega_{\mu\eta})  = 0, \nonumber\\
&& \Box \Sigma_{\mu\nu} 
+ \frac {3}{2}\,\partial_\mu (\partial^\eta \Sigma_{\nu\eta}) 
- \frac {3}{2}\, \partial_\nu (\partial^\eta \Sigma_{\mu\eta}) = 0, \nonumber\\
&&\Box \Omega_{\mu\nu} 
+ \frac {3}{2}\,\partial_\mu (\partial^\eta \Omega_{\nu\eta}) 
- \frac {3}{2}\,\partial_\nu (\partial^\eta \Omega_{\mu\eta}) = 0,
\end{eqnarray}
the Lagrangian densities ${\cal L}_{(\phi, K)}$ and ${\cal L}_{(\phi, \bar K)}$ remain invariant.
In the next section, we shall
see that these restrictions would {\it not} be required to be imposed
on the theory (from outside) within the framework of BRST formalism.

A close look at the 2D Abelian 1-form gauge theory, 4D Abelian 2-form gauge theory and 6D Abelian 
3-form gauge theory allows  us to generalize our results to an arbitrary Abelian $p$-form
gauge theory. We note that such general $D$-dimensional Abelian theories would have dual-gauge symmetry 
[and corresponding (anti-)dual-BRST symmetries] {\it whenever the condition}: 
$* \,d \big(*\,A^{(p)}\big) \,=\, \delta \,A^{(p)}$
{\it is satisfied}. This happens only when the dimension of the spacetime turns out to be exactly equal to 
$[(2p - 1) +1]$ (i.e. $D = 2p$). 
In a very concise manner, we discuss the self-duality condition for a general Abelian $p$-form gauge 
field in $D= 2p$ dimensions of spacetime for the  gauge-fixed Lagrangian density in our Appendix {\bf D}.
Of course the existence of the discrete symmetries would force us to pick up {\it only} the Feynman gauge.

\section {(Anti-)BRST symmetries: Conserved charges}

The most general forms of the gauge-fixed coupled (but equivalent) Lagrangian densities
(43) and (44) can be obtained by incorporating the Faddeev-Popov ghost terms as  [22] 
\begin {eqnarray}
{\cal L}_{(b)} &=& \frac{1}{2} \;{\cal K}_{\mu\nu}\; {\cal K}^{\mu\nu}
- {\cal K}^{\mu\nu} \bigg(\frac{1}{4!} \; \varepsilon_{\mu\nu\eta\kappa\rho\sigma} H^{\eta\kappa\rho\sigma}
+ \frac{1}{2} \left[\partial_\mu \phi_\nu^{(2)} - \partial_\nu \phi_\mu^{(2)}\right]\bigg) \nonumber\\
&-& \frac{1}{2}\;K^{\mu\nu} K_{\mu\nu} - B B_2 + K^{\mu\nu} \left(\partial^\eta A_{\eta\mu\nu} 
+ \frac{1}{2} \left[\partial_\mu \phi_\nu^{(1)} 
- \partial_\nu \phi_\mu^{(1)}\right ]\right) \nonumber\\
&+& \Bigl(\partial_\mu \bar C_{\nu\eta} + \partial_\nu \bar C_{\eta\mu} 
+ \partial_\eta \bar C_{\mu\nu}\Bigr) \Bigl (\partial^\mu C^{\nu\eta}\Bigr)  
+ \left(\partial \cdot \phi^{(1)} \right)B_1 -\frac{1}{2}\; B_1^2 
 +\frac{1}{2}\;B^2_3\nonumber\\
&-& \left(\partial \cdot \phi^{(2)}\right)B_3  
+ (\partial \cdot\bar \beta) B - (\partial_\mu \bar \beta_\nu 
- \partial_\nu \bar \beta_\mu) (\partial^\mu \beta^\nu)
- (\partial \cdot \beta) B_2 - 2 \bar F^\mu f_\mu \nonumber\\
&+&(\partial_\mu \bar C^{\mu\nu} + \partial^\nu \bar C_1) f_\nu 
-  (\partial_\mu  C^{\mu\nu} + \partial^\nu C_1) \bar F_\nu - \partial_\mu \bar C_2 \partial^\mu C_2,
\end{eqnarray}
\begin {eqnarray}
{\cal L}_{(\bar b)} &=& \frac{1}{2} \;\bar {\cal K}_{\mu\nu}\; \bar {\cal K}^{\mu\nu}
+ \bar {\cal K}^{\mu\nu} \bigg( \frac{1}{4!} \; \varepsilon_{\mu\nu\eta\kappa\rho\sigma}\,
 H^{\eta\kappa\rho\sigma}
- \frac{1}{2} \left[\partial_\mu \phi_\nu^{(2)} - \partial_\nu \phi_\mu^{(2)} \right]\bigg)\nonumber\\  
&-& \frac{1}{2}\;\bar K^{\mu\nu} \bar K_{\mu\nu} - B B_2 
- \bar K^{\mu\nu}\left(\partial^\eta A_{\eta\mu\nu} - \frac{1}{2} \left[\partial_\mu \phi_\nu^{(1)} 
- \partial_\nu \phi_\mu^{(1)}\right ]\right)\nonumber\\ 
&+& \Bigl(\partial_\mu \bar C_{\nu\eta} 
+ \partial_\nu \bar C_{\eta\mu} + \partial_\eta \bar C_{\mu\nu}\Bigr)
\Bigl(\partial^\mu C^{\nu\eta}\Bigr ) + \left(\partial \cdot \phi^{(1)}\right)B_1 
- \frac{1}{2}\; B_1^2 + \frac{1}{2}\;B^2_3\nonumber\\
&-& \left(\partial \cdot \phi^{(2)}\right)B_3   + (\partial \cdot\bar \beta) B - (\partial_\mu \bar \beta_\nu 
- \partial_\nu \bar \beta_\mu) (\partial^\mu \beta^\nu)
-(\partial \cdot \beta) B_2 - 2 \bar f^\mu F_\mu \nonumber\\
&-&  (\partial_\mu \bar C^{\mu\nu} - \partial^\nu \bar C_1) F_\nu
+  (\partial_\mu  C^{\mu\nu} - \partial^\nu C_1) \bar f_\nu  - \partial_\mu \bar C_2 \partial^\mu C_2,
\end{eqnarray} 
where the fermionic antisymmetric tensor (anti-)ghost fields ($\bar C_{\mu\nu} $)$C_{\mu\nu}$ [with ghost number
equal to $(-1)+1$], the bosonic Lorentz vector (anti-)ghost fields ($\bar \beta_\mu$)$\beta_\mu$ [with ghost number
$(-2)+2$], the Lorentz scalar fermionic (anti-)ghost fields ($\bar {C}_2$)$C_2$ 
[with ghost number $(-3)+3$] are required for the validity of unitarity in the theory. Furthermore, we have fermionic
auxiliary (anti-)ghost fields ($\bar F_\mu$)$F_\mu$ and ($\bar f_\mu$)$f_\mu$ in the theory 
together with the (anti-)ghost fields ($\bar C_1$)$C_1$. All these fields have ghost number 
equal to $(-1)+1$. We have auxiliary fields $B, B_1, B_2, B_3$ also in our complete 
theory which are used for the specific  linearizations.

Our present coupled and equivalent Lagrangian densities (53) and (54) differ slightly from such
Lagrangian densities in [29].  This is due to the fact that there are extra pieces in (53) and (54)
that were absent in the corresponding Lagrangian densities in [29].  These terms are 
[$- (\partial \cdot \phi^{(2)}) B_3 + (1/2)B^2_3$], [$- 2 \bar F^\mu f_\mu$] and [$- 2 \bar f^\mu F_\mu$]
that are present in (53) and (54). The term [$- (\partial \cdot \phi^{(2)}) B_3 + (1/2)B^2_3$] is required
for the gauge-fixing of the vector field $\phi_\mu^{(2)}$ and other two terms
[$- 2 \bar F^\mu f_\mu$] and [$- 2 \bar f^\mu F_\mu$] are required so that the CF-type conditions
(47) and (66) (see below) could be derived from (53) and (54). We discuss more about 
these issues in our Appendix {\bf A} and establish  the root-cause of this difference.

We note that, under the following off-shell nilpotent ($s_b^2 = 0$) supersymmetric type
BRST symmetry transformations ($s_b$) (see, e.g., [21,22] for details)
\begin {eqnarray}
&&s_b A_{\mu\nu\eta} = \partial_\mu C_{\nu\eta} + \partial_\nu C_{\eta\mu} + \partial_\eta C_{\mu\nu},
\quad s_b \bar C_{\mu\nu} = K_{\mu\nu}, \quad 
s_b C_{\mu\nu} = \partial_\mu \beta_\nu - \partial_\nu \beta_\mu, \nonumber\\ 
&& s_b \bar K_{\mu\nu} = \partial_\mu f_\nu - \partial_\nu f_\mu,
\quad s_b \bar \beta_\mu = \bar F_\mu, \qquad
s_b \beta_\mu = \partial_\mu C_2, \;\;\qquad s_b F_\mu = - \;\partial_\mu B, \nonumber\\
&&s_b {\bar C}_2 = B_2, \qquad s_b C_1 = - B, \qquad s_b \bar C_1 = B_1, \quad
s_b \phi^{(1)}_\mu = f_\mu, \quad s_b \bar f_\mu = \partial_\mu B_1,\nonumber\\
&& s_b \bigl[ C_2,\; f_\mu,\; {\bar F}_\mu,\; B,
B_1,\; B_2,\;B_3,\; \phi^{(2)}_\mu,\; K_{\mu\nu}, \;{\cal K}_{\mu\nu}, \;\bar {\cal K}_{\mu\nu}, \; 
H_{\mu\nu\eta\kappa} \bigr ] = 0,
\end{eqnarray}
the Lagrangian density ${\cal L}_{(b)}$ transforms to a total spacetime derivative as given below
\begin{eqnarray}
s_b {\cal L}_ {(b)} &=&  \partial_\mu \Bigl [ (\partial^\mu C^{\nu\eta} + \partial^\nu C^{\eta\mu}
+ \partial^\eta C^{\mu\nu}) \; K_{\nu\eta}  
- (\partial^\mu \beta^\nu - \partial^\nu \beta^\mu)\; \bar F_\nu \nonumber\\
&+&   K^{\mu\nu}\; f_\nu + B_1 \;f^\mu + B \;\bar F^\mu - B_2 \;(\partial^\mu C_2) \Bigr ].
\end{eqnarray}
Hence, the action integral $S = \int d^6x \;{\cal L}_{(b)}$ would remain invariant for the physically well-defined
fields of the theory which vanish off at infinity (due to Gauss's divergence theorem).

Like the above BRST symmetry transformations, the Lagrangian density  ${\cal L}_{(\bar b)}$ respects
an off-shell nilpotent ($s_{ab}^2 = 0$) anti-BRST symmetry transformations ($s_{ab}$):
\begin{eqnarray}
&&s_{ab} A_{\mu\nu\eta} = \partial_\mu \bar C_{\nu\eta} + \partial_\nu \bar C_{\eta\mu}
+ \partial_\eta \bar C_{\mu\nu}, \;\; s_{ab} \bar C_{\mu\nu} = \partial_\mu \bar \beta_\nu
- \partial_\nu \bar \beta_\mu,  \;\; s_{ab}  C_{\mu\nu} = \bar K_{\mu\nu}, \nonumber\\
&& s_{ab} K_{\mu\nu} = \partial_\mu \bar f_\nu - \partial_\nu \bar f_\mu, \;\;\quad
 s_{ab}  \beta_\mu =  F_\mu,\; \;\quad
s_{ab} \bar \beta_\mu = \partial_\mu \bar C_2, \;\;\quad s_{ab} \bar F_\mu = - \;\partial_\mu B_2, \nonumber\\
&&s_{ab} C_2 = B, \quad s_{ab} f_\mu = - \partial_\mu B_1, \quad s_{ab} C_1 = - B_1,
\quad s_{ab} \bar C_1 = - B_2, \quad s_{ab} \phi^{(1)}_\mu = \bar f_\mu,\nonumber\\
&& s_{ab} \bigl [ \bar C_2, \;\bar f_\mu, \;F_\mu,\; B,
B_1, \;B_2,\;B_3,\; \phi^{(2)}_\mu, \;\bar K_{\mu\nu}, 
\;{\cal K}_{\mu\nu}, \;\bar {\cal K}_{\mu\nu},\; H_{\mu\nu\eta\kappa}  \bigr ] = 0.
\end{eqnarray}
It can be checked that the Lagrangian density ${\cal L}_{(\bar b)}$ transforms, under the
above transformations (57), to a spacetime
total derivative as given below
\begin{eqnarray}
s_{ab} {\cal L}_{(\bar b)} &=& \partial_\mu \Bigl [- (\partial^\mu {\bar C}^{\nu\eta} 
+ \partial^\nu {\bar C}^{\eta\mu}
+ \partial^\eta {\bar C}^{\mu\nu}) \;\bar K_{\nu\eta} 
+ {\bar K}^{\mu\nu} \;{\bar f}_\nu \nonumber\\
&-& (\partial^\mu {\bar \beta}^\nu -
\partial^\nu {\bar \beta}^\mu ) \;F_\nu 
+ B_1 \;{\bar f}^\mu - B_2 \;F^\mu + B\; (\partial^\mu {\bar C}_2) \Bigr ].
\end{eqnarray}
Thus, the action integral in 6D spacetime $\big(S \;=\; \int d^6x \;$ ${\cal L}_{(\bar b)}\big)$ 
remains invariant for the physically
well-defined fields of the theory which fall off rapidly at infinity.
To be explicit, it is Gauss's divergence theorem that implies that all  fields of the r.h.s. 
of (58) will be evaluated at infinity and, because of the physical arguments, 
these fields would go to zero at infinity.

According to Noether's theorem, the invariance of an action under the continuous symmetry
transformations, leads to the existence of conserved currents. From the action principle, it turns out that 
these Noether currents are conserved because of the validity of Euler-Lagrange equations of motion 
(that also ensue from the least action principle). Ultimately, we have the following
expressions for the conserved currents  $J^\mu_{(ab)}$ and $J^\mu_{(b)}$ 
for the (anti-)BRST symmetry invariance
of the Lagrangian densities ${\cal L}_{(\bar b)}$, 
and ${\cal L}_{(b)}$, namely;
\begin{eqnarray}
J^{\mu}_{(ab)} &=&  \frac {1}{2!}\; \varepsilon^{\mu\nu\eta\kappa\lambda\rho}\;
(\partial_\nu {\bar C}_{\eta\kappa}) \;{\bar{\cal K}}_{\lambda\rho} 
+ {\bar K}^{\mu\nu}\; {\bar f}_\nu 
- (\partial^{\mu} {\bar C}^{\nu\eta} +\partial^{\nu} {\bar C}^{\eta\mu} 
+ \partial^{\eta} {\bar C}^{\mu\nu} ) \;{\bar K}_{\nu\eta} \nonumber\\ &+& 
 (\partial^{\mu}  C^{\nu\eta} + \partial^{\nu} C^{\eta\mu} 
+ \partial^{\eta} C^{\mu\nu})\;(\partial_\nu \bar \beta_\eta  
- \partial_\eta \bar \beta_\nu ) - (\partial^{\mu} \beta^{\nu} 
- \partial^{\nu} \beta^{\mu})\;(\partial_\nu \bar C_2)\nonumber\\
&-&   B_2 \;F^{\mu} + B \;(\partial^{\mu} \bar C_2) - (\partial^{\mu} \bar \beta^{\nu} 
- \partial^{\nu} \bar \beta^{\mu}) \;F_\nu + B_1 \;\bar f^{\mu},
\end{eqnarray}
\begin{eqnarray}
J^{\mu}_{(b)} &=& - \frac {1}{2!} \;\varepsilon^{\mu\nu\eta\kappa\lambda\rho}\; (\partial_\nu C_{\eta\kappa})
\; {\cal K}_{\lambda\rho} + (\partial^{\mu} C^{\nu\eta}
+\partial^{\nu} C^{\eta\mu} + \partial^{\eta} C^{\mu\nu} )\;K_{\nu\eta} + K^{\mu\nu} \;f_\nu  \nonumber\\ 
&-& (\partial^{\mu} \bar C^{\nu\eta} + \partial^{\nu} \bar C^{\eta\mu} 
+ \partial^{\eta} \bar C^{\mu\nu})\;(\partial_\nu \beta_\eta 
- \partial_\eta \beta_\nu )  
- (\partial^{\mu} \bar\beta^{\nu} - \partial^{\nu} \bar\beta^{\mu})\; (\partial_\nu C_2)\nonumber\\
&+&  B_1 \;f^{\mu} - B_2 \;(\partial^{\mu} C_2) - (\partial^{\mu} \beta^{\nu} 
- \partial^{\nu} \beta^{\mu})\;\bar F_\nu + B\;\bar F^{\mu}.
\end{eqnarray}
The conservation law [$\partial_\mu J^\mu _{((a)b)} = 0$] for the (anti-)BRST currents  could be proven by exploiting
the following Euler-Lagrange (E-L) equations of motion from ${\cal L}_{(b)}$, namely;
\begin{eqnarray}
&& K_{\mu\nu} = \partial^\eta A_{\eta\mu\nu} + \frac{1}{2} \left(\partial_\mu \phi_\nu^{(1)} 
- \partial_\nu \phi_\mu^{(1)}\right),
\;\;\quad\quad \Box \phi_\mu^{(1)} 
+ \partial_\mu \Big(\partial \cdot \phi^{(1)}\Big) = 0,  \nonumber\\
&& {\cal K}_{\mu\nu} = \frac{1}{4!}\; \varepsilon_{\mu\nu\eta\kappa\lambda\rho}\; H^{\eta\kappa\lambda\rho} 
+ \frac {1}{2}\;\left(\partial_\mu \phi_\nu^{(2)} - \partial_\nu \phi_\mu^{(2)}\right), \quad \Box B_1 = 0, \quad \Box B_3 = 0, \nonumber\\
&&  \frac {1}{2!} \; \varepsilon_{\mu\nu\eta\kappa\lambda\rho}\; (\partial^\mu K^{\nu\eta}) 
=  \partial _\kappa {\cal K}_{\lambda\rho} + \partial _\lambda {\cal K}_{\rho\kappa}
 + \partial _\rho {\cal K}_{\kappa\lambda}, \quad \partial_\mu K^{\mu\nu} + \partial^\nu B_1 = 0,  \nonumber\\
&&  \frac {1}{2!} \; \varepsilon_{\mu\nu\eta\kappa\lambda\rho}\; (\partial^\mu {\cal K}^{\nu\eta}) 
=  \partial _\kappa  K_{\lambda\rho} + \partial _\lambda  K_{\rho\kappa}
 + \partial _\rho  K_{\kappa\lambda},  \quad \partial_\mu {\cal K}^{\mu\nu} + \partial^\nu B_3 = 0, \nonumber\\
&&\Box \bar C_{\mu\nu} - \frac {3}{2}\left(\partial_\mu \bar F_\nu 
- \partial_\nu \bar F_\mu \right) = 0, \qquad \Box \bar C_1 = 0, \qquad \Box C_2 = 0, \qquad \Box \bar C_2 = 0,   \nonumber\\
&& \Box \phi_\mu^{(2)} + \partial_\mu \Big(\partial \cdot \phi^{(2)}\Big) = 0, \quad \Box C_1 = 0,
\quad \Box \bar \beta_\mu = 0,\quad \Box \beta_\mu = 0, \quad \Box \bar F_\mu = 0, \nonumber\\ &&
 \Box C_{\mu\nu} - \frac {3}{2}\left(\partial_\mu f_\nu - \partial_\nu f_\mu\right) = 0,
\quad B = - (\partial \cdot \beta),  \quad B_2 = (\partial \cdot \bar \beta), \quad \partial \cdot f = 0,\nonumber\\
&&\partial_\mu C^{\mu\nu} + \partial^\nu C_1 - 2 f^\nu = 0, \qquad \partial \cdot \bar F = 0, 
\quad B_3 = \left(\partial \cdot \phi^{(2)}\right), \quad \Box {\cal K}_{\mu\nu} = 0,  \nonumber\\
&&\partial_\mu \bar C^{\mu\nu} + \partial^\nu \bar C_1 - 2 \bar F^\nu = 0, 
\;\; B_1 = \left(\partial \cdot \phi^{(1)}\right),
 \;\; \Box K_{\mu\nu} = 0, \quad \Box f_\mu = 0,  
\end{eqnarray}
and the ones that emerge from the Lagrangian density ${\cal L}_{(\bar b)}$ are:
\begin{eqnarray}
&& \bar K_{\mu\nu} = - \partial^\eta A_{\eta\mu\nu} + \frac{1}{2} \left(\partial_\mu \phi_\nu^{(1)} 
- \partial_\nu \phi_\mu^{(1)}\right), 
\qquad\quad \Box \phi_\mu^{(1)} + \partial_\mu \Big(\partial \cdot \phi^{(1)}\Big) = 0,\nonumber\\
&& \bar {\cal K}_{\mu\nu} = - \frac{1}{4!}\; \varepsilon_{\mu\nu\eta\kappa\lambda\rho}\; H^{\eta\kappa\lambda\rho} 
+ \frac {1}{2}\;\left(\partial_\mu \phi_\nu^{(2)} - \partial_\nu \phi_\mu^{(2)}\right), \quad \Box B_1= 0, \quad \Box B_3 = 0, \nonumber\\
&&  \frac {1}{2!} \; \varepsilon_{\mu\nu\eta\kappa\lambda\rho}\; (\partial^\mu \bar K^{\nu\eta}) 
=  \partial _\kappa \bar {\cal K}_{\lambda\rho} + \partial _\lambda  \bar {\cal K}_{\rho\kappa}
+ \partial _\rho \bar {\cal K}_{\kappa\lambda}, \quad\quad \partial_\mu \bar {\cal K}^{\mu\nu} + \partial^\nu B_3 = 0,\nonumber\\
&&  \frac {1}{2!} \; \varepsilon_{\mu\nu\eta\kappa\lambda\rho}\; (\partial^\mu \bar {\cal K}^{\nu\eta}) 
=  \partial _\kappa  \bar K_{\lambda\rho} + \partial _\lambda  \bar K_{\rho\kappa}
+ \partial _\rho  \bar K_{\kappa\lambda}, 
\quad\quad \partial_\mu \bar K^{\mu\nu} + \partial^\nu B_1 = 0,\nonumber\\
&& \Box \phi_\mu^{(2)} + \partial_\mu \Big(\partial \cdot \phi^{(2)}\Big) = 0,\;
\quad \Box \bar C_1 = 0, \;\quad \Box C_1 = 0,
\;\quad \Box C_2 = 0, \quad \Box \bar C_2 = 0,  \nonumber\\
&& \Box C_{\mu\nu} + \frac {3}{2} \left(\partial_\mu F_\nu - \partial_\nu F_\mu\right) = 0,  
\quad B_2 = (\partial \cdot \bar \beta), \;\quad \Box F_\mu = 0, 
\;\quad  B = - (\partial \cdot \beta),\nonumber\\
&& \Box \bar C_{\mu\nu} + \frac {3}{2}\left(\partial_\mu \bar f_\nu - \partial_\nu \bar f_\mu\right) = 0,
 \; \quad B_1 = \left(\partial \cdot \phi^{(1)}\right), \;\quad
 \Box \beta_\mu = 0, \;\quad \Box \bar \beta_\mu = 0,\nonumber\\
&&\partial_\mu C^{\mu\nu} - \partial^\nu C_1 + 2F^\nu = 0, \qquad 
 \partial \cdot \bar f = 0, \;\quad \Box K_{\mu\nu} = 0, \qquad \Box {\cal K}_{\mu\nu} = 0,\nonumber\\
&&\partial_\mu \bar C^{\mu\nu} - \partial^\nu \bar C_1 + 2 \bar f^\nu = 0, 
\quad \partial \cdot F = 0, \quad \Box \bar f_\mu  = 0, 
\quad B_3 = \left(\partial \cdot \phi^{(2)}\right).
\end{eqnarray}
The above conserved currents $J^\mu_{((a)b)}$ lead to the derivation of the following 
off-shell nilpotent ($Q_{(a)b}^2 = 0$) and
conserved ($\dot Q_{(a)b} =  0$) (anti-)BRST charges $Q_{(a)b}$:
\begin{eqnarray}
Q_{ab} &=& \int d^5 x \; \bigg[\frac {1}{2!} \; \varepsilon^{0ijk lm} \,(\partial_i \bar C_{jk})\, 
{\bar {\cal K}}_{lm} + \bar K^{0 i} \,\bar f_i + B_1 \,\bar f^{0}  
  + B \,{\dot {\bar C}}_2 - B_2 \, F^{0}\nonumber\\ 
&-& (\partial^{0} \bar C^{\nu\eta} + \partial^{\nu} \bar C^{\eta 0} 
+ \partial^{\eta} \bar C^{0 \nu})\,\bar K_{\nu\eta} 
- (\partial^{0} \beta^{i} - \partial^{i} \beta^{0}) \;(\partial_i \bar C_2) \nonumber\\ 
&+& (\partial^{0}  C^{\nu\eta} + \partial^{\nu}  C^{\eta 0} 
+ \partial^{\eta} C^{0 \nu})(\partial_\nu \bar \beta_\eta 
- \partial_\eta \bar \beta_\nu ) 
-(\partial^0 \bar \beta^{i} - \partial^{i} \bar \beta^{0})\; F_i  \bigg ],
\end{eqnarray}
\begin{eqnarray}
Q_b &=& \int d^5 x \; \bigg[- \frac {1}{2!}\, \varepsilon^{0ijklm} (\partial_i C_{jk})\, {\cal K}_{lm} 
 + K^{0i}\, f_i + B_1 \,f^{0} - B_2 \,{\dot C}_2 + B \,\bar F^{0}\nonumber\\
&+&(\partial^{0} C^{\nu\eta} \partial^{\nu} C^{\eta 0} + \partial^{\eta} C^{0 \nu} )\,K_{\nu\eta}  
- (\partial^{0} \bar\beta^{i} - \partial^{i} \bar\beta^{0})\;(\partial_i C_2) \nonumber\\ 
&-& (\partial^{0} \bar C^{\nu\eta} + \partial^{\nu} \bar C^{\eta 0} 
+ \partial^{\eta} \bar C^{0 \nu})(\partial_\nu \beta_\eta 
- \partial_\eta \beta_\nu )  
-(\partial^0 \beta^{i} - \partial^{i} \beta^{0})\;\bar F_i  \bigg ].
\end{eqnarray}
The above conserved charges  $Q_{(a)b}$ are the generators of the continuous and infinitesimal 
(anti-)BRST symmetry transformations, as it
can be checked that
\begin{eqnarray}
s_{(a)b}\;\Psi \;=\; \pm\; i\; \left[\Psi,\; Q_{(a)b}\right]_{\pm},
\end{eqnarray}
where the $(\pm)$ signs, present as the subscripts on the square bracket, correspond to the
(anti)commutators for the generic field $\Psi$ being (fermionic)bosonic in nature. Similarly,
the ($\pm$) signs, in front of the square bracket, are chosen appropriately (see, e.g., [30]).
Furthermore, one has to use the appropriate canonical brackets
(that are derived from the Lagrangian densities ${\cal L}_{(b)}$ and 
${\cal L}_{(\bar b)}$) in the evaluations of the  above (anti)commutators.

We would like to mention that there are {\it four} CF-type restrictions [21,22] on the theory
which have been derived by exploiting the theoretical potential and power of superfield
approach to BRST formalism [31,32]. All of these (that are bosonic 
as well as  fermionic in nature) can also be derived
by exploiting the equations of motion (61) and (62). The fermionic CF-type restrictions
[amongst the fermionic (anti-)ghost fields] were originally derived by exploiting the 
superfield technique [21]. These useful and interesting restrictions are 
\begin{eqnarray}
f_\mu + F_\mu = \partial_\mu C_1, \qquad\quad \bar f_\mu + \bar F_\mu = \partial_\mu \bar C_1.
\end{eqnarray}
Concentrating on the appropriate relationships in (61) and (62), it is evident that these 
Euler-Lagrange equations of motion produce the 
above CF-type restrictions (66).
Exploiting the off-shell nilpotent and absolutely anticommuting
(anti-)BRST symmetry transformations (55) and (57), it can be checked
that the restrictions in (47) and (66) are (anti-) BRST invariant quantities and, hence, they
are physical restrictions on the theory. We re-emphasize that the coupled Lagrangian densities 
(53) and (54) produce all the CF-type restrictions (47) and (66) as a set of off-shoots from the equations of 
motion (61) and (62).

The roles of the (anti-)BRST invariant CF-type restrictions (47) and (66) are two folds.
First, these allow us to obtain a set of two coupled (but equivalent) Lagrangian densities (53) and (54)
for the theory. Second, we observe that the above nilpotent (anti-)BRST symmetry
transformations ($s_{(a)b}$) obey perfect absolute anticommutativity property $\{s_b,\;s_{ab} \} = 0$ for
all the fields of the theory except the following:
\begin{eqnarray}
&&\{s_b,\;s_{ab} \}\;A_{\mu\nu\eta}\; \ne\; 0, \;\quad \{s_b,\;s_{ab} \}\;C_{\mu\nu} \;\ne\; 0, \;\quad\
\{s_b,\;s_{ab} \}\;\bar C_{\mu\nu} \;\ne \;0.
\end{eqnarray}
If we compute the above anticommutators, in a straightforward manner, they turn out to be non-zero.
However, the above fields ($A_{\mu\nu\eta}, C_{\mu\nu}, \bar C_{\mu\nu}$) also respect the absolute anticommutativity
property on the constrained hypersurface (embedded in the 6D spacetime manifold) where the CF-type restrictions
(47) and (66) are valid. For instance, only due to (47), we have $\{s_b, \; s_{ab}\}A_{\mu\nu\eta} = 0$.
In addition, the validity of $\{s_b, \; s_{ab}\}C_{\mu\nu} = 0$ and $\{s_b, \; s_{ab}\}\bar C_{\mu\nu} = 0$
is true only when (66) is satisfied.  We re-emphasize, once again, that restrictions (47) and
(66) have been derived due to the superfield approach to Abelian 3-form gauge 
theory [21] and, to the best of our knowledge, they cannot be derived by using any other method.
However, it is straightforward to note that (47) and (66) can be derived from the equations of motion (61) and 
(62) as well. The interesting point to be emphasized is that the coupled Lagrangian 
densities (53) and (54), which produce (61) and (62), have been derived from the knowledge of $s_{(a)b}$.
The off-shell nilpotent ($s^2_{(a)b} = 0$) and absolutely anticommuting
($s_b s_{ab} + s_{ab} s_b = 0 $) symmetries $s_{(a)b}$, however,  have been derived 
from the superfield formalism (see, e.g., [21] for details). Hence, it is the superfield formalism which
is more {\it basic} as far as the derivation of CF-type restrictions (47) and (66) is concerned.

We wrap up this section with some remarks. First, the absolute anticommutativity of
the (anti-)BRST charges $Q_{(a)b}$ is satisfied only on the constrained hypersurface defined by the field
equations corresponding to the CF-type restrictions (47) and (66). Second, the physicality
criteria: $Q_{(a)b} |phys\rangle = 0$ leads to the annihilation of the physical states by the operator form
of the first-class constraints of the theory (cf. Sect. 7 for details). Third, both the Lagrangian densities 
${\cal L}_{(b)}$ and ${\cal L}_{(\bar b)}$ respect both the off-shell nilpotent (anti-)BRST symmetry
transformations on the hypersurface (in the 6D spacetime manifold) which is described by the 
CF-type field equations. This statement can be succinctly expressed in the 
following mathematical form  [cf. (47), (66)], namely;
\begin{eqnarray}
s_{ab}{\cal L}_{(b)} &=& \partial_\mu \Big[(\partial^\mu \bar C^{\nu\eta} 
+ \partial^\nu \bar C^{\eta\mu} + \partial^\eta \bar C^{\mu\nu})K_{\nu\eta} 
+ A^{\mu\nu\eta}\,(\partial_\nu \bar f_\eta - \partial_\eta \bar f_\nu ) + B_1 \,\bar f^\mu \nonumber\\
&+& (\partial^\mu \bar \beta^\nu - \partial^\nu \bar \beta^\mu )f_\nu 
 +  \bar C^{\mu\nu} \,(\partial_\nu B_1) 
- C^{\mu\nu} \,(\partial_\nu B_2) 
+ B\,(\partial^\mu \bar C_2) -B_2\, F^\mu  \nonumber\\ 
&-& \bar K^{\mu\nu} \,\bar F_\nu  \Big ]
- (\partial^\mu \bar C^{\nu\eta} + \partial^\nu \bar C^{\eta\mu} + \partial^\eta \bar C^{\mu\nu})\;
\partial_\mu \Big[\bar K_{\nu\eta} + K_{\nu\eta} \nonumber\\ 
&-& \left(\partial_\nu \phi_\eta^{(1)} 
- \partial_\eta \phi_\nu^{(1)}\right)\Big]
- \Big[\bar K_{\mu\nu} + K_{\mu\nu} - \left(\partial_\mu \phi_\nu^{(1)} 
- \partial_\nu \phi_\mu^{(1)}\right)\Big]\,(\partial^\mu \bar f^\nu) \nonumber\\
&+& \bar K^{\mu\nu} \, \partial_\mu \Big[\bar f_\nu + \bar F_\nu - \partial_\nu \bar C_1\Big]
-(\partial^\mu \bar \beta^\nu - \partial^\nu \bar \beta^\mu) 
\partial_\mu\Big[f_\nu + F_\nu - \partial_\nu C_1\Big] \nonumber\\
&+& (\partial^\mu B_2) \Big[f_\mu + F_\mu - \partial_\mu C_1\Big]
- (\partial^\mu B_1) \,\Big[\bar f_\mu + \bar F_\mu - \partial_\mu \bar C_1\Big],
\end{eqnarray}
\begin{eqnarray}
s_b{\cal L}_{(\bar b)} &=& - \partial_\mu \Big[(\partial^\mu C^{\nu\eta} 
+ \partial^\nu  C^{\eta\mu} + \partial^\eta  C^{\mu\nu}) \bar K_{\nu\eta} 
+ A^{\mu\nu\eta}\,(\partial_\nu f_\eta - \partial_\eta  f_\nu ) \nonumber\\
&-& (\partial^\mu \beta^\nu - \partial^\nu \beta^\mu ) \bar f_\nu 
+ K^{\mu\nu} \, F_\nu +  \bar C^{\mu\nu} (\partial_\nu B) 
+ C^{\mu\nu} (\partial_\nu B_1)  - B \;\bar F^\mu \nonumber\\
&+& B_2 \;(\partial^\mu  C_2) - B_1\, f^\mu \Big ]
+ (\partial^\mu C^{\nu\eta} + \partial^\nu C^{\eta\mu} + \partial^\eta C^{\mu\nu})\,
\partial_\mu \Big[\bar K_{\nu\eta} +  K_{\nu\eta} \nonumber\\ 
&-& \left(\partial_\nu \phi_\eta^{(1)} 
- \partial_\eta \phi_\nu^{(1)}\right)\Big]
-\Big[\bar K_{\mu\nu} + K_{\mu\nu} - \left(\partial_\mu \phi_\nu^{(1)} 
- \partial_\nu \phi_\mu^{(1)}\right)\Big ]\,(\partial^\mu  f^\nu) \nonumber\\ 
&+& K^{\mu\nu} \; \partial_\mu \Big[f_\nu + F_\nu - \partial_\nu  C_1\Big]
-(\partial^\mu \beta^\nu - \partial^\nu  \beta^\mu) 
\partial_\mu \Big[\bar f_\nu + \bar F_\nu - \partial_\nu \bar C_1\Big] \nonumber\\
&-& (\partial^\mu B_1)\Big[f_\mu + F_\mu - \partial_\mu C_1\Big] 
- (\partial^\mu B) \,\Big[\bar f_\mu + \bar F_\mu - \partial_\mu \bar C_1\Big].
\end{eqnarray} 
Thus, we note that the coupled Lagrangian densities (53) and (54) are equivalent as far as the off-shell nilpotent
(anti-)BRST symmetry transformations on the constrained hypersurface, defined by 
the CF-type field equations (47) and (66), are concerned.
Finally, under the off-shell nilpotent (anti-)BRST
transformations, it is interesting to point out that the curvature tensor 
$H_{\mu\nu\eta\kappa}$ (owing its origin to the exterior derivative) remains invariant.

It appears that the (anti-)BRST symmetries provide a physical realization 
of the exterior derivative. However, the absolute anticommutativity property
($s_b s_{ab} + s_{ab} s_b = 0$) of the nilpotent
(anti-)BRST symmetry transformations imply that {\it only one} of them could be identified with the exterior
derivative of differential geometry because the BRST and anti-BRST symmetry transformations are 
linearly independent of each-other. In fact, as it turns out, it is the BRST symmetry transformations $s_b$
(and corresponding charge $Q_b$) that provide the physical realization of the exterior derivative.
An extensive discussion on these issues could be found in our Sect. 7 [see, equations (98) and (102)]
where we have shown the explicit mappings between the cohomological operators 
of differential geometry and conserved charges of our present gauge theory. 
Of course, there is another physical quantity which also provides a realization 
of the exterior derivative but that happens only in the six ($5 + 1$)-dimensions of spacetime
where we obtain the off-shell nilpotent (anti-)co-BRST symmetry transformations and their corresponding
conserved and nilpotent charges. These nilpotent charges, in fact, provide physical realizations
of the (co-)exterior derivatives of differential geometry (see, Sect. 7).

\section {(Anti-)co-BRST symmetries: Conserved charges}

The Lagrangian density ${\cal L}_{(b)}$ remains quasi-invariant under the
following off-shell nilpotent ($s_d^2 = 0$) co-BRST/dual-BRST symmetry transformations ($s_d$):
\begin{eqnarray}
&& s_d A_{\mu\nu\eta} = \frac{1}{2}\, \varepsilon_{\mu\nu\eta\kappa\rho\sigma} \partial^{\kappa} {\bar C}^{\rho\sigma}, 
\;\;\quad s_d {\bar C}_{\mu\nu} = \partial_\mu {\bar \beta}_\nu - \partial_\nu {\bar \beta}_\mu,\;\;\quad
 s_d {\bar \beta}_\mu = \partial_\mu {\bar C}_2,  \nonumber\\
&& s_d {\bar C}_1 =  B_2, \quad s_d \beta_\mu = - f_\mu, \;\quad
 s_d C_1 =  B_3, \;\quad
s_d \phi_\mu^{(2)} = \bar F_\mu, \;\quad s_d C_2 = B,\nonumber\\
&& s_d C_{\mu\nu} = {\cal K}_{\mu\nu}, \quad s_d \bar f_\mu = \partial_\mu B_2,
\quad s_d F_\mu = \partial_\mu B_3 ,\quad
s_d {\bar {\cal K}}_{\mu\nu} = \partial_\mu \bar F_\nu -  \partial_\nu \bar F_\mu, \nonumber\\
&& s_d [\partial^{\eta} A_{\eta\mu\nu}, \;\phi_\mu^{(1)},\; K_{\mu\nu}, \quad
\bar K_{\mu\nu},\; {\cal K}_{\mu\nu},\;B, \;B_1,\; B_2, \;B_3, \;\bar C_2,\; f_\mu, \;\bar F_\mu] = 0,
\end{eqnarray}
because the Lagrangian density ${\cal L}_{(b)}$ transforms to a total spacetime derivative as
\begin {eqnarray} 
s_d {\cal L}_{(b)} &=& - \;\partial_\mu \Bigl [(\partial^\mu \bar C^{\nu\eta} + \partial^\nu \bar C^{\eta\mu}
 + \partial^\eta \bar C^{\mu\nu})\;{\cal K}_{\nu\eta} 
- B \;(\partial^\mu \bar C_2) \nonumber\\ &-& B_2\; f^\mu  + {\cal K}^{\mu\nu} \;\bar F_\nu   
+ B_3 \; \bar F^\mu - (\partial^\mu \bar\beta^\nu - \partial^\nu \bar\beta^\mu) \;f_\nu \Bigr ].
\end{eqnarray}
One of the decisive features of the dual-BRST symmetry is the invariance of the total
gauge-fixing term [i.e. $s_d (\partial^\mu A_{\mu\nu\eta}) = 0,\; s_d K_{\mu\nu} = 0, 
\;s_d\; \phi_\mu^{(1)}= 0$] which owes its origin to the co-exterior derivative $\delta = - *\;d\;*$ as we have 
$\delta A^{(3)} = (1/2!)(dx^\mu \wedge dx^\nu)(\partial^\eta A_{\eta\mu\nu})$.

Analogous to the transformations (70), we have the following off-shell nilpotent ($s^2_{ad} = 0$)
anti-co-BRST (or anti-dual-BRST) symmetry in the theory $(s_{ad})$, namely;
\begin {eqnarray}
&&s_{ad} A_{\mu\nu\eta} = \frac{1}{2}\; \varepsilon_{\mu\nu\eta\kappa\rho\sigma} \partial^{\kappa} C^{\rho\sigma}, 
\quad s_{ad} \beta_\mu = -\partial_\mu  C_2,\quad
 s_{ad}  C_{\mu\nu} = - \bigl(\partial_\mu \beta_\nu - \partial_\nu \beta_\mu \bigr),  \nonumber\\
&& s_{ad} {\bar C}_{\mu\nu} = {\bar {\cal K}}_{\mu\nu},\;\;\;
 s_{ad}  C_1 =  - B, \;\;\; s_{ad} {\bar\beta}_\mu = \bar f_\mu, \;\;\; s_{ad} {\bar C}_2 = -B_2, \;\;\;
 s_{ad} {\bar C}_1 = -  B_3, \nonumber\\ 
&&s_{ad} \phi_\mu^{(2)} = F_\mu, \quad s_{ad} \bar F_\mu = - \partial_\mu B_3, \quad
 s_{ad} f_\mu = - \partial_\mu B, \quad  s_{ad}  {\cal K}_{\mu\nu} = \partial_\mu F_\nu 
-  \partial_\nu F_\mu, \nonumber\\
&&s_{ad} [\partial^{\eta} A_{\eta\mu\nu}, \;\phi_\mu^{(1)},\; K_{\mu\nu}, \; {\bar K}_{\mu\nu}, \;
{\bar {\cal K}}_{\mu\nu},\; B,\; B_1, \;B_2,\;
B_3,\; C_2, \;\bar f_\mu, \;F_\mu] = 0,
\end{eqnarray}
under which, once again, the total gauge-fixing term remains invariant 
[i.e. $ s_{ad} \phi_\mu^{(1)}= 0,\; s_{ad} (\partial^\mu A_{\mu\nu\eta}) = 0,\;  
s_{ad} \bar K_{\mu\nu} = 0$] and the Lagrangian density 
${\cal L}_{(\bar b)}$ transforms as follows:
\begin{eqnarray}
s_{ad} {\cal L}_{(\bar b)} &=& \;\partial_\mu \Bigl [(\partial^\mu C^{\nu\eta} + \partial^\nu C^{\eta\mu}
 + \partial^\eta C^{\mu\nu})\;\bar {\cal K}_{\nu\eta} 
- \bar {\cal K}^{\mu\nu}\; F_\nu  + B\;\bar f^\mu \nonumber\\
&+& B_2 \;(\partial^\mu C_2) - B_3 \;F^\mu   
 - (\partial^\mu \beta^\nu - \partial^\nu \beta^\mu)\; \bar f_\nu \Bigr ].
\end{eqnarray}
As a consequence of the above, 
the action integral $S = \int d^6 x \;{\cal L}_{(\bar b)}$ remains invariant for the well-defined
physical fields of the theory that fall off rapidly at infinity.

According to Noether's theorem, the above continuous symmetry invariances lead to
the derivation of the following Noether's conserved currents:
\begin{eqnarray}
J^{\mu}_{(d)} &=& \frac {1}{2!} \,\varepsilon^{\mu\nu\eta\kappa\lambda\rho} 
\,(\partial_\nu \bar C_{\eta\kappa}) \,K_{\lambda\rho} - {\cal K}^{\mu\nu} \,\bar F_\nu 
- (\partial^{\mu} \bar C^{\nu\eta} + \partial^{\nu} \bar C^{\eta\mu} 
+ \partial^{\eta} \bar C^{\mu\nu}) \,{\cal K}_{\nu\eta} \nonumber\\ 
&+& B\, (\partial^{\mu} \bar C_2) + B_2\; f^{\mu} 
+ (\partial^{\mu}  C^{\nu\eta} + \partial^{\nu} C^{\eta\mu} 
+ \partial^{\eta} C^{\mu\nu})\;(\partial_\nu \bar \beta_\eta 
- \partial_\eta \bar \beta_\nu )  \nonumber\\
&+& (\partial^{\mu} \bar \beta^{\nu} - \partial^{\nu} \bar \beta^{\mu}) \;f_\nu 
- B_3 \;\bar F^{\mu} 
- (\partial^{\mu} \beta^{\nu} - \partial^{\nu} \beta^{\mu})\;(\partial_\nu \bar C_2),
\end{eqnarray}
\begin{eqnarray}
J^{\mu}_{(ad)} &=& - \frac {1}{2!}\, \varepsilon^{\mu\nu\eta\kappa\lambda\rho} 
\,(\partial_\nu C_{\eta\kappa})\, \bar K_{\lambda\rho} -{\bar {\cal K}}^{\mu\nu}\, F_\nu  
+ (\partial^{\mu} C^{\nu\eta} + \partial^{\nu} C^{\eta\mu} 
+ \partial^{\eta} C^{\mu\nu})\,{\bar {\cal K}}_{\nu\eta} \nonumber\\ 
&+& B_2 \,(\partial^{\mu} C_2) + B \;\bar f^{\mu}
+ (\partial^{\mu} \bar C^{\nu\eta} 
+ \partial^{\nu} \bar C^{\eta\mu} + \partial^{\eta} \bar C^{\mu\nu})\;(\partial_\nu \beta_\eta 
- \partial_\eta  \beta_\nu )  \nonumber\\
&-& (\partial^{\mu}  \beta^{\nu} - \partial^{\nu} \beta^{\mu}) \;\bar f_\nu 
- B_3 \; F^{\mu}
+ (\partial^{\mu} \bar \beta^{\nu} - \partial^{\nu} \bar \beta^{\mu}) \;(\partial_\nu C_2).
\end{eqnarray}
The conservation law [$ \partial_\mu J^\mu_{(a)d} = 0$] can be proven by exploiting the Euler-Lagrange equations
of motions (61) and (62) that emerge from the Lagrangian densities ${\cal L}_{(b)}$ and ${\cal L}_{(\bar b)}$.
The explicit  expressions for the  conserved charges $(Q_r = \int d^5 x\; J^0_{(r)},\; r = d, ad$), which
are the generators of the (anti-)co-BRST symmetry transformations (72) and (70), are 
\begin{eqnarray}
Q_d &=& \int d^5 x \; \bigg[ \frac {1}{2!} \, \varepsilon^{0ijklm} \,(\partial_i \bar C_{jk})\, K_{lm} 
- (\partial^{0} \bar C^{\nu\eta} + \partial^{\nu} \bar C^{\eta 0} 
+ \partial^{\eta} \bar C^{0\nu} )\;{\cal K}_{\nu\eta} \nonumber\\ 
&-& {\cal K}^{0i} \,\bar F_i + B \;{\dot{\bar C}}_2  
+ (\partial^0  C^{\nu\eta} + \partial^\nu  C^{\eta 0} 
+ \partial^{\eta}  C^{0 \nu})\;(\partial_\nu \bar \beta_\eta 
- \partial_\eta \bar \beta_\nu )  \nonumber\\ 
&-&(\partial^0 \beta^{i} - \partial^{i} \beta^{0})\;(\partial_i \bar C_2) + B_2\; f^{0} 
- B_3 \,\bar F^{0} + (\partial^{0} \bar\beta^{i} - \partial^{i} \bar\beta^{0}) \;f_i \bigg],
\end{eqnarray}
\begin{eqnarray}
Q_{ad} &=& \int d^5 x \;\bigg[ - \frac {1}{2!} \, \varepsilon^{0ijk lm} \,(\partial_i  C_{jk})\, \bar K_{lm} 
+ (\partial^{0}  C^{\nu\eta} + \partial^{\nu}  C^{\eta 0} + \partial^{\eta}  C^{0\nu} )\;{\bar {\cal K}}_{\nu\eta} 
\nonumber\\ &-& \bar {\cal K}^{0i}\,  F_i + B_2 \;{\dot C}_2  
+ (\partial^0  \bar C^{\nu\eta} + \partial^\nu  \bar C^{\eta 0} 
+ \partial^{\eta} \bar C^{0 \nu})\;(\partial_\nu  \beta_\eta 
- \partial_\eta \beta_\nu )  \nonumber\\ 
&+&(\partial^0 \bar \beta^{i} - \partial^{i} \bar \beta^{0})\;(\partial_i C_2) + B \;\bar f^{0}
- B_3 \,F^{0} - (\partial^{0} \beta^{i} - \partial^{i} \beta^{0})\; \bar f_i  \bigg].
\end{eqnarray}
Thus, we conclude that, in addition to the nilpotent (anti-)BRST symmetry transformations, we have another set
of nilpotent (anti-)co-BRST symmetry transformations in the theory. However, it is to be 
emphasized that the off-shell nilpotent (anti-)BRST and (anti-)co-BRST symmetries co-exist together 
for the Abelian 3-form gauge theory {\it only} in six ($5 + 1$)-dimensions of spacetime.  
All these symmetry transformations
are fermionic in nature. The distinguishing feature of these symmetries is the invariance of the {\it total kinetic}
and {\it total gauge-fixing} terms under the (anti-)BRST and (anti-)co-BRST symmetry transformations, respectively.
We would like to add that these fermionic symmetry transformations are the most fundamental symmetries in
our theory (which would turn out to be a model for the Hodge theory as we shall discuss later).

We close this section with the following observations. First of all, it can be noted that 
\begin{eqnarray}
s_{ad} {\cal L}_{(b)} &=& \partial_\mu \bigg[(\partial^\mu \beta^\nu  - \partial^\nu \beta^\mu)\; \bar F_\nu
- C^{\mu\nu} \;(\partial_\nu B_3) + B \;\bar f^\mu - \bar C^{\mu\nu} \;(\partial_\nu B) \nonumber\\ 
&+& \bar {\cal K}^{\mu\nu} \; f_\nu - B_3 \;F^\mu 
- \frac{1}{3} \; \varepsilon^{\mu\nu\eta\kappa\rho\sigma} \; F_\nu (\partial_\eta A_{\kappa\rho\sigma}) 
+ B_2 \;(\partial^\mu C_2) \nonumber\\ 
&-& (\partial^\mu C^{\nu\eta} + \partial^\nu C^{\eta\mu} 
+ \partial^\eta C^{\mu\nu})\; {\cal K}_{\nu\eta} \bigg] 
+ (\partial_\mu B_3)\;\big[f^\mu + F^\mu - \partial^\mu C_1\big]\nonumber\\  
&+& (\partial^\mu C^{\nu\eta} + \partial^\nu C^{\eta\mu} 
+ \partial^\eta C^{\mu \nu} )\;\partial_\mu \Big[{\cal K}_{\nu\eta} 
+ \bar {\cal K}_{\nu\eta} - \left(\partial_\nu \phi_\eta^{(2)}
- \partial_\eta \phi_\nu^{(2)}\right)\Big ] \nonumber\\
&-& \bar {\cal K}^{\mu\nu} \; \partial_\mu \big[f_\nu + F_\nu - \partial_\nu C_1 \big] 
-(\partial_\mu B)\;\big[\bar f^\mu + \bar F^\mu 
- \partial^\mu \bar C_1\big] \nonumber\\
&+& \Bigl[{\cal K}_{\mu\nu} + \bar {\cal K}_{\mu\nu} 
- \left(\partial_\mu \phi^{(2)}_\nu - \partial_\nu \phi^{(2)}_\mu \right)\Bigr ]\;(\partial^\mu F^\nu)
\nonumber\\
&-&(\partial^\mu \beta^\nu - \partial^\nu \beta^\mu)\; \partial_\mu \big[\bar f_\nu 
+ \bar F_\nu - \partial_\nu \bar C_1 \big], 
\end{eqnarray}
\begin{eqnarray}
s_d {\cal L}_{(\bar b)} &=& \partial_\mu \bigg[- (\partial^\mu \bar \beta^\nu  
- \partial^\nu \bar \beta^\mu)\;F_\nu + \bar C^{\mu\nu} \;(\partial_\nu B_3) + B_2 \; f^\mu
- C^{\mu\nu} \;(\partial_\nu B_2) \nonumber\\ 
&+& {\cal K}^{\mu\nu} \; \bar f_\nu - B_3 \; \bar F^\mu 
+ \frac{1}{3} \; \varepsilon^{\mu\nu\eta\kappa\rho\sigma} \; \bar F_\nu (\partial_\eta A_{\kappa\rho\sigma}) 
   + B \;(\partial^\mu \bar C_2) \nonumber\\ 
&+& (\partial^\mu \bar C^{\nu\eta} + \partial^\nu \bar C^{\eta\mu} 
+ \partial^\eta \bar C^{\mu\nu})\; \bar {\cal K}_{\nu\eta} \bigg] 
+ (\partial_\mu B_3)\;[\bar f^\mu + \bar F^\mu - \partial^\mu \bar C_1] \nonumber\\ 
&-& (\partial^\mu \bar C^{\nu\eta} + \partial^\nu \bar C^{\eta\mu} + \partial^\eta \bar C^{\mu\nu}) 
\,\partial_\mu \Big[{\cal K}_{\nu\eta}
+ \bar {\cal K}_ {\nu\eta} - \left(\partial_\nu \phi_\eta^{(2)} - \partial_\eta \phi_\nu^{(2)}\right)\Big] \nonumber\\
&-& {\cal K}^{\mu\nu} \; \partial_\mu \big[\bar f_\nu + \bar F_\nu - \partial_\nu \bar C_1 \big] 
- (\partial_\mu B_2)\;[f^\mu + F^\mu - \partial^\mu C_1] \nonumber\\
&+& \Bigl[{\cal K}_{\mu\nu} + \bar {\cal K}_{\mu\nu} 
- \left(\partial_\mu \phi^{(2)}_\nu - \partial_\nu \phi^{(2)}_\mu \right)\Bigr ]\;(\partial^\mu \bar F^\nu) 
\nonumber\\
&+&(\partial^\mu \bar \beta^\nu - \partial^\nu \bar \beta^\mu)\; \partial_\mu[f_\nu 
+ F_\nu - \partial_\nu C_1],
\end{eqnarray}
which shows that  ${\cal L}_{(b)}$ and ${\cal L}_{(\bar b)}$ are {\it equivalent} because 
both of them respect the off-shell nilpotent 
(anti-)co-BRST symmetry transformations {\it together} on the constrained hypersurface 
[where the field equations (47) and (66) are true]. Furthermore, the off-shell nilpotent (anti-)co-BRST 
symmetry transformations are absolutely anticommuting in nature {\it only} on the hypersurface defined by the 
CF-type field equations. For instance, $\{s_d, s_{ad}\}A_{\mu\nu\eta} = 0, \;
\{s_d, s_{ad}\}C_{\mu\nu} = 0,\; \{s_d, s_{ad}\}\bar C_{\mu\nu} = 0$ only when CF-type restrictions 
(47) and (66) are taken into account. Thus, we conclude that the hallmark of a $p$-form $(p = 1, 2,3,...)$
gauge theory, within the framework of BRST formalism, is the existence of CF-type restrictions. In fact, in our 
earlier works [21,22], we have provided  mathematical basis for the existence  of CF-type
restrictions and their connection with the concept of geometrical object called gerbes. 
Finally, it can be checked that (anti-)co-BRST symmetry transformations (72) and (70) can 
be derived from the analogue of (65) where $Q_{(a)b}$ should be replaced by $Q_{(a)d}$.
Thus, conserved (anti-)co-BRST charges $Q_{(a)d}$ [of (77) and (76)] 
are the generator of transformations (72) and (70).

\section{Bosonic symmetry: Conserved charge}

As has been  pointed out earlier, there are four fermionic (i.e. nilpotent) symmetry transformations
($s_{(a)b},\, s_{(a)d}$)   in the theory. The following anticommutators, between two of the
above fermionic symmetry transformations, define a new bosonic symmetry in the theory. 
These anticommutators (between suitable fermionic symmetries) are as follows:
\begin{eqnarray}
\{s_b, \; s_d\} \;=\; s_\omega, \qquad\qquad \qquad  \{s_{ab}, \; s_{ad}\} \;=\; s_{\bar \omega}.
\end{eqnarray}
Rest of the anticommutators of the fermionic symmetry transformations do not define any new symmetry. 
For instance, we have already seen that $\{s_b, s_{ab}\} = 0,\; \{s_d, s_{ad}\} = 0$ 
on the constrained hypersurface  described by the CF-type field equations. 
In exactly similar fashion, as it turns out, the anticommutators $\{s_b, s_{ad}\}$ $= 0$ and
$\{s_d,  s_{ab}\} = 0$ are also zero modulo a $U(1)$ vector 
gauge transformations (see, e.g., Appendix E below and (90)).
Thus, for all practical purposes, the absolute anticommutativity amongst the fermionic symmetry transformations 
$s_{(a)b}$ and $s_{(a)d}$ is true except the anticommutators in (80). 
The latter anticommutators define an infinitesimal continuous single bosonic symmetry [cf. (87) below] for our present theory.

The bosonic symmetry transformations $s_\omega$ yield the following infinitesimal transformations 
for the relevant fields of the theory, namely;
\begin{eqnarray}
&&s_\omega A_{\mu\nu\eta} = \frac {1}{2}\; \varepsilon_{\mu\nu\eta\kappa\rho\sigma}\,
(\partial^\kappa K^{\rho\sigma}) + \Bigl(\partial_\mu {\cal K}_{\nu\eta} + \partial_\nu {\cal K}_{\eta\mu} 
+ \partial_\eta {\cal K}_{\mu\nu}\Bigr), \nonumber\\
&& s_\omega C_{\mu\nu} = - \;(\partial_\mu f_\nu - \partial_\nu f_\mu), \qquad
s_\omega \bar C_{\mu\nu} = \partial_\mu \bar F_\nu - \partial_\nu \bar F_\mu, \qquad
 s_\omega \bar \beta_\mu  = \partial_\mu B_2, \nonumber\\ 
&& s_\omega \beta_\mu = \partial_\mu B, \qquad 
s_\omega [ B,  B_1,   B_2, B_3, C_1,  \bar C_1, C_2, \bar C_2, \phi_\mu^{(1)}, \phi_\mu^{(2)}, f_\mu, 
\bar f_\mu, F_\mu, \bar F_\mu, \nonumber\\
&& K_{\mu\nu}, \bar K_{\mu\nu},{\cal K}_{\mu\nu}, \bar {\cal K}_{\mu\nu}] = 0. 
\end{eqnarray} 
The above transformations are symmetry transformations because the Lagrangian density ${\cal L}_{(b)}$ 
remains quasi-invariant as it transforms to a total spacetime derivative: 
\begin{eqnarray}
s_\omega {\cal L}_{(b)} &=& \partial_\mu \Bigl[(\partial^\mu {\cal K}^{\nu\eta} + \partial^\nu {\cal K}^{\eta\mu}
+ \partial^\eta {\cal K}^{\mu\nu}) \;K_{\nu\eta} +  B \;(\partial^\mu B_2)   \nonumber\\
&-& (\partial^\mu K^{\nu\eta} + \partial^\nu K^{\eta\mu}
+ \partial^\eta K^{\mu\nu})\;{\cal K}_{\nu\eta} -  B_2\; (\partial^\mu B) \nonumber\\ 
&+& (\partial^\mu f^\nu - \partial^\nu f^\mu)\; \bar F_\nu  
+ (\partial^\mu \bar F^\nu - \partial^\nu \bar F^\mu) \;f_\nu   \Bigr].
\end{eqnarray}   
Similarly, under the infinitesimal bosonic symmetry transformations $(s_{\bar \omega})$:
\begin{eqnarray}
&&s_{\bar \omega} A_{\mu\nu\eta} = \frac {1}{2}\; 
\varepsilon_{\mu\nu\eta\kappa\rho\sigma}\,(\partial^\kappa \bar K^{\rho\sigma}) 
+ \Bigl(\partial_\mu \bar {\cal K}_{\nu\eta} 
+ \partial_\nu \bar {\cal K}_{\eta\mu} + \partial_\eta \bar {\cal K}_{\mu\nu}\Bigr),\nonumber\\
&&s_{\bar \omega} C_{\mu\nu} = - \;(\partial_\mu F_\nu - \partial_\nu F_\mu), \quad
s_{\bar \omega} \bar C_{\mu\nu} = \partial_\mu \bar f_\nu - \partial_\nu \bar f_\mu, \quad
s_{\bar \omega} \bar \beta_\mu  = - \partial_\mu B_2, \nonumber\\
&&s_{\bar\omega} \beta_\mu = - \;\partial_\mu B,  \qquad
s_{\bar \omega} [ B, B_1, B_2, B_3, C_1, \bar C_1,  C_2, \bar C_2, \phi_\mu^{(1)}, \phi_\mu^{(2)}, f_\mu, 
\bar f_\mu, F_\mu, \bar F_\mu,\nonumber\\
&& K_{\mu\nu}, {\bar K}_{\mu\nu}, {\cal K}_{\mu\nu}, {\bar {\cal K}}_{\mu\nu}] = 0,
\end{eqnarray} 
the Lagrangian density ${\cal L}_{(\bar b)}$ transforms to a total spacetime derivative as follows:
\begin{eqnarray}
s_{\bar \omega} {\cal L}_{(\bar b)} &=& - \partial_\mu\Bigl[(\partial^\mu \bar {\cal K}^{\nu\eta} 
+ \partial^\nu \bar {\cal K}^{\eta\mu}
+ \partial^\eta \bar {\cal K}^{\mu\nu}) \;\bar K_{\nu\eta} + B \;(\partial^\mu B_2) \nonumber\\
&-& (\partial^\mu \bar K^{\nu\eta} 
+ \partial^\nu \bar K^{\eta\mu} + \partial^\eta \bar K^{\mu\nu}) \;\bar{\cal K}_{\nu\eta} 
- B_2\; (\partial^\mu B)  \nonumber\\
&+& (\partial^\mu \bar f^\nu - \partial^\nu \bar f^\mu) \;F_\nu  
+ (\partial^\mu  F^\nu - \partial^\nu F^\mu)\; \bar f_\nu  \Bigr].
\end{eqnarray}  
We note [from (82) and (84)] that the equivalent action integrals 
($S_1 = \int d^6x\; {\cal L}_{(b)}$, $S_2 = \int d^6x\; {\cal L}_{(\bar b)}$) remain invariant under the
transformations $s_\omega$ and $s_{\bar \omega}$ for the physical fields of the theory which 
fall off rapidly at infinity (due to the Gauss divergence theorem).

It is to be pointed out that the following transformations are also true, namely;
\begin{eqnarray}
s_\omega {\cal L}_{(\bar b)} &=& \partial_\mu \Bigl[(\partial^\mu K^{\nu\eta} 
+ \partial^\nu  K^{\eta\mu} + \partial^\eta K^{\mu\nu}) \bar {\cal K}_{\nu\eta} + B\;(\partial^\mu B_2) \nonumber\\
&-& (\partial^\mu {\cal K}^{\nu\eta} + \partial^\nu  {\cal K}^{\eta\mu} 
+ \partial^\eta  {\cal K}^{\mu\nu}) \bar K_{\nu\eta} - B_2 \;(\partial^\mu B)  \nonumber\\
&-& (\partial^\mu f^\nu - \partial^\nu f^\mu )\; \bar f_\nu 
- (\partial^\mu \bar F^\nu - \partial^\nu \bar F^\mu )\; F_\nu  \Bigr ] \nonumber\\
&+& \partial_\mu \Big[K_{\nu\eta} + \bar K_{\nu\eta} - \left(\partial_\nu \phi^{(1)}_\eta
- \partial_\eta \phi^{(1)}_\nu \right) \Big](\partial^\mu {\cal K}^{\nu\eta} 
+ \partial^\nu {\cal K}^{\eta\mu} + \partial^\eta {\cal K}^{\mu\nu}) \nonumber\\
&-& \partial_\mu \Big[\bar f_\nu + \bar F_\nu - \partial_\nu \bar C_1 \Big](\partial^\mu f^\nu
- \partial^\nu f^\mu),
\end{eqnarray}
\begin{eqnarray}
s_{\bar \omega} {\cal L}_{(b)} &=& \partial_\mu \Bigl[(\partial^\mu \bar {\cal K}^{\nu\eta} 
+ \partial^\nu \bar {\cal K}^{\eta\mu} + \partial^\eta \bar {\cal K}^{\mu\nu}) K_{\nu\eta} 
+ B_2 \;(\partial^\mu B) \nonumber\\
&-& (\partial^\mu \bar K^{\nu\eta} 
+ \partial^\nu \bar K^{\eta\mu} + \partial^\eta \bar K^{\mu\nu}) {\cal K}_{\nu\eta} 
- B \;(\partial^\mu B_2) \nonumber\\
&+& (\partial^\mu \bar f^\nu - \partial^\nu \bar f^\mu )\; f_\nu 
+ (\partial^\mu F^\nu - \partial^\nu F^\mu )\; \bar F_\nu  \Bigr ] \nonumber\\ 
&-& \partial_\mu  \Big[{\cal K}_{\nu\eta} + \bar {\cal K}_{\nu\eta} 
- \left(\partial_\nu \phi^{(2)}_\eta -  \partial_\eta \phi^{(2)}_\nu \right) \Big] (\partial^\mu K^{\nu\eta}
+ \partial^\nu K^{\eta\mu} + \partial^\eta K^{\mu\nu})\nonumber\\
&-& \partial_\mu \Big[\bar f_\nu + \bar F_\nu 
- \partial_\nu \bar C_1\Big](\partial^\mu f^\nu - \partial^\nu f^\mu).
\end{eqnarray}
Thus, both the coupled Lagrangian densities  ${\cal L}_{(b)} $ and ${\cal L}_{(\bar b)}$ respect the 
bosonic symmetries $s_\omega$ and $s_{\bar \omega}$ on the constrained hypersurface defined by the CF-type
field equations (47) and (66). Mention can {\it also} be made of the key observation that, even though
$s_\omega$ and $s_{\bar \omega}$ look quite different [cf. (81), (83)], they are actually equivalent 
on the constrained hypersurface where the CF-type restrictions (47) and (66) are true.
In fact, as it turns out,  we have 
\begin{eqnarray}
\qquad s_\omega \;+ \; s_{\bar \omega}  \;=\; 0\;\; \Longrightarrow \;s_\omega \;=\; -\; s_{\bar \omega},
\end{eqnarray}
which shows that {\it only one} of $s_\omega$ and $s_{\bar \omega}$ is an independent bosonic symmetry
transformation in the theory. Thus, henceforth, we shall take $s_\omega$ as the 
{\it only} bosonic symmetry in our present theory.

According to Noether's theorem, the above infinitesimal and continuous symmetry transformation 
leads to the derivation of the following  conserved current 
(w.r.t. the Lagrangian density ${\cal L}_{(b)}$), namely;
\begin{eqnarray}
J^{\mu}_{(\omega)} &=&  \frac {1}{2!} \; \varepsilon^{\mu\nu\eta\kappa\lambda\rho} 
\; (\partial_\nu K_{\eta\kappa})\; K_{\lambda\rho}
- \frac {1}{2!} \;  \varepsilon^{\mu\nu\eta\kappa\lambda\rho} 
\;(\partial_\nu {\cal K}_{\eta\kappa})\;{\cal K}_{\lambda\rho} \nonumber\\ 
&-& (\partial^{\mu} \bar \beta^{\nu} - \partial^{\nu} \bar \beta^{\mu})\;(\partial_{\nu} B) - (\partial^{\mu}  \beta^{\nu} - \partial^{\nu} \beta^{\mu})\;(\partial_{\nu} B_2) \nonumber\\
&-& (\partial^{\mu} \bar C^{\nu\eta} + \partial^{\nu} \bar C^{\eta\mu} 
+ \partial^{\eta} \bar C^{\mu\nu}) \;(\partial_{\nu} f_{\eta} - \partial_{\eta} f_{\nu}) \nonumber\\
&-& (\partial^{\mu} C^{\nu\eta} + \partial^{\nu} C^{\eta\mu} 
+ \partial^{\eta} C^{\mu\nu})\;(\partial_\nu \bar F_\eta 
- \partial_\eta  \bar F_\nu ).
\end{eqnarray}
The conservation law ($\partial_\mu J^{\mu}_{(\omega)}  = 0$)
can be proven by exploiting the equations of motion (61) derived from 
Lagrangian density ${\cal L}_{(b)}$. Exploiting the usual tricks of quantum field theory, it 
can be checked that the exact expression for the conserved charge is
\begin{eqnarray}
Q_{\omega} &=& \int d^5 x \; \bigg[\frac {1}{2!} \; \varepsilon^{0ijklm} \;(\partial_i  K_{jk}) \; K_{lm} 
-\frac {1}{2!} \; \varepsilon^{0ijklm} \;(\partial_i  {\cal K}_{jk}) \;{\cal K}_{lm} \nonumber\\
&-& (\partial^0 \bar \beta^{i} 
- \partial^{i} \bar \beta^{0})\;(\partial_i B) 
- (\partial^{0} \beta^{i} - \partial^{i} \beta^{0}) \;(\partial_i B_2) \nonumber\\ 
&-& (\partial^{0}  \bar C^{\nu\eta} + \partial^{\nu}  \bar C^{\eta 0} + \partial^{\eta}  \bar C^{0\nu})\;
(\partial_\nu f_\eta - \partial_\eta f_\nu )  \nonumber\\ 
&-& (\partial^0   C^{\nu\eta} + \partial^\nu   C^{\eta 0} 
+ \partial^{\eta}  C^{0 \nu})\;(\partial_\nu  \bar F_\eta 
- \partial_\eta \bar F_\nu ) \bigg]. 
\end{eqnarray}
The interesting point, to be stressed, is the origin of the derivation of conserved charge $Q_\omega$
that has emerged out from the bosonic symmetry transformations corresponding to 
$s_\omega$.  The latter is equal to the anticommutator of two nilpotent 
(fermionic) symmetry  transformations. As pointed out earlier, the above nilpotent symmetry 
transformations are the analogue of  the (co-)exterior derivatives of differential geometry. 
As a consequence, the anticommutator of the suitable nilpotent symmetry transformations 
(that is equivalent to a bosonic symmetry transformation)  is 
the analogue of the Laplacian operator of differential geometry. We shall see, in Sect. 7, that the conserved
charge $Q_\omega$  provides an accurate physical realization of the Laplacian operator of differential geometry
from the point of view of the ghost number consideration as well as the specific algebra obeyed by it. 
Finally, it can be verified that, besides being the Casimir operator for the whole algebra,
$Q_\omega$ is also the generator of transformations (81) if we exploit the analogue of (65) appropriately.

Before we wrap up this section, we would like to state that, for the present higher dimensional (i.e. $D = 6$ ) 
and higher-form (i.e. $p = 3$) Abelian gauge theory, we have:
\begin{eqnarray}
&& s_{\omega}^{(1)} \bar \beta_\mu = \{s_b, \; s_{ad}\} \bar \beta_\mu = \partial_\mu (B_1 - B_3), \quad
s_{\omega}^{(1)} \phi^{(1)}_\mu = \{s_b, \; s_{ad}\} \phi^{(1)}_\mu = -  \partial_\mu B, \nonumber\\
&& s_{\omega}^{(1)} \phi^{(2)}_\mu =\{s_b, \; s_{ad}\} \phi^{(2)}_\mu = -  \partial_\mu B, \quad
 s_{\omega}^{(2)} \beta_\mu = \{s_{ab}, \; s_d\} \beta_\mu = \partial_\mu (B_1 + B_3), \nonumber\\
&& s_{\omega}^{(2)} \phi^{(1)}_\mu =\{s_{ab}, \; s_d\} \phi^{(1)}_\mu = \partial_\mu B_2,\quad
 s_{\omega}^{(2)} \phi^{(2)}_\mu = \{s_{ab}, \; s_d\} \phi^{(2)}_\mu = -  \partial_\mu B_2.
\end{eqnarray}
For the rest of the fields of the theory, it can be checked that 
$\{s_b, \, s_{ad}\} = 0, \; \{s_d, \, s_{ab}\}$ $= 0$. As mentioned earlier, we already know that 
$\{s_b, \, s_{ab}\} = 0,\; \{s_d, \, s_{ad}\} = 0$ on the hypersurface defined by the CF-type field equations 
(47) and (66). Thus, we conclude that, out of all the possible anticommutators between $s_{(a)b}$ and
$s_{(a)d}$, only two of them define a bosonic symmetry (i.e. $s_\omega = \{s_b, \, s_d\} 
= - \{s_{ab}, \, s_{ad}\}$) and rest of them have absolute anticommutativity property only up to a $U(1)$
vector gauge transformations. We would like to lay emphasis on the fact that this observation, in the 
context of 6D Abelian 3-form gauge theory, is totally {\it different} from our experiences 
in 2D Abelian 1-form and 4D Abelian 2-form gauge theories (see, e.g., [10--16]) where there exists an absolute 
anticommutativity between the nilpotent transformations $s_{(a)b}$ and
$s_{(a)d}$ except $s_\omega = \{s_b, \, s_d\} 
= - \{s_{ab}, \, s_{ad}\}$ which defines the bosonic symmetry.
Furthermore, it can be checked explicitly that the transformations $\{s_b, \,s_{ad}\}$ and $\{s_d,\, s_{ab}\}$
[cf. (90)] are, even though,  the symmetry transformations for the 
Lagrangian densities ${\cal L}_{(b)}$ and/or ${\cal L}_{(\bar b)}$, these  {\it do not}
commute with $s_g$ [see, e.g., (95) below]. We discuss about these symmetries and connected 
issues in our Appendix {\bf E}. Our {\it final remark} is the fact that one can choose $B = B_1 = B_2 = B_3 = 0$ so that
$s_\omega^{(1)} = s_\omega^{(2)} = 0$ in equation (90). This choice, it may be emphasized, does not
spoil the existence of the fundamental fermionic (i.e. nilpotent) symmetries $s_{(a)b}$ and $s_{(a)d}$ 
which are the analogues of the nilpotent (co-)exterior derivatives of the differential geometry.

\section{Discrete and ghost-scale symmetries}

A close look at the Lagrangian densities (53) and (54) demonstrates that the non-ghost part of ${\cal L}_{(b)}$
and $ {\cal L}_{(\bar b)}$ remains invariant under the following discrete symmetry transformations:
\begin{eqnarray}
&& A_{\mu\nu\eta} \to \pm  \frac{i}{3!}\; \varepsilon_{\mu\nu\eta\kappa\rho\sigma}\; A^{\kappa\rho\sigma}, 
\qquad K_{\mu\nu} \to \pm i {\cal K}_{\mu\nu},\qquad
 {\cal K}_{\mu\nu} \to \pm i K_{\mu\nu}, \nonumber\\ && \bar K_{\mu\nu} \to \pm i \bar {\cal K}_{\mu\nu}, 
\quad  \bar {\cal K}_{\mu\nu} \to \pm i \bar K_{\mu\nu},\quad
\phi^{(1)}_\mu \to \pm i \phi^{(2)}_\mu, 
\quad \phi^{(2)}_\mu \to \pm i \phi^{(1)}_\mu, \nonumber\\ &&
 B_1 \to \pm i B_3, \qquad\qquad B_3 \to \pm i B_1. 
\end{eqnarray}
In exactly similar fashion, the ghost part of the Lagrangian densities 
(53) and (54) respect the following discrete symmetry transformations:
\begin{eqnarray}
&& C_{\mu\nu} \to \pm i \bar C_{\mu\nu}, \quad \bar C_{\mu\nu} \to \pm i C_{\mu\nu}, 
\quad \beta_\mu \to \pm i \bar \beta_\mu,\quad
 \bar \beta_\mu \to \mp i \beta_\mu,\quad C_2 \to \pm i \bar C_2,\nonumber\\ &&
\bar C_2 \to \pm i C_2, \qquad
 C_1 \to \pm i \bar C_1, \qquad 
\bar C_1 \to \pm i C_1, \qquad B \to \mp i B_2, \quad
 B_2 \to \pm i B,\nonumber\\ &&  f_\mu \to \pm i \bar F_\mu, \qquad  \bar F_\mu \to \pm i f_\mu,\quad
 \bar f_\mu \to  \pm i F_\mu, \qquad  F_\mu \to \pm i \bar f_\mu.
\end{eqnarray} 
It is obvious now that the total Lagrangian densities (53) and (54) remain invariant under the combined 
discrete symmetry transformations listed in  (91) and (92).  We shall  see, in the next section, 
that the above discrete symmetry transformations play a key role in providing the physical 
realization of  Hodge duality ($*$) operation of the differential geometry.

It is to be noted that the decisive feature of the discrete symmetry transformations, 
in the gauge sector [cf. (91)] of the Lagrangian densities in the Feynman gauge, 
has been the {\it self-duality} condition on the gauge field [cf. (37)] which is intimately connected with 
the Hodge duality ($*$) operation. This is the reason that the generalization of 
$A_{\mu\nu\eta} \to \pm \frac{i}{3!} \; \varepsilon_{\mu\nu\eta\kappa\rho\sigma}\; A^{\kappa\rho\sigma}$  in
the gauge sector (together with the  discrete symmetry transformations in the ghost sector) provides the 
physical realization of the Hodge duality ($*$) operation of differential geometry. 
In fact, the discrete symmetry transformations (91) and (92) are intimately connected with the 
original discrete symmetry transformation in (38) (which is nothing but the self-duality condition
for the Abelian 3-form gauge field).

The ghost part of the Lagrangian densities (53) and (54), in addition to the discrete
transformations (92), respect a continuous scale
symmetry transformations as 
\begin{eqnarray}
&&C_{\mu\nu} \to e^{+\Lambda} C_{\mu\nu}, \quad \bar C_{\mu\nu} \to e^{- \Lambda} \bar C_{\mu\nu}, 
\quad \beta_\mu \to e^{+2\Lambda}\beta_\mu,\quad
 \bar \beta_\mu \to e^{-2\Lambda} \bar \beta_\mu,\nonumber\\ && C_2 \to e^{+3\Lambda}C_2, \quad
\quad \bar C_2 \to e^{-3\Lambda}\bar C_2, \;\quad
 C_1 \to e^{+\Lambda} C_1, \;\;\quad 
\bar C_1 \to e^{-\Lambda} \bar C_1, \nonumber\\ && B \to e^{+2\Lambda} B, \qquad
 B_2 \to e^{-2\Lambda}B_2, \qquad  f_\mu \to e^{+\Lambda} f_\mu, 
\qquad  \bar f_\mu \to e^{-\Lambda} \bar f_\mu, \nonumber\\
&&  F_\mu \to e^{+\Lambda} F_\mu, \qquad\qquad  
\bar F_\mu \to e^{-\Lambda} \bar F_\mu,
\end{eqnarray}  
where $\Lambda$ is  a global (spacetime independent) parameter and numerals $(\mp1, \mp2, \mp3)$ in the 
exponentials stand for the ghost number of the (anti-)ghost fields. According to the basic 
tenets of  BRST formalism, 
all the rest of the fields (in the gauge sector of the Lagrangian densities) are endowed with 
the ghost number equal to zero.
As a consequence, the ghost-scale symmetry transformations, on the generic field $\Psi$ of this sector, is:
\begin{eqnarray}
\Psi \;\longrightarrow \; \Psi ^{'} \;= \;e^{0 \cdot \Lambda}\; \Psi \;\;\Longrightarrow\;\;  \Psi^{'} \;= \;  \Psi,
\end{eqnarray}  
where the generic field $\Psi$ stands for 
$\Psi = A_{\mu\nu\eta}, \;K_{\mu\nu},$ $ \;{\cal K}_{\mu\nu}, \;\bar K_{\mu\nu}, \; \bar {\cal K}_{\mu\nu},\;
B_1,\; B_3$, $\;\phi^{(1)}_\mu, \;\phi^{(2)}_\mu$. 
Choosing $\Lambda = 1$ in the scale symmetry transformations (93) and (94),
we obtain the following infinitesimal ghost-scale symmetry transformations:
\begin{eqnarray}
&&s_g C_{\mu\nu} = + C_{\mu\nu}, \quad s_g \bar C_{\mu\nu} = - \bar C_{\mu\nu}, 
\qquad s_g \beta_\mu = +2\beta_\mu, \qquad
 s_g \bar \beta_\mu  = -2 \bar \beta_\mu, \nonumber\\ && s_g C_2 = +3C_2, \;\;\qquad s_g \bar C_2 = -3 \bar C_2, \;\;\qquad
 s_g C_1 = + C_1, \;\;\qquad s_g \bar C_1 = - \bar C_1, \nonumber\\ && s_g B =+2 B,\quad
 s_g B_2 = -2 B_2, \quad s_g f_\mu = + f_\mu, \quad  s_g \bar f_\mu = - \bar f_\mu,\quad
 s_g F_\mu = + F_\mu, \nonumber\\ &&  s_g \bar F_\mu = - \bar F_\mu,\quad
s_g [A_{\mu\nu\eta}, \, K_{\mu\nu},\, {\cal K}_{\mu\nu},\, \bar K_{\mu\nu}, \,\bar {\cal K}_{\mu\nu},
\,\phi^{(1)}_\mu, \phi^{(2)}_\mu,\, B_1,\, B_3] = 0, 
\end{eqnarray}  
where $s_g$ is actually the infinitesimal version of the ghost-scale  transformations (93) and (94).
The latter equations are valid for the finite value of $\Lambda$.

According  to Noether's theorem, the continuous symmetry transformations (95) lead to the derivation of the 
Noether ghost conserved current as given below
\begin{eqnarray}
J^{\mu}_{(g)} &=& (\partial^{\mu} \bar C^{\nu\eta} + \partial^{\nu} \bar C^{\eta\mu} 
+ \partial^{\eta} \bar C^{\mu\nu})\; C_{\nu\eta}
+ (\partial^{\mu} C^{\nu\eta}  
+ \partial^{\eta} C^{\mu\nu}+ \partial^{\nu} C^{\eta\mu}) \;\bar C_{\nu\eta}   \nonumber\\
&-& 2 (\partial^{\mu} \bar \beta^{\nu} - \partial^{\nu} \bar \beta^{\mu}) \;\beta_\nu 
+ 2 (\partial^{\mu}  \beta^{\nu} - \partial^{\nu} \beta^{\mu})\; \bar  \beta_\nu 
- 2  B_2 \;\beta^\mu - \bar C ^{\mu\nu}\; f_\nu  \nonumber\\ &-& 2 B \;\bar \beta^\mu - C_1 \;\bar F^\mu 
- \bar C_1 \;f^\mu + 3 C_2 \;(\partial^\mu \bar C_2) + 3 \bar C_2 \;(\partial^\mu C_2) - C^{\mu\nu} \;\bar F_\nu.
\end{eqnarray}
The conservation law $\partial_\mu J^\mu_{(g)} = 0$ can be proven by exploiting the equations of motion
(61) and (62) for the (anti-)ghost fields of our present theory. The conserved current $J^\mu_{(g)}$ leads 
to the derivation of the conserved ghost charge  ($Q_{g} = \int d^5x\; J^0_{(g)}$) as:
\begin{eqnarray}
Q_{g} &=& \int d^5 x \; \Bigl[(\partial^{0} \bar C^{\nu\eta} + \partial^{\nu} \bar C^{\eta 0} 
+ \partial^{\eta} \bar C^{0\nu})\; C_{\nu\eta} +(\partial^{0} C^{\nu\eta}  
+ \partial^{\eta} C^{0\nu}+ \partial^{\nu} C^{\eta 0}) \;\bar C_{\nu\eta}   \nonumber\\
&-& 2 (\partial^{0} \bar \beta^{i} - \partial^{i} \bar \beta^{0})\;\beta_i 
+ 2 (\partial^{0}  \beta^{i} - \partial^{i} \beta^{0}) \;\bar  \beta_i 
- 2  B_2 \;\beta^0 - C ^{0 i} \;\bar F_i \nonumber\\
&-& 2 B \;\bar \beta^0 - C_1 \;\bar F^0 
- \bar C_1\; f^0 + 3 C_2 \;{\dot {\bar C}}_2
+ 3 \bar C_2 \;\dot C_2 - \bar C^{0 i}\; f_i\Bigr].
\end{eqnarray}
The charge $Q_g$ is the generator of infinitesimal transformations (95) if we 
exploit the power and potential of the analogue of relationship (65) by using the canonical 
brackets, that are derived from the Lagrangian density 
${\cal L}_{(b)}$, in the evaluation of the commutators.

\section{Cohomological aspects: Algebraic structures}

We have noted, thus far, that there exist {\it six} continuous symmetries in the theory. Four of them 
(i.e. $s_{(a)b}$, $s_{(a)d}$) are fermionic (nilpotent) in nature and two of them ($s_\omega, s_g$)
are bosonic in nature. We can verify, in a straightforward manner, that the operator form of these 
continuous symmetry transformations satisfy the following algebraic structures, namely;
\begin{eqnarray}
&&s^2_{(a)b} = 0, \qquad s^2_{(a)d} = 0,\qquad  \{ s_b , s_{ab}\} = 0,\qquad
 \{ s_d , s_{ad}\} = 0, \nonumber\\ && \{ s_b , s_{ad}\} = 0, \;\;\quad\qquad  [s_g, s_b] = + s_b, \;\;\quad\qquad
 [s_g, s_{ab}] = - s_{ab}, \nonumber\\ && [s_g, s_d] = - s_d, \;\quad\qquad  [s_g, s_{ad}] = +s_{ad}, \;\quad\qquad
 \{ s_{ab} , s_d\} = 0, \nonumber\\ && \{ s_b , s_d\} = s_\omega = - \{ s_{ab} , s_{ad}\}, \quad
[ s_\omega , s_r ] = 0, \quad r = g, b, ab, d, ad, 
\end{eqnarray}
where we have taken the infinitesimal versions of the continuous symmetry 
transformations (55), (57), (70), (72), (81) and (95).
We also note that we mean by $s^2_b =0$ (in the operator form) as $\frac{1}{2} \{s_b,  s_b\} \Psi  \equiv\; 
\frac{1}{2} (s_b s_b  + s_b s_b) \Psi = 0,\; \{ s_b , s_{ab}\} \Psi \equiv (s_b s_{ab} + s_{ab} s_b) \Psi = 0,
\;[ s_\omega , s_r ]\Psi$ $\equiv (s_\omega s_r - s_r s_\omega) \Psi = 0$, etc., 
where $\Psi$ is the generic field of our present  theory.

A close look at the algebra (98) establishes the fact that this algebra is the analogue of the 
algebra satisfied by the de Rham cohomological operators $(d, \delta , \Delta)$ of differential geometry
as one knows that the standard algebra, obeyed by the exterior derivative $d$, co-exterior derivative
$\delta$ and the Laplacian operator $\Delta = (d + \delta)^2 = (d \delta + \delta d)$, is [17--20]:
\begin{eqnarray}
&&d^2 = \delta^2 = 0, \quad \{d, \delta\} = \Delta =  (d +  \delta)^2, \quad
 [\Delta, d] = 0, \quad [\Delta, \delta] = 0.
\end{eqnarray}
An accurate comparison of equations (98) and (99) establishes the fact the set ($s_b, s_d, s_\omega$)
and ($s_{ab},s_{ad},- s_\omega$) are the analogues of the de Rham cohomological operators 
$(d, \delta , \Delta)$ of differential geometry where there exists a two-to-one mapping because
$(s_{b}, s_{ad}) \longrightarrow \;d,\; (s_{d}, s_{ab})$ $\longrightarrow \delta$ and  
$\{s_{b}, \; s_{d}\} = - \; \{s_{ab},\; s_{ad}\}  = s_\omega \Longrightarrow  \Delta$, at the algebraic level.

Even though, there is a perfect matching between ($d, \,\delta,\, \Delta$) and the 
transformations $s_{(a)d},\; s_{(a)b}$ and $s_\omega$ at the algebraic level, 
there are a couple of points which are missing as far as the perfect analogy is concerned. 
First, as we know, the co-exterior derivative $\delta$ is connected to the exterior derivative $d$ by the relation: 
$\delta = \pm\; *\;d\;*$ where ($*$) is the Hodge duality operation. Thus, we have to provide the physical realization
of ($*$) in the language of symmetry properties. Second, we know that the (co-)exterior derivatives [i.e. $(\delta)d$]
(lower)raise the degree of a form by one when they operate on the latter whereas the action of the Laplacian operator
does not change the degree of a form at all. We should be able to provide the analogy of 
the above observations in the language of symmetry properties if we have to prove that the 
6D Abelian 3-form gauge theory is a perfect field theoretic model for the Hodge theory where 
there exist appropriate physical realizations of the cohomological operators $(d, \delta, \Delta)$ in terms
of the symmetry properties of our present theory.

We address the first of the above issues in the following manner. As stated earlier, the combination of the discrete
symmetry transformations (91) and (92) provide the analogue of the Hodge duality ($*$) operation
because the following relationships:
\begin{eqnarray}
s_d \,\Psi = \pm\, *\, s_b \, * \,\Psi, \qquad\qquad \qquad s_{ad} \,\Psi = \pm\, *\,s_{ab} \, * \, \Psi,
\end{eqnarray}
are true where $\Psi$ is the generic field of the theory and $*$ is the combined 
discrete symmetry transformations. Thus, we note that the  
interplay of the continuous and discrete symmetry transformations of the theory provide the analogue of 
connection between the co-exterior derivative ($\delta$) and exterior derivative ($d$) of differential geometry
(i.e. $\delta = \pm\, *\, d \,*$). The $(\pm)$ signs in the relation $\delta = \pm \,* d \,*$ are determined by the
dimensions of the spacetime manifold and the degree of the forms  that are involved in an inner product 
(in the realm of differential geometry [17--20]). We have to provide the physical origin for the 
$(\pm)$ signs in the above.

The $(\pm)$ signs of equation (100) are decided by a couple of successive  operations of the discrete 
symmetry transformations (91) and (92) on the generic field [i.e. $*(*\Psi) = \pm\, \Psi$] (see, e.g., [20]).
As it turns out, only four fields $\beta_\mu,\, \bar \beta_\mu,\, B, \, B_2$ 
are the ones that  possess $(+)$ sign after 
two successive operations of discrete symmetry transformations. The rest of the fields 
carry ($-$) sign after the above successive operations. Furthermore, the CF-type of restrictions of (47)
and (66) remain invariant under the combined discrete symmetry operations (91) and (92). 
This shows the fact that our present theory is a {\it self-dual} theory where the CF-type
restrictions (which are responsible for the absolute anticommutativity and the existence of 
the coupled Lagrangian densities) are physical and duality-invariant. We would like to 
add that our Lagrangian density ${\cal L}_b$ also remains invariant under two successive
operations of discrete symmetry transformations [i.e. $*(* {\cal L}_b) = +{\cal L}_b$].

We address now the second of the issues that have been 
raised earlier. The analogue of the degree of a form (in differential geometry) is
the ghost number defined as: $i\; Q_g\; |\chi\rangle_n = n \; |\chi\rangle_n$ where $|\chi\rangle_n$ is a non-trivial
state  with ghost number equal to $n$ (in the total quantum Hilbert space of states). It can be checked that 
the following relations are true:
\begin{eqnarray}
&&i\;Q_g\; Q_b|\chi\rangle_n = (n + 1) \;Q_b \;|\chi\rangle_n, \quad
i\;Q_g\; Q_{ad}|\chi\rangle_n = (n + 1) \; Q_{ad} \;|\chi\rangle_n, \nonumber\\
&&i\;Q_g\; Q_d|\chi\rangle_n = (n - 1) \;Q_d \;|\chi\rangle_n, \quad
i\;Q_g\; Q_{ab}|\chi\rangle_n = (n - 1) \; Q_{ab} \;|\chi\rangle_n, \nonumber\\
&&i\;Q_g\; Q_{\omega}|\chi\rangle_n = \;n \;Q_{\omega} \;|\chi\rangle_n, 
\end{eqnarray}
where the conserved charges $Q_{(a)b}, \,Q_{(a)d}, \,Q_{\omega}$ and $Q_{g}$ satisfy the analogue 
of algebraic relations,
that are satisfied by the operator form of the symmetry transformations (98), as
\begin{eqnarray}
&&Q_{(a)b}^2 = 0, \qquad Q_{(a)d}^2 = 0, \qquad \{Q_b,\; Q_{ab}\} = 0,\qquad
\{ Q_d,\; Q_{ad} \} = 0, \nonumber\\ && \{ Q_d,\; Q_{ab}\} = 0, \;\qquad
i\;[Q_{g}, \; Q_{b}] = +\; Q_b, \;\qquad i\;[Q_{g}, \; Q_{ab}] = - \; Q_{ab}, \nonumber\\ &&
i\;[Q_{g}, \; Q_{ad}] = +\; Q_{ad}, \;\;\qquad  i\;[Q_{g}, \; Q_d] = -\; Q_d, \;\;\qquad [ Q_{\omega},\; Q_r ] = 0, \nonumber\\ &&
  \{Q_b,\; Q_d\} = Q_\omega = -\; \{Q_{ad}, \; Q_{ab}\}, 
\qquad (r = b, ab, d, ad, g, {\omega} ).
\end{eqnarray}
Thus, we note that the ghost numbers of $Q_b |\chi\rangle_n,\, Q_d |\chi\rangle_n$ 
and $Q_\omega\, |\chi\rangle_n$
are $(n+1),\; (n-1)$ and $n$, respectively. Similarly, the states $Q_{ad}\, |\chi\rangle_n, \;
Q_{ab}\, |\chi\rangle_n $ and 
$Q_{\omega}\, |\chi\rangle_n$ are endowed with ghost numbers equal to $(n+1),\; (n-1)$ and $n$, respectively.
In our Appendices {\bf B} and {\bf C}, we have given simple proofs of the validity of $Q^2_{(a)b} = 0, 
\{Q_b, \; Q_{ab}\} = 0$, etc., by using 
the concept of conserved charges as the generators for the continuous nilpotent transformations. 
This way, one can also prove the rest of the relations of (102).
Finally, we point out that, if the ghost number of a state is identified with the degree 
of a form, then, we have two-to-one mappings: $(Q_b, Q_{ad}) \to d$, $(Q_{ab}, Q_d) \to \delta$ and
$Q_\omega = \{Q_b, Q_d\} = - \{Q_{ab}, Q_{ad}\} \to \Delta$ between the
conserved charges and the cohomological operators of differential geometry.

There is yet another link that we have  {\it not} been able to establish, so far, within the framework of BRST
formalism in the language of symmetry properties and their corresponding conserved 
charges. This issue is connected with the Hodge decomposition theorem 
which states that, on a manifold without a boundary, one can uniquely decompose a given $n$-form 
($f_n$) as a unique sum of an exact form  ($de_{n-1}$), a co-exact form ($\delta c_{n+1}$) 
and a harmonic form $w_n$ (i.e. $dw_n = 0,\; \delta w_n = 0 \;\Rightarrow \;  \Delta w_n = 0$) as [17--20]
\begin{eqnarray}  
f_n = w_n + d \;e_{n-1} + \delta \;c_{n+1},
\end{eqnarray}
where the de Rham cohomological operators ($d, \delta, \Delta$) of differential geometry play a very decisive 
role (in the above celebrated Hodge decomposition theorem).

The above issue can also be addressed within the framework of BRST formalism in the space of quantum Hilbert space 
of states. In fact, any arbitrary state $|\xi \rangle_n$ with a ghost 
number $n$ (i.e. $i \;Q_g |\xi \rangle_n = n \;|\xi \rangle_n$)
can be decomposed uniquely, in the quantum Hilbert space of states,
in terms of a BRST exact state $ Q_b|\zeta  \rangle_{n-1}$, a co-exact state   
$ Q_d|\kappa  \rangle_{n+1}$ and a harmonic state $|w \rangle_n$ $\big(Q_w |w \rangle_n = 0 
\Rightarrow  Q_b |w \rangle_n = 0, \;Q_d |w \rangle_n = 0\big)$ as 
\begin{eqnarray}  
|\xi \rangle_n = |w \rangle_n + Q_b \; |\zeta  \rangle_{n-1} + Q_d\; |\kappa \rangle_{n+1}.
\end{eqnarray}
Due to two-to-one mapping between the conserved charges  and the cohomological operators:
$(Q_b,\; Q_{ad})\rightarrow d, \;(Q_d,\; Q_{ab})\rightarrow \delta,$ 
$(Q_w, - Q_\omega)\rightarrow \Delta$, it is straightforward to re-express (104) in an alternative way 
(in the total quantum Hilbert space of states) as 
\begin{eqnarray}  
|\xi \rangle_n = |w \rangle_n + Q_{ad}\; |\zeta \rangle_{n-1} + Q_{ab} \; |\kappa \rangle_{n+1},
\end{eqnarray}
where $|w \rangle_n$ is the harmonic state (i.e. the most {\it symmetric} state in the whole theory)
as it is (anti-)BRST as well as (anti-)co-BRST invariant. In other words, it obeys the following:
\begin{eqnarray}  
Q_{a(b)}\; |w \rangle_n = 0,\qquad\qquad\qquad \quad Q_{a(d)} \;|w \rangle_n = 0.
\end{eqnarray} 
Thus, we note that $|w \rangle_n$ (i.e. the harmonic state) can be chosen to be the physical state 
$|phys \rangle$ of the theory which respects all the {\it four} basic symmetries ($s_{(a)b}, s_{(a)d}$). 
We have chosen the same states in (104) and (105) because there is exactly two-to-one
mapping between the conserved charges and de Rham cohomological operators.

Now we dwell a bit on the constraints of the theory. First of all, we know that the original Lagrangian
density $\big({\cal L}_0 = \frac{1}{24} H^{\mu\nu\eta\kappa} H_{\mu\nu\eta\kappa}\big)$ is a singular Lagrangian 
density that supports a set of primary and secondary constraints which are of first-class variety in the language of
Dirac's prescription for the classification scheme [33,34].  
These are $\Pi^{0\mu\nu} = (1/3) H^{00\mu\nu} \approx 0$, and
$\partial_i H^{0ijk} \approx 0$ which finally imply $\Pi^{ij} \equiv (1/3) H^{00ij} \approx 0, \Pi^{0i} =
(1/3) H^{000i} \approx 0$ and $\dot \Pi^{ij} = - \partial_k H^{0ijk} \approx 0$.
Taking the physical state as the harmonic  state (i.e. $|w \rangle = |phys \rangle$), we check that 
the following operator form of the first-class constraints of the original Lagrangian density 
annihilate the physical state     
$|phys \rangle$:
\begin{eqnarray}
&&Q_b\; |phys \rangle = \;0\; \Longrightarrow  \; K^{0i}\; |phys \rangle = 0,
\;\;\qquad K^{ij}\; |phys \rangle = 0, \nonumber\\
&& \varepsilon^{0ijklm} (\partial_k {\cal K}_{lm}) \; |phys \rangle = 0, \nonumber\\
&&Q_d \;|phys \rangle = \;0\; \Longrightarrow  \; {\cal K}^{0i}\; |phys \rangle = 0,
\;\;\qquad {\cal K}^{ij}\; |phys \rangle = 0, \nonumber\\
&& \varepsilon^{0ijklm} (\partial_k K_{lm}) \; |phys \rangle = 0. 
\end{eqnarray}
We note that the above constraint conditions on the physical states are nothing but
$\Pi^{ij}\, |phys\rangle = 0,\; \Pi^{0i}$ $\, |phys\rangle = 0$ and $\dot \Pi^{ij} |phys\rangle = 0$. 
This can be seen from the equations
of motion quoted in (61). It is, furthermore, interesting to note that $Q_d |phys\rangle = 0$  yields the constraints 
on the physical states that are {\it dual} to whatever we obtain from $Q_b |phys\rangle = 0$. This 
statement can be verified  from the discrete symmetry transformations (91) as well.

The other two requirements $Q_{ab} \,|phys \rangle = 0$ and $Q_{ad} \,|phys \rangle = 0$ 
(that emerge from the requirement: $Q_w |phys \rangle = 0$) lead to the same 
restrictions as (107). Thus, we finally note that the primary first-class
constraints $(\Pi ^{ij} \approx  0, \Pi^{0i} \approx 0)$ and the time 
derivative $(\dot\Pi ^{ij} \approx  0)$ annihilate the physical state 
($|phys \rangle$) of the theory which emerge from $Q_b |phys\rangle = 0$. As a consequence, 
our quantization scheme is consistent with the Dirac's prescription for the quantization of 
physical systems with constraints. In our present theory, we also obtain the dual-version 
of the above constraints on the physical states from $\,Q_d\,|phys\rangle = 0$.
These conditions $\,Q_d \,|phys\rangle = 0,\; Q_b \,|phys\rangle = 0$ might force the 6D Abelian 3-form gauge 
theory, to be a model for the q-TFT. We plan to dwell on this issue in the next section.

\section{Physical aspects of nilpotent symmetries: A few observations}

In our earlier works on 2D (non-)Abelian 1-form gauge theories [10--12],  we have 
exploited the {\it on-shell} nilpotent (anti-)BRST and (anti-)co-BRST symmetry 
transformations to prove that these theories are exact models of topological 
field theories (TFTs). We have been able to express the Lagrangian densities 
as well as symmetric energy momentum tensors of these 2D theories as the sum of 
BRST and co-BRST exact terms. Furthermore, we have obtained {\it four} sets of 
topological invariants  $(\bar I_k)I_k$ and $(\bar J_k)J_k$ w.r.t. the {\it off-shell} nilpotent 
(anti-) BRST as well as (anti-)co-BRST transformations which obey proper 
recursion relations (see, e.g., [33] for details). These invariants are succinctly 
expressed, respectively, as
\begin{eqnarray}
\bar I_{k} = {\displaystyle \oint}_{C_{k}} \overline {V}_{k}, 
\;\;\quad I_{k} = {\displaystyle \oint}_{C_{k}} V_{k}, \;\;\quad
\bar J_{k} = {\displaystyle \oint}_{C_{k}} \overline W_{k},\;\;\quad  J_{k} = {\displaystyle \oint}_{C_{k}} W_{k},   
\end{eqnarray}  
where $(\overline V_k)V_k$ and $(\overline W_k)W_k$ are the k-forms $(k = 0, 1, 2)$ and $C_k$ 
are the homology cycles on the 2D  closed Riemann surface (that is to be an Euclidean version 
of 2D non-compact Minkowskian  spacetime manifold) on which the 2D theory is defined. It is essential 
to have the Euclidean version of the 2D non-compact Minkowskian manifold so that the 
topological invariants, homology cycles, etc., could find their proper physical/geometrical meaning [35].

As far as the physical application of our present (anti-)BRST and (anti-)co-BRST symmetry transformations is concerned,
we would like to study, first of all, the nature of Lagrangian density (53) in terms of the following {\it on-shell}
nilpotent symmetries 
\begin{eqnarray}
&& s_b A_{\mu\nu\eta} = \partial_\mu C_{\nu\eta} + \partial_\nu C_{\eta\mu} 
+ \partial_\eta C_{\mu\nu},\quad {s}_b \beta_\mu = \partial_\mu C_2, \;\quad
s_b C_{\mu\nu} = \partial_\mu \beta_\nu
- \partial_\nu \beta_\mu, \nonumber\\ &&
 {s}_b \bar \beta_\mu = \frac {1}{2} \left(\partial^\eta \bar C_{\eta\mu} + \partial_\mu \bar C_1\right),\qquad\qquad
 {s}_b \bar C_{\mu\nu} = \partial^\eta A_{\eta\mu\nu} + \frac{1}{2} \left(\partial_\mu \phi_\nu^{(1)} 
- \partial_\nu \phi_\mu^{(1)}\right ), 
\nonumber\\
&& {s}_b \phi_\mu^{(1)} = \frac {1}{2} \left(\partial^\eta  C_{\eta\mu} 
+ \partial_\mu  C_1\right), \quad\qquad 
{s}_b C_1 = \big(\partial \cdot \beta\big), \quad\qquad
 {s}_b \bar C_2 = \big(\partial \cdot \bar \beta\big),\nonumber\\
&&{s}_b \bar C_1 = \left(\partial \cdot \phi^{(1)}\right), \qquad\qquad
{s}_b \left [C_2,\; \phi_\mu^{(2)},\; H_{\mu\nu\eta\kappa}\right] = 0,
\end{eqnarray}
\begin{eqnarray}
&&{s}_d A_{\mu\nu\eta} = \frac{1}{2}\; \varepsilon_{\mu\nu\eta\kappa\rho\sigma}\; 
\partial^\kappa \bar C^{\rho\sigma}, \;\quad {s}_d \bar \beta_\mu 
= \partial_\mu \bar C_2, \qquad
 {s}_d \bar C_{\mu\nu} = \partial_\mu \bar \beta_\nu
- \partial_\nu \bar \beta_\mu, \nonumber\\ && {s}_d  \beta_\mu = - \frac {1}{2} \left(\partial^\eta  C_{\eta\mu} 
+ \partial_\mu  C_1\right), \qquad{s}_d \bar C_1 = \big(\partial \cdot \bar \beta\big), \qquad
 {s}_d C_1 = \left(\partial \cdot \phi^{(2)}\right), \nonumber\\
&& {s}_d  C_{\mu\nu} 
= \frac{1}{4!} \; \varepsilon_{\mu\nu\eta\kappa\rho\sigma} H^{\eta\kappa\rho\sigma}
 + \frac{1}{2} \left(\partial_\mu \phi_\nu^{(2)} - \partial_\nu \phi_\mu^{(2)}\right), 
\qquad {s}_d C_2 = - \big(\partial \cdot \beta\big),\nonumber\\
&& {s}_d \phi_\mu^{(2)} = \frac {1}{2} \left(\partial^\eta \bar C_{\eta\mu} + \partial_\mu  \bar C_1\right), 
 \quad \qquad {s}_d \left [\bar C_2,\; \phi_\mu ^{(1)},\; \partial^\eta A_{\eta\mu\nu} \right] = 0,
\end{eqnarray}
which are derived from the off-shell nilpotent BRST and co-BRST symmetry transformations [cf. (55), (70)]
by the following substitutions
\begin{eqnarray}
&&K_{\mu\nu} = \partial^\eta A_{\eta\mu\nu} 
+ \frac {1}{2} \left(\partial_\mu \phi_\nu^{(1)} - \partial_\nu \phi_\mu^{(1)}\right), \;\quad B_2 =  \big(\partial \cdot \bar \beta\big), \;\quad
    B_3 = \left(\partial \cdot \phi^{(2)}\right), \nonumber\\
&& {\cal K}_{\mu\nu} = \frac {1}{4!}\; \varepsilon_{\mu\nu\eta\kappa\lambda\rho}\; H^{\eta\kappa\lambda\rho} 
+ \frac {1}{2} \left(\partial_\mu \phi_\nu^{(2)} - \partial_\nu \phi_\mu^{(2)}\right), \quad\qquad
  B_1 = \left(\partial \cdot \phi^{(1)}\right),  \nonumber\\ && f_\mu = \frac{1}{2}\; \big(\partial^\eta C_{\eta\mu} +  \partial_\mu C_1 \big),\quad
\bar F_\mu = \frac{1}{2}\;\big(\partial^\eta \bar C_{\eta\mu} + \partial_\mu \bar C_1 \big), 
\quad B = - \big(\partial \cdot \beta \big).
\end{eqnarray}
The above relations emerge, as the equations of motion, from the Lagrangian density (53). 
We note that $ s_{(b)d}$ are on-shell
$\Big(\Box C_1 =0,\, \Box {\bar C}_1 =0,\, \Box C_2 =0,\, \Box {\bar C}_2 =0,\, \Box \beta_\mu =0,\, 
\Box \bar \beta_\mu = 0,\,
\Box \phi^{(1)}_\mu \,+\, \partial_\mu (\partial \cdot \phi^{(1)})= 0, \, 
\Box \phi^{(2)}_\mu \,+\, \partial_\mu (\partial \cdot \phi^{(2)})= 0,\;
\Box C_{\mu\nu} - (3/4)\big[\partial_\mu \big(\partial^\eta C_{\eta\nu}\big) 
- \partial_\nu \big(\partial^\eta C_{\eta\mu} \big) \big]$ $= 0,\,
\Box \bar C_{\mu\nu} - (3/4)\big[\partial_\mu \big(\partial^\eta \bar C_{\eta\nu}\big) 
- \partial_\nu \big(\partial^\eta \bar C_{\eta\mu} \big) \big]$ $= 0 \Big)$
nilpotent (i.e. ${ s}^2_{(b)d} = 0$). Furthermore, these are  actual 
symmetry transformations for the following Lagrangian density
\begin {eqnarray}
\tilde {\cal L}_{(b)} &=& \frac{1}{2} \bigg[\partial^\eta A_{\eta\mu\nu} 
+ \frac{1}{2} \left(\partial_\mu \phi_\nu^{(1)} 
- \partial_\nu \phi_\mu^{(1)}\right )\bigg] \bigg[\partial_\kappa A^{\kappa\mu\nu} 
+ \frac{1}{2} \left(\partial^\mu \phi^{\nu (1)} 
- \partial^\nu \phi^{\mu (1)}\right )\bigg]\nonumber\\
&-& \frac{1}{2} \left[\frac{1}{4!} \; \varepsilon_{\mu\nu\eta\kappa\rho\sigma} H^{\eta\kappa\rho\sigma}
 + \frac{1}{2} \left(\partial_\mu \phi_\nu^{(2)} - \partial_\nu \phi_\mu^{(2)}\right)\right]
\bigg[\frac{1}{4!} \; \varepsilon^{\mu\nu\alpha\gamma\lambda\zeta} H_{\alpha\gamma\lambda\zeta} \nonumber\\
&+& \frac{1}{2} \left(\partial^\mu \phi^{\nu (2)} - \partial^\nu \phi^{\mu (2)}\right)\bigg]  
+  \Bigl(\partial_\mu \bar C_{\nu\eta} + \partial_\nu \bar C_{\eta\mu} 
+ \partial_\eta \bar C_{\mu\nu}\Bigr) \bigl (\partial^\mu C^{\nu\eta}\bigr) \nonumber\\
&-& \big(\partial \cdot \bar \beta \big)\big(\partial \cdot  \beta \big) 
- \partial_\mu \bar C_2 \,\partial^\mu C_2 + \frac{1}{2} \left (\partial \cdot \phi^{(1)} \right)^2 
- \left(\partial_\mu \bar \beta_\nu - \partial_\nu \bar \beta_\mu \right)\big( \partial^\mu \beta^\nu \big)\nonumber\\
&+& \frac {1}{2} \left(\partial_\mu \bar C^{\mu\nu} + \partial^\nu \bar C_1\right)
\Big( \partial^\eta  C_{\eta\nu} + \partial_\nu  C_1 \Big) 
- \frac{1}{2} \Big(\partial \cdot \phi^{(2)} \Big)^2,
\end{eqnarray}
which is derived from (53) by the appropriate substitutions from (111). For an exact topological field theory, 
it is essential that this Lagrangian density should be able to be expressed 
as the sum of BRST and co-BRST exact terms (especially for a field theoretic model for 
the Hodge theory as is the case with our earlier works on 2D (non)Abelian theories [10]).

We demonstrate that our present model of 6D Abelian 3-form gauge theory is a model 
for a quasi-topological field theory (q-TFT) because one can express (112) as the 
sum of BRST and co-BRST exact terms modulo a total spacetime derivative term plus a 
single {\it extra} term. To corroborate the above statement, it can be checked that, we have the following 
\begin{eqnarray}
\tilde {\cal L}_{(b)} &=& {s}_b\; \big[T_1 + T_2 + T_3 + T_4 + T_5 \big] 
+{s}_d \;\big[P_1 + P_2 + P_3 + P_4 + P_5\big] \nonumber\\ 
&-& \frac{1}{2}\,\big(\partial_\mu \bar C^{\mu\nu}\big)\;\big( \partial^\eta  C_{\eta\nu} \big)  - \partial_\mu Z^\mu.
\end{eqnarray}
Here the exact expressions for $T_i, \;P_i\; (i = 1, 2,...,5)$ and $Z^\mu$ are as listed below
\begin{eqnarray}
&& T_1 = \frac{1}{2} \left [ \partial^\eta A_{\eta\mu\nu} + \frac{1}{2} 
\left(\partial_\mu \phi^{(1)}_\nu - \partial_\nu \phi^{(1)}_\mu \right) \right ]\;{\bar C}^{\mu\nu},\quad
T_2 = \frac{1}{2} \left(\partial\cdot \phi^{(1)} \right) \bar C_1,\nonumber\\ &&
T_3 = - \frac{1}{2}\; \big(\partial\cdot \beta \big)\;\bar C_2,\;\quad
T_4 = - \frac{1}{2} \left(\partial_\mu \bar\beta_\nu \right) C^{\mu\nu},\;\quad
T_5 = - \frac{1}{2} \left(\partial\cdot \bar\beta \right)  C_1, 
\end{eqnarray}
\begin{eqnarray}
&& P_1 = - \frac{1}{2}\bigg[\frac{1}{4!} \varepsilon_{\mu\nu\eta\kappa\rho\sigma} H^{\eta\kappa\rho\sigma}
 + \frac{1}{2} \Big(\partial_\mu \phi_\nu^{(2)} - \partial_\nu \phi_\mu^{(2)} \Big)\bigg] C^{\mu\nu},\;\;
 P_5 =   \frac{1}{2}\, \big(\partial\cdot \beta \big)\,\bar C_1, \nonumber\\ &&
P_3 = \frac{1}{2} \left(\partial\cdot \bar\beta \right) C_2, \;\quad
P_4 = - \frac{1}{2}\, \big(\partial_\mu \beta_\nu \big)\,\bar C^{\mu\nu}, \;\quad
P_2 = - \frac{1}{2} \left(\partial\cdot \phi^{(2)} \right) C_1,
\end{eqnarray}
\begin{eqnarray}
Z^\mu &=& \frac{1}{2}\,\bigg[\Big( \partial^\mu  C^{\nu\eta} + \partial^\nu C^{\eta\mu} 
+ \partial^\eta  C^{\mu\nu}\Big) \,\bar C_{\nu\eta} -\big(\partial^\mu C_2 \big)\, \bar C_2
 - \bar C^{\mu\nu} \big(\partial^\eta  C_{\eta\nu}\big) \quad\nonumber\\
&-& \big(\partial_\eta \bar C^{\eta\mu} + \partial^\mu \bar C_1\big)\, C_1   
- \Big( \partial^\mu \bar C^{\nu\eta} + \partial^\nu \bar C^{\eta\mu} 
+ \partial^\eta \bar C^{\mu\nu}\Big) \, C_{\nu\eta}  \nonumber\\
&-& \big(\partial^\eta \bar C_{\eta\nu} + \partial_\nu \bar C_1\big)\, C^{\mu\nu}
+ \big(\partial^\mu \bar C_2 \big)\, C_2  \bigg].
\end{eqnarray}
Thus, we note that the Lagrangian density (113) looks very much like the model for an exact TFT 
{\it but} for the term $\frac{1}{2}\; \big(\partial_\mu \bar C^{\mu\nu} \big)\big( \partial^\eta  C_{\eta\nu} \big)$.
This is why, we have christened our present theory as a model for the q-TFT because it misses by a 
single term to be an exact model for TFT. The same kind of observation has been made 
in our earlier work on 4D free Abelian 2-form gauge theory where we have proven 
its quasi-topological nature [24] in an explicit manner.

We now focus on the topological invariants like (108).
For our present 6D Abelian 3-form gauge theory, the invariants like (108), are as follows
\begin{eqnarray}
V_0 (+3) &=& \big(B_1 \,C_2 \big), \nonumber\\
V_1(+2) &=& \Bigl[\big(\partial_\mu \bar C_1 \big)\; C_2 + 
B_1 \;\beta_\mu \Bigr ] \,dx^\mu, \nonumber\\
V_2(+1) &=& \left [\big(\partial_\mu {\bar C}_1 \big)\; \beta_\nu 
+ \frac{1}{2!} \;B_1 \;C_{\mu\nu} \right] \big(dx^\mu \wedge dx^\nu \big), \nonumber\\
V_3(0) &=& \bigg [\frac{1}{2!}\;\big(\partial_\mu \bar C_1 \big)
\; C_{\nu\eta} + \frac{1}{3!} \;B_1 \; A_{\mu\nu\eta} \bigg ]
\big(dx^\mu \wedge dx^\nu\wedge dx^\eta \big), \nonumber\\
V_4(-1) &=& \bigg [\frac{1}{3!}\;(\partial_\mu \bar C_1 \big) \; A_{\nu\eta\kappa} 
+ \frac{1}{4!} \; \bar C_1 \; H_{\mu\nu\eta\kappa} \bigg ] 
\big(dx^\mu \wedge dx^\nu\wedge dx^\eta \wedge dx^\kappa \big) \nonumber\\
&\equiv&  d\left[\frac{1}{3!}\;\bar C_1 \; A_{\nu\eta\kappa} \right ] 
\big(dx^\nu\wedge dx^\eta \wedge dx^\kappa \big),\nonumber\\
V_5 (-2) &=& 0, \qquad\qquad \big(d^2 = 0 \big).
\end{eqnarray}
The above invariants are defined (w.r.t. $Q_b$) on the Euclidean version of 
the non-compact 6D Minkowskian 
spacetime manifold where our present theory is considered. 
At this stage, a few useful  comments are in order. First, the zero-form invariant $V_0(+3)$ 
is BRST invariant (i.e. $s_b V_0 = 0$). Second, the numbers ($+3, +2, +1,$ $0, -1$) in 
the round brackets correspond to the ghost numbers that uniquely characterize the 
topological invariants. Third, the invariants terminate at $k= 4$ as, $V_4(-1)$ with 
ghost number ($-1$), turns out to be an exact form. Fourth, the topological invariants 
${\bar I}_k$ (w.r.t. anti-BRST symmetries) can be obtained from $I_k$ ($k = 0, 1, 2, 3, 4, 5$) by the replacements: 
$C_2 \rightarrow \bar C_2,\, C_{\mu\nu} \rightarrow \bar C_{\mu\nu},
\, K_{\mu\nu} \rightarrow \bar K_{\mu\nu},\,
\bar C_{\mu\nu} \rightarrow  C_{\mu\nu},\, A_{\mu\nu\eta} \rightarrow A_{\mu\nu\eta}$ 
and $H_{\mu\nu\eta\kappa} \rightarrow  H_{\mu\nu\eta\kappa}$.
Fifth, it is clear that the ghost numbers for the existing invariants $\bar I_k$ w.r.t. anti-BRST 
symmetries  would be in the order $(-3, -2, -1, 0,$ $+1)$, respectively.  
Finally, as a key signature of the topological properties, the above invariants obey 
the following recursion relations:
\begin{eqnarray}
s_b\, I_k \;= \;d\, I_{k-1}, \qquad s_{ab}\,\bar I_k \;=\; d\, \bar I_{k-1}, \qquad k = 1, 2, 3, 4, 5.
\end{eqnarray}
Thus, it is clear that our present model of 6D Abelian 3-form gauge theory captures 
one of the key features of an exact TFT (see, e.g., [10--12] for details).

Analogous to the topological invariants $(\bar I_k)I_k$ w.r.t. the off-shell nilpotent symmetries and 
corresponding conserved (anti-)BRST charges, one can write down the invariants with respect to the 
(anti-)co-BRST charges. The ones, w.r.t. the co-BRST charge, are 
\begin{eqnarray}
{\overline W}_0 (-3) &=&  \big(B_3\, \bar C_2\big), \nonumber\\
{\overline W}_1 (-2) &=&  \Bigl[\big(\partial_\mu  C_1 \big)\; \bar C_2 
+ B_3 \; \bar\beta_\mu \Bigr ] \;dx^\mu, \nonumber\\
{\overline W}_2 (-1) &=&  \left [\big(\partial_\mu C_1 \big)\; \bar\beta_\nu 
+ \frac{1}{2!} \;B_3\; \bar C_{\mu\nu} \right ]\big(dx^\mu \wedge dx^\nu\big), \nonumber\\
{\overline W}_3 (0) &=&  \bigg[\frac{1}{3!} \;\big(\partial_\mu \bar C_1 \big)\;C_{\nu\eta} 
+ B_3 \left(\frac{1}{3!}\;\varepsilon_{\mu\nu\eta\kappa\rho\sigma} 
\;A^{\kappa\rho\sigma }\right) \bigg ] \big(dx^\mu \wedge dx^\nu\wedge dx^\eta\big), \nonumber\\
{\overline W}_4 (+1) &=&  \bigg [\frac{1}{3!}\;\big(\partial_\mu  C_1 \big)\;
\bigg(\frac{1}{3!}\;\varepsilon_{\nu\eta\kappa\lambda\rho\sigma}\; A^{\lambda\rho\sigma}\bigg) 
\nonumber\\ &+& \frac{1}{3!} B_3 \bigg(\frac{1}{3!}\varepsilon_{\nu\eta\kappa\lambda\rho\sigma}
\partial_\mu A^{\lambda\rho\sigma}\bigg) \bigg]
\big(dx^\mu \wedge dx^\nu\wedge dx^\eta \wedge dx^\kappa \big) \nonumber\\
&\equiv&  d \left[ \frac{1}{3!}\; C_1
\left(\frac{1}{3!}\;\varepsilon_{\nu\eta\kappa\lambda\rho\sigma}\; A^{\lambda\rho\sigma}\right) 
 \right ] \big(dx^\nu\wedge dx^\eta \wedge dx^\kappa \big),\nonumber\\
{\overline W}_5 (+2) &=& 0, \qquad\qquad(d^2 = 0).
\end{eqnarray} 
A few comments are in order, at this juncture. First, it can be checked that $\overline W_0(-3)$
is a co-BRST invariant quantity [i.e. $s_d \overline W_0(-3) = 0$]. 
Second, the {\it five} numbers ($-3, -2, -1, 0, +1$) in the round brackets correspond to 
the ghost numbers which provide the accurate characterization of a specific invariant.
Third, one can obtain the 
k-forms $W_k$ ($k = 0, 1, 2, 3, 4$)  w.r.t. anti-co-BRST charge by the replacements: 
$\bar C_2 \rightarrow  C_2,\, \bar C_{\mu\nu} \rightarrow  C_{\mu\nu},\,
{\cal K}_{\mu\nu} \rightarrow \bar{\cal K}_{\mu\nu},\, 
C_{\mu\nu} \rightarrow  \bar C_{\mu\nu},\, \bar \beta_\mu \rightarrow \beta_\mu$ and
$A_{\mu\nu\eta} \rightarrow A_{\mu\nu\eta}$.
Fourth, the ghost numbers of invariants $W_k$ (with $k = 0, 1, 2, 3, 4, 5$) would be in the 
order $(+3, +2, +1, 0,$ $-1)$. 
Fifth, the forms ${\overline W}_4(+1)$ and $W_4(-1)$ turn out to 
be exact forms. As a consequence, we find that $\overline W_5 = 0, W_5 = 0$ due to $d^2 = 0$.
Finally, the above invariants follow the recurrence relations 
\begin{eqnarray}
s_d {\overline W}_k \;=\; d\, {\overline W}_{k-1}, \qquad
 s_{ad} W_k \;=\; d\, W_{k-1}, \qquad k = 1, 2, 3, 4, 5.
\end{eqnarray}   
We note  that our present theory does capture one of the key features of TFT. We have purposely denoted
the invariants w.r.t. the co-BRST charge with a bar (${\overline W}_k$) because the anti-ghost fields appear
in the co-BRST symmetry transformations [cf. (70), (110)]. As a consequence, the invariants, w.r.t. the 
anti-co-BRST charge, are denoted without a bar. This observation should be contrasted with the 
invariants w.r.t. BRST and anti-BRST charges
where we have taken the opposite convention for the notations of these invariants.

Now we focus on the invariants starting with the ghost number $(+2)$ w.r.t. the BRST charge.
A set of such quantities [that follows the recursion relations (118)] are 
\begin{eqnarray}
V_0 (+2) &=& K^{\mu\nu} \big(\partial_\mu \beta_\nu \big),\nonumber\\
V_1 (+1) &=& \left[\big(\partial_\mu \bar C^{\nu\eta}\big) \big(\partial_\nu \beta_\eta \big) 
+ \frac{1}{2!}\; K^{\nu\eta}\big(\partial_\mu C_{\nu\eta}\big)\right] dx^\mu, \nonumber\\
V_2 (0) &=& \left[\frac{1}{2!}\; \big(\partial_\mu \bar C^{\eta\kappa}\big)
\big(\partial_\nu C_{\eta\kappa}\big)\right] \big(dx^\mu \wedge dx^\nu)\nonumber\\
&\equiv& d \left[\frac{1}{2!}\; \bar C^{\eta\kappa} \big(\partial_\nu C_{\eta\kappa}\big)\right] dx^\nu,\nonumber\\
V_3 (-1) &=& 0, \quad\qquad (d^2 = 0).
\end{eqnarray}
As discussed earlier, we can obtain the invariants starting with ghost number $(-2)$ w.r.t. the conserved 
and nilpotent 
anti-BRST charge by the replacements: $K_{\mu\nu} \to \bar K_{\mu\nu},\, 
C_{\mu\nu} \to \bar C_{\mu\nu},\, \bar C_{\mu\nu} \to C_{\mu\nu}$ and $\beta_\mu \to \bar \beta_\mu.$
To compute the invariants ($\overline W_k$)  (with $k = 0, 1, 2, 3$) w.r.t. the co-BRST charge, we have to replace:
$K_{\mu\nu} \to {\cal K}_{\mu\nu}, \,\beta_\mu \to \bar \beta_\mu, \,C_{\mu\nu}$ $\to \bar C_{\mu\nu}, 
\bar C_{\mu\nu} \to C_{\mu\nu}$ in the above set of invariants. As is evident, these
invariants will be characterized by the ghost numbers $(-2, -1, 0)$. 
These invariants, w.r.t. the nilpotent (anti-)co-BRST charges, obey the recursion relations (120). Similarly, the 
invariants $W_k$ (with $k = 0, 1, 2, 3$), w.r.t. the anti-co-BRST charge, can be obtained from $V_k$ by 
the {\it only} the replacement
$K_{\mu\nu} \to \bar {\cal K}_{\mu\nu}$. As a consequence, these invariants would be characterized  
by the ghost numbers ($+2, +1, 0$). Thus, we conclude that if we know the set of invariants w.r.t. the 
conserved and nilpotent BRST charge, we can obtain all the other invariants w.r.t. anti-BRST and (anti-)co-BRST
charges by exploiting the appropriate discrete symmetry transformations listed in Sect. 6.

We venture, now, to obtain the invariants w.r.t. the nilpotent and conserved BRST charge corresponding to
the ghost number $(+1)$. These invariants, obeying the recursion relations (118), are as follows
\begin{eqnarray}
V_0 (+1) &=& K^{\mu\nu} C_{\mu\nu} - \bar C^{\mu\nu} 
\big(\partial_\mu \beta_\nu - \partial_\nu \beta_\mu \big),\nonumber\\
V_1 (0) &=& \Big[\big(\partial_\mu \bar C^{\nu\eta}\big) C_{\nu\eta} 
+ \bar C^{\nu\eta} \big(\partial_\mu C_{\nu\eta} \big) \Big] dx^\mu \nonumber\\
&\equiv& d \Big[\bar C^{\nu\eta} C_{\nu\eta}\Big],\nonumber\\
V_2 (-1) &=& 0, \quad\qquad (d^2 = 0).
\end{eqnarray}
It is evident, from our previous discussions, that the invariants [starting with the ghost number 
$(-1)$] w.r.t. the anti-BRST charge can be obtained from the above by the replacements: 
$K_{\mu\nu} \to \bar K_{\mu\nu},\, C_{\mu\nu} \to \bar C_{\mu\nu},\, 
\bar C_{\mu\nu} \to  C_{\mu\nu},\, \beta_\mu \to \bar \beta_\mu$. 
These invariants would also obey the recursion relations (118). From the set of invariants listed in (122),
we can obtain all the invariants of theory w.r.t. anti-BRST and (anti-)co-BRST charges by appropriate 
use of discrete symmetry transformation of Sect. 6. This has already been done for the invariants
starting with ghost number ($+2$) from equation (121). Exactly the same substitutions, from 
the discrete symmetry transformations of Sect. 6, have to be exploited here, too, for the 
complete list of invariants.

We make, in the following, some  remarks that are very decisive. These are connected
with some invariants which obey exactly the same recursion relations as (118) and (120)
but they are {\it not} physically interesting on various grounds. 
First, we discuss about the alternative to the BRST invariants 
with the characteristic ghost number $(+1)$. These can be also constructed as follows
\begin{eqnarray}
V_0 (+1) &=& \big(\partial_\mu K^{\mu\nu}\big) f_\nu,\nonumber\\
V_1 (0) &=& \Big[\partial_\mu \big(\partial_\nu \bar C^{\nu\eta}\big) f_\eta 
+ \big(\partial_\nu K^{\nu\eta}\big) \left(\partial _\mu \phi^{(1)}_\eta \right) \Big ] dx^\mu,\nonumber\\
V_2 (-1) &=& \Big[\partial_\mu \big(\partial_\eta \bar C^{\eta\kappa}\big)
\left(\partial _\nu \phi^{(1)}_\kappa \right) \Big ] dx^\mu \wedge dx^\nu \nonumber\\
&\equiv&  d \Big[\big(\partial_\eta \bar C^{\eta\kappa}\big)
\left(\partial _\nu \phi^{(1)}_\kappa \right) \Big ] dx^\nu, \nonumber\\
V_3 (-2) &=& 0, \qquad\quad (d^2 = 0).
\end{eqnarray} 
However, one can check that the mass dimension of $V_0 (+1)$, in the above equation, 
is {\it seven} (i.e. $[V_0 (+1)] = [M]^7$) in the natural units where $\hbar = c = 1$. 
In a 6D theory, an invariant of mass dimension  {\it seven} is {\it not}
allowed. Thus, we do not include (123) in the list of appropriate  invariants. Similarly, a set of BRST invariants
starting with the ghost number ($+2$) can be constructed as 
\begin{eqnarray}
V_0 (+2) &=& \big( \partial \cdot \bar F\big)C_2,\nonumber\\
V_1 (+1) &=& \Big [\partial_\mu \big( \partial \cdot \bar \beta\big)C_2 
- \big(\partial \cdot \bar F\big) \beta_\mu \Big] dx^\mu,\nonumber\\
V_2 (0) &=& \left [ \frac{1}{2!}\,  \big(\partial \cdot \bar F\big) C_{\mu\nu}
- \partial_\mu \big( \partial \cdot \bar \beta\big) \beta_\nu \right] \big(dx^\mu \wedge dx^\nu \big),\nonumber\\
V_3 (-1) &=& \bigg [ \frac{1}{2!} \;\partial_\mu \big(\partial \cdot \bar \beta\big) C_{\nu\eta}
- \frac {1}{3!} \,\big( \partial \cdot \bar F\big) A_{\mu\nu\eta} \bigg] 
\big(dx^\mu \wedge dx^\nu \wedge dx^\eta \big),\nonumber\\
V_4 (-2) &=& - \bigg [ \frac{1}{3!} \;\partial_\mu \big(\partial \cdot \bar \beta\big) A_{\nu\eta\kappa}
+ \frac {1}{4!} \,\big( \partial \cdot \bar \beta\big) H_{\mu\nu\eta\kappa}  \bigg] 
\big(dx^\mu \wedge dx^\nu \wedge dx^\eta \wedge dx^\kappa \big)\nonumber\\
&\equiv & d \left[- \frac{1}{3!}\, \big(\partial \cdot \bar \beta\big)\, A_{\nu\eta\kappa} \right] 
\big(dx^\nu \wedge dx^\eta \wedge dx^\kappa \big),\nonumber\\
V_5 (-3) &=& 0, \qquad\quad (d^2 = 0), 
\end{eqnarray} 
which obey the correct recursion relation. As discussed earlier, other invariants corresponding to (123) and (124)
can also be obtained by exploiting the appropriate discrete symmetry transformations of
Sect. 6.

We note that the mass dimension of $V_0 (+2)$ is six (i.e. $[V_0 (+2)]$ $= [M]^6$) in natural units 
(where $\hbar = c = 1$). Thus, we have {\it two} invariants carrying the  mass-dimension six
[cf. (121), (124)]. However, we discard  (124) as  
physical invariant because only ghost fields are present in 
(124). We prefer invariant $V_0 (+2)$ in (121) as it contains some physical fields because 
$K_{\mu\nu} = \partial^\eta A_{\eta\mu\nu} + \frac{1}{2}[\partial_\mu \phi_\nu^{(1)} 
- \partial_\nu \phi_\mu^{(1)}]$. Physically, all the invariants of this section are ``confined" 
because they form composite with the (anti-)ghost fields of our present theory.
For the sake of completeness, we would like to point out that the following objects
\begin{eqnarray}
&& \sum^k_{i=1} (- 1)^i \, V_i (k - i), \qquad\qquad \sum^k_{i=1} (- 1)^i \, {\overline V}_i (k - i),\nonumber\\
&& \sum^k_{i=1} (- 1)^i \, W_i (k - i), \qquad\qquad \sum^k_{i=1} (- 1)^i \, {\overline W}_i (k - i),
\end{eqnarray} 
define a class of the $k^{th}$-cohomology group w.r.t. BRST, anti-BRST, co-BRST and anti-co-BRST charges.
As a consequence, the above objects carry  some 
topological information about the (super)manifold.

We wrap up this section with a couple of remarks.  First, we have discussed the quasi-topological nature
of 6D Abelian 3-form gauge theory by exploiting the on-shell nilpotent BRST and 
co-BRST symmetry transformations [cf. (113)]. However, one can discuss the above 
properties in terms of the on-shell nilpotent anti-BRST and anti-co-BRST symmetry as well 
(where the Lagrangian density (54) 
would play very important role). Second, it is known in literature that, even if there are
propagating degrees of freedom in a theory, the theory can still capture some of the key features of TFT 
(see, e.g., [36,37] for details). We have shown that, exactly above kind of situation is prevalent 
in the case of 4D Abelian 2-form [24]  as well as 6D Abelian 3-form gauge theories
which are perfect models for the Hodge theory. This observation should be contrasted 
with the 2D Abelian 1-form gauge theory which is a perfect model for a new kind of TFT as well as a 
Hodge theory [10]. In all the above claims, the dual-BRST (i.e. co-BRST) symmetry plays a very significant role.

\section{Conclusions}

In our present endeavor, we have been able to establish that the ``classical"
dual-gauge symmetry transformations would be always associated with any arbitrary 
Abelian $p$-form ($ p = 1, 2, 3, ...$) gauge theory in some specific $D$-dimensions of
spacetime (when $D = 2 p$). We have been able to show concisely that this is true in the cases 
of Abelian 1-form gauge theory in two (1 + 1)-dimensions of spacetime [10--12], Abelian 2-form 
gauge theory in four (3 + 1)-dimensions of spacetime [13--16] and, in our present investigation, 
we have shown the existence of dual-gauge symmetry transformations  for the Abelian 
3-form gauge theory in higher $(D > 4)$ six (5 + 1)-dimensions of spacetime in great 
detail.

 In all the above discussions, we have considered
the gauge-fixed Lagrangian densities in the Feynman gauge {\it only} because the discrete
symmetry transformations (i.e. the analogue of  Hodge duality operator) are respected when this ``gauge"
is chosen. In other words, the most symmetric theory (i.e. the one respecting the discrete as well 
as continuous symmetries) itself chooses the ``Feynman gauge"
 out of all the gauges available in the literatures on quantization of gauge theories.
Thus, we firmly believe that a $D = 2p$ 
dimensional Abelian $p$-form gauge theory would {\it always} be endowed with the dual-gauge 
symmetry transformations at the ``classical'' level in the Feynman gauge which 
can be promoted to the quantum level.

We know  that the {\it usual} gauge symmetries are 
generated by the first-class constraints (in the language of Dirac's prescription [33,34]
for the classification scheme) of any arbitrary $p$-form gauge theory in any
arbitrary $D$-dimensions of spacetime. However, the dual-gauge symmetries exist for any 
arbitrary Abelian $p$-form  gauge theory {\it only} in $ D = 2 p$ dimensions of 
spacetime where the origin for such an existence lies in the {\it self-duality} 
condition [cf. (4), (8), (37)]. In other words, the dual-gauge symmetry exists for the 
Abelian $p$-form gauge theories where there is a self-duality condition which is 
mathematically dictated by the Levi-Civita tensor of the $D = 2p$ dimensions of spacetime. 
There is yet another distinguishing feature that differentiates the gauge- and  
dual-gauge symmetry transformations for an Abelian theory. Whereas the curvature tensor
(owing its origin to the exterior derivative) remains invariant under the continuous 
gauge symmetry transformations, it is the gauge-fixing term (owing its origin to the 
co-exterior derivative) that remains invariant under the continuous dual-gauge 
symmetry transformations.

For the $2p$-dimensional Abelian $p$-form gauge theories, one can generalize the 
above ``classical" dual-gauge symmetry transformations to the ``quantum" level in the 
language of  off-shell nilpotent (anti-)dual BRST [or (anti-)co-BRST] symmetry 
transformations. The latter off-shell nilpotent and absolutely anticommuting
symmetry transformations should be contrasted with the usual off-shell nilpotent 
and absolutely anticommuting (anti-)BRST symmetry transformations. In fact, it is 
the {\it total} kinetic term that remains invariant under the proper (anti-)BRST symmetry 
transformations whereas it is the {\it total} gauge-fixing term that remains unchanged 
under the proper (anti-)co-BRST symmetry transformations. For the Abelian 
$p$-form gauge theory, the curvature tensor for the gauge field (owing its origin 
to the exterior derivative) remains {\it certainly} invariant under the  nilpotent 
(anti-)BRST transformations. However, for such theories, it is the gauge-fixing 
term for the same gauge field (owing its origin to the co-exterior derivative)
that remains {\it definitely} unchanged under the (anti-)co-BRST transformations. Hence, the 
nomenclatures are very appropriate for the above nilpotent symmetry transformations.

It is of utmost importance to point out that {\it exactly} similar kind of restrictions 
[cf. (3), (6), (36)] must be imposed on the (dual-)gauge parameters for the existence of
(dual-)gauge invariance of  the gauge-fixed Lagrangian densities of any arbitrary
$D = 2p$ dimensional Abelian $p$-form gauge theory. 
Within the framework of the BRST formalism, however, there are  {\it no} such restrictions
on any parameters of the theory. The coupled Lagrangian densities (53), (54) respect both
the off-shell nilpotent (anti-)BRST and (anti-)co-BRST symmetry transformations [cf. (57), (55), (72) and (70)]
on the hypersurface defined by the CF-type restrictions [cf. (47) and (66)]. These Lagrangian densities provide
a tractable field theoretic models for the Hodge theory in the Feynman gauge.

For all the $D =2p$ dimensional Abelian $p$-form gauge theories, one can define a bosonic symmetry in the theory
that emerges from the suitable anticommutators between the above nilpotent (anti-)BRST and (anti-)co-BRST
symmetry transformations.  These nilpotent symmetry transformations 
provide the physical realizations of the (co-)exterior derivatives and their appropriate anticommutators 
provide the physical realization of the Laplacian operator. There always exists a ghost-scale symmetry 
in the above theories which is needed for the
definition of the ghost number of a state in the quantum Hilbert space. Together, the algebraic structures 
of all the six continuous symmetries (and their corresponding generators) provide the physical realizations of 
the algebra obeyed by the de Rham cohomological operators (as well as the Hodge decomposition theorem that is
defined in terms of the above cohomological operators [cf. (103)-(105)]).

There exists a discrete set of symmetries in the above $D = 2p$ dimensional Abelian $p$-form gauge theories that is
connected with the self-duality conditions [cf. (4), (8), (37)]. This condition, in turn, is very intimately related
with the Hodge duality $(*)$ operation of differential geometry. As a consequence, these symmetries provide the physical
realization of the Hodge duality ($*$) operation (of differential geometry). Thus, the spacetime $D = 2p$
dimensional Abelian $p$-form gauge theories (in the Feynman gauge)
automatically provide the field theoretic models for the Hodge theory
where all the de Rham cohomological operators, Hodge duality $(*)$ operation and Hodge decomposition theorem, etc.,
find their physical realizations in the language of discrete and continuous symmetry properties of these theories. 
Furthermore, it turns out that the mathematical condition $* ~d\, *\,A^{(p)} = \delta \, A^{(p)}$ is always satisfied
for the Abelian $p$-form gauge theories in $D = 2p$ dimensions of flat spacetime. Such theories are always endowed with
the dual-gauge and off-shell nilpotent and absolutely anticommuting
(anti-)co-BRST symmetry transformations (see, e.g., Appendix {\bf D} for details).

One of the novel observations in our present investigation is the fact that $\{s_b, s_{ad} \} = s_\omega^{(1)}$
and $\{s_d, s_{ab} \} = s_\omega^{(2)}$ define {\it new} bosonic symmetries in the theory. 
However, these new bosonic symmetries cannot be identified with the Laplacian operator 
of differential geometry because these  do {\it not}
commute with the ghost-scale symmetry transformations. Hence, these do not correspond to the Casimir operators
for the whole algebra. Further, a close look and a careful observation of the transformations in (90) shows that
$s_d$ and $s_{ab}$ (as well as $s_b$ and $s_{ad}$) anticommute with each-other up to a $U(1)$ vector gauge
transformations. In the context of 2D Abelian 1-form [10--12] and 4D Abelian 2-form gauge theory [13--16],
 it has been found that $\{s_b, s_{ad} \} = 0$ and $\{s_d, s_{ab} \} = 0$ (which imply the absolute 
anticommutativity between the
above nilpotent (fermionic) transformations). We have discussed, in detail, about these new bosonic
symmetries in our Appendix {\bf E} and established that these can {\it not} be identified with the 
Laplacian operator of differential geometry.

In our earlier works on 2D (non-)interacting  Abelian 1-form [10] and 4D Abelian 2-form
gauge theories [24], we have shown
that the former theories turn out to be a perfect model for a new TFT and the latter theory is proven
to be a model for q-TFT. Both varieties of theories are, however, perfect field theoretic 
models for the Hodge theory. We have proven, in our present investigation (cf. Sect. 8) 
that the 6D Abelian 3-form gauge theory, besides being a perfect model for the Hodge theory, 
is also a model for the q-TFT where the physical application of the
dual-BRST symmetry turns up in a very convincing manner. In other words, as is evident from
the d.o.f. counting, we have demonstrated that the 6D Abelian 3-form gauge theory is {\it not} a perfect
model for the TFT. We note, in passing, that the physical contents of the
4D Abelian 2-form and 6D Abelian 3-form gauge theories
are very much similar to each-other within the framework of BRST formalism as both of them are the 
perfect models for the Hodge theory as well as q-TFT.

We have not discussed, in our present investigation, anything about the interacting $p$-form gauge theories
(with matter fields) as well as the more general non-Abelian $p$-form gauge theories. In this connection, we 
would like to state that, so far, we have been able to establish that the interacting 2D Abelian 1-form 
gauge theory with Dirac fields [10] and the 2D (non-)Abelian 1-form gauge theories (without any interaction with 
matter fields) also provide the field theoretic models for the Hodge theory (see, e.g., [11,12]).
It would be very challenging endeavor to obtain the tractable field theoretic models for the Hodge theory in the
cases of {\it interacting} (non-)Abelian $p$-form ($p \geq 2$) gauge theories in any arbitrary dimension of spacetime
(where matter fields would be also present). These are some of the issues that are presently under 
investigation and our results would be reported in the future.\\

\noindent
{\bf Acknowledgements}\\

\vskip 0.3cm

\noindent
Three of us (RK, SK, AS) would like to thank UGC, Govt. of India, New Delhi, 
for financial support under the SRF, RGNF and RFSMS schemes, respectively.\\

\begin{center}
\bf{\Large Appendix A: Coupled Lagrangian densities}
\end{center}


In our earlier works [21,22], we have derived the gauge-fixing and Faddeev-Popov ghost terms for 
the starting Lagrangian density (${\cal L}_{(0)} = \frac{1}{24} H^{\mu\nu\eta\kappa} H_{\mu\nu\eta\kappa}$),
within the framework of BRST formalism, by exploiting the following standard expressions (see, e.g., [21,22]):  
\[
{\cal L}_{(b)}^{(0)} = \frac{1}{24}\; H^{\mu\nu\eta\kappa}\, H_{\mu\nu\eta\kappa} 
+ s_b \,s_{ab}\;\bigg[\frac{1}{2} \;\bar C_2\, C_2  - \frac{1}{2}\;\bar C_1\, C_1 
- \frac{1}{2} \;\bar C_{\mu\nu} \,C^{\mu\nu}\]
\[~~ - \frac{1}{2} \phi_\mu^{(1)}\, \phi^{\mu(1)}
 - \frac{1}{2} \;\phi_\mu^{(2)}\, \phi^{\mu(2)} - \bar\beta_\mu\, \beta^\mu 
- \frac{1}{6}\; A_{\mu\nu\eta}\, A^{\mu\nu\eta} \bigg ],
\eqno{(A.1)}\]
\[
{\cal L}_{(\bar b)}^{(0)} = \frac{1}{24}\; H^{\mu\nu\eta\kappa} \,H_{\mu\nu\eta\kappa} 
- s_{ab}\, s_b\;\bigg[\frac{1}{2} \;\bar C_2\, C_2  - \frac{1}{2}\;\bar C_1\, C_1 
- \frac{1}{2} \;\bar C_{\mu\nu}\, C^{\mu\nu} \]
\[~~ - \frac{1}{2} \phi_\mu^{(1)}\, \phi^{\mu(1)} 
- \frac{1}{2} \;\phi_\mu^{(2)}\, \phi^{\mu(2)} - \bar\beta_\mu\, \beta^\mu  
- \frac{1}{6}\; A_{\mu\nu\eta} \,A^{\mu\nu\eta}\bigg ],
\eqno{(A.2)}\]
where $s_b$ and $s_{ab}$ are the off-shell nilpotent symmetry transformations (55) and (57), respectively,
and $\Big({\cal L}_{(b)}^{(0)}, \;$ ${\cal L}_{(\bar b)}^{(0)}\Big)$
 are the coupled (but equivalent) Lagrangian densities that
respect the nilpotent (anti-)BRST symmetry transformations $s_{(a)b}$. 
Within the above square brackets in $(A.1)$ and $(A.2)$, we have chosen the  
combinations of fields that have mass dimensions equal to four
and ghost numbers equal to zero for the derivation of the 6D (anti-)BRST invariant Lagrangian densities. 
The emerging  Lagrangian densities, however, 
do {\it not} produce the  fermionic CF-type of conditions (66)
from the equations of motion. Thus, we have altered a bit the above combinations
of fields in the square brackets of $(A.1)$ and $(A.2)$ as 
\[
{\cal L}_{(b)} = \frac{1}{24}\; H^{\mu\nu\eta\kappa}\, H_{\mu\nu\eta\kappa} 
+ s_b\, s_{ab}\;\bigg[\frac{1}{2} \;\bar C_2\, C_2  - \frac{1}{2}\;\bar C_1\, C_1 
- \frac{1}{2} \;\bar C_{\mu\nu} \,C^{\mu\nu} \]
\[- \frac{1}{2}\; \phi_\mu^{(1)}\, \phi^{\mu(1)} 
- \frac{1}{2} \;\phi_\mu^{(2)}\, \phi^{\mu(2)} + \bar\beta_\mu \,\beta^\mu 
- \frac{1}{6}\; A_{\mu\nu\eta}\, A^{\mu\nu\eta}\bigg ],
\eqno{(A.3)}\]
\[
 {\cal L}_{(\bar b)} = \frac{1}{24}\; H^{\mu\nu\eta\kappa}\, H_{\mu\nu\eta\kappa} 
- s_{ab} \,s_b\;\bigg[\frac{1}{2} \;\bar C_2\, C_2  - \frac{1}{2}\;\bar C_1\, C_1 
- \frac{1}{2} \;\bar C_{\mu\nu}\, C^{\mu\nu} \]
\[- \frac{1}{2}\; \phi_\mu^{(1)}\, \phi^{\mu(1)} 
- \frac{1}{2} \;\phi_\mu^{(2)}\, \phi^{\mu(2)} + \bar\beta_\mu\, \beta^\mu 
- \frac{1}{6}\; A_{\mu\nu\eta} \,A^{\mu\nu\eta}\bigg ],
\eqno{(A.4)}\]
which leads to the derivation of appropriate coupled and equivalent Lagrangian densities (53) 
and (54) (see, Sect. 3). The Lagrangian densities [i.e. (53), (54)] yield CF-type of restrictions 
(47) and (66) from equations of motion [cf. (61), (62)].

We would like to lay stress on the fact that the difference between {$(A.1)$ as well as $(A.2)$] and  [$(A.3)$
as well as $(A.4)$] is {\it only} the term $\bar \beta^\mu \beta_\mu$ which carries ($-$) sign in the former pair
and ($+$) sign in the latter pair. The key advantage of the choice in {$(A.3)$ and $(A.4)$] leans heavily on the
derivation of CF-type restrictions (47) and (66) which automatically turn out to be (anti-)BRST 
as well as (anti-)co-BRST invariant.  This happens because (47) and (66) are 
derived from the equations of motion which always respect the basic symmetries
of the theory.  As a consequence, these (anti-)BRST and (anti-)co-BRST invariant 
restrictions are {\it physical} 
in some sense  (particularly in the light of our theory being a model for the Hodge theory).
\\

\begin{center}
\bf{\Large Appendix B: Nilpotency of $Q_{(a)b}$}
\end{center}


In this Appendix, we shall explicitly show that the BRST charge $Q_b$ is nilpotent of 
order two (i.e. $Q_b^2 = 0 \Rightarrow  \frac{1}{2}\; \{Q_b, \; Q_b\} = 0$) from the symmetry principle 
$s_b Q_b = - i \{Q_b, \; Q_b \} = 0 $ where the conserved and nilpotent BRST charge $Q_b$ plays
the key role of a generator. For this purpose, we use here the 
BRST symmetry transformations $s_b$ [cf. (55)] and the expression for the conserved charge $Q_b$
[cf. (64)]. It can be explicitly checked that 
\[
s_b\, Q_b \;=\; - \int d^5x \;\Big[(\partial^0 \bar F^i - \partial^i \bar F^0)(\partial_i C_2)\qquad\quad \]
\[~~~~~~~~~~~~~~~~~~~~~~~~~~~~~~~~~~~~~~~~~~~~~~ + (\partial^0 K^{\nu\eta} + \partial^\nu K^{\eta 0} + \partial^\eta K^{0 \nu})(\partial_\nu \beta_\eta 
- \partial_\eta \beta_\nu)  \Big].\qquad 
\eqno{(B.1)} \]
The following E-L equations of motion [cf. (61)]
\[
\Box \;\bar C_{\mu\nu} - \frac {3}{2}(\partial_\mu \bar F_\nu 
- \partial_\nu \bar F_\mu) = 0, \qquad
\partial_\mu \bar C^{\mu\nu} + \partial^\nu \bar C_1 - 2 \bar F^\nu = 0, \]
\[\frac {1}{2} \; \varepsilon_{\mu\nu\eta\kappa\lambda\rho}\; (\partial^\mu {\cal K}^{\nu\eta}) 
=  \partial _\kappa  K_{\lambda\rho} + \partial _\lambda  K_{\rho\kappa}
 + \partial _\rho  K_{\kappa\lambda},
\eqno{(B.2)} \]
can be exploited to re-express $(B.1)$ in a different form. For instance, we obtain the following 
useful relations: 
\[
(\partial^0 \bar F^i - \partial^i \bar F^0) = \frac {2}{3}\; \Box \; \bar C^{0i},\qquad
\partial_i \bar C^{0i} =  \partial^0 \bar C_1 - 2 \bar F^0,\]
\[\partial^0 K^{\nu\eta} + \partial^\nu  K^{\eta 0}  + \partial^\eta   K^{0 \nu} 
= -\frac {1}{2} \varepsilon^{0\mu\lambda\rho\nu\eta} (\partial_\mu {\cal K}_{\lambda\rho}),
\eqno{(B.3)}\]
from $(B.2)$ as special cases.
Using the above equations of motion $(B.3)$ in equation $(B.1)$ and performing a partial integration, 
we can re-express $(B.1)$  as  follows:
\[
s_b\, Q_b = \int d^5x \;\partial_i \bigg[ - \frac {2}{3} \left(\Box\; \bar C^{0i}\right) C_2 
+ \frac {1}{2}\; \varepsilon^{0ijklm}\; {\cal K}_{jk} (\partial_l \beta_m - \partial_m \beta_l) \bigg] \]
\[+ \frac {2}{3} \int d^5x \;\Big [\partial^0 (\Box \; \bar C_1)\; C_2
- 2 (\Box \;\bar F^0)\; C_2\Big ].\qquad\qquad\quad\;
\eqno{(B.4)}\]
It is clear from the above expression that the first integral vanishes  at infinity 
(due to  Gauss's divergence theorem) and the second 
integral is equal to zero due to the Euler-Lagrange equations of motion $\Box \,\bar C_1 = 0$ 
and $\Box \, \bar F^0 = 0$ . 
Therefore, $s_b Q_b = - i\; \{Q_b, \; Q_b\} =  0$  implies that the BRST charge $Q_b$ is
nilpotent.

In an exactly similar fashion, 
we prove the nilpotency of the anti-BRST charge $Q_{ab}$ by exploiting the 
principle of symmetry transformations where the concept of generator plays an 
important role. For instance, taking the help from (63) and (67), it can be checked that  
\[s_{ab}\, Q_{ab} = \int d^5x\; \Big[- (\partial^0 F^i - \partial^i  F^0)(\partial_i \bar C_2) \qquad\qquad\qquad\qquad\;\;\]
\[+ (\partial^0 \bar K^{\nu\eta}
+ \partial^\nu \bar K^{\eta 0} + \partial^\eta \bar K^{0 \nu})(\partial_\nu \bar \beta_\eta 
- \partial_\eta \bar \beta_\nu)  \Big ].
\eqno{(B.5)}\]
The above expression can be re-expressed by using the following Euler-Lagrange equations of motion [cf. (62)]
that emerge from the Lagrangian density ${\cal L}_{(\bar b)}$, namely; 
\[\frac {1}{2} \, \varepsilon_{\mu\nu\eta\kappa\lambda\rho}\, (\partial^\mu \bar {\cal K}^{\nu\eta}) 
=  \partial _\kappa  \bar K_{\lambda\rho} + \partial _\lambda  \bar K_{\rho\kappa}
 + \partial _\rho  \bar K_{\kappa\lambda}, \]
\[\Box \,C_{\mu\nu} + \frac {3}{2}\,(\partial_\mu  F_\nu 
- \partial_\nu  F_\mu) = 0.
\eqno{(B.6)}\]
Actually, it can be seen that the above equations lead to 
\[\partial^0 \bar K^{\nu\eta} + \partial^\nu  \bar K^{\eta 0}  + \partial^\eta   \bar K^{0 \nu} 
= - \;\frac {1}{2} \; \varepsilon^{0\mu\lambda\rho\nu\eta}\; (\partial_\mu \bar {\cal K}^{\lambda\rho}),\]
\[(\partial^0 F^i - \partial^i  F^0) = - \frac {2}{3}\;\Box \;C^{0i},
\eqno{(B.7)}\]
as special cases. Substitution of $(B.7)$ into $(B.5)$, yields
\[s_{ab}\, Q_{ab} =  \int d^5x \;\partial_i \bigg [\frac {2}{3} \;(\Box \;C^{0i}) \; \bar C_2 
- \frac {1}{2}\; \varepsilon^{0ijklm}\; \bar {\cal K}_{jk} (\partial_l \bar \beta_m 
- \partial_m \bar \beta_l) \bigg] \]
\[- \frac{2}{3}\;\int d^5x \;\Big [\Box \;(\partial_i C^{0i}) \Big ]\;\bar C_2. \hskip 3.3cm
\eqno{(B.8)}\]
This expression can be further simplified by using the equation of motion 
$\partial_\mu C^{\mu\nu} - \partial^\nu C_1 + 2 F^\nu = 0$ which leads to   
$\partial_i C^{0i} = - (\partial^0 C_1) + 2 F^0.$ The final form of $(B.8)$ is: 
\[
s_{ab}\, Q_{ab} = \int d^5x \;\partial_i \bigg[\frac {2}{3}(\Box \;C^{i0}) \;\bar C_2 
- \frac {1}{2}\; \varepsilon^{0ijklm}\; \bar {\cal K}_{jk} (\partial_l \bar \beta_m 
- \partial_m \bar \beta_l) \bigg] \]
\[+ \frac{2}{3}\,\int d^5x \,\Big[\partial^0 (\Box \,C_1)\; \bar C_2 
- 2(\Box\, F^0) \, \bar C_2\Big].\qquad\qquad
\eqno{(B.9)}\]
The above equation explicitly implies that $s_{ab} Q_{ab} = - i \; \{Q_{ab}, \; Q_{ab}\} 
\Longrightarrow Q^2_{ab} = 0$ when we use the Euler-Lagrange equations of motion 
$\Box \;C_1 = 0$ and $\Box\; F^0 = 0$ and throw away the total space derivative terms.
\\

\begin{center}
\bf{\Large Appendix C: Anticommutativity check}
\end{center}
Here we provide  the key ingredients for the proof of absolute anticommutativity  
of the conserved (anti-)BRST charges $Q_{(a)b}$ [cf. (63), (64)] with the help of  
symmetry transformations [cf. (57), (55)] and the concept of generators for the theory. Let us take as an example  
the proof of $s_b Q_{ab} = - i \{Q_{ab}, \; Q_b\}  = 0$. Using (63) and (55), we obtain the following:
\[s_b \,Q_{ab} = \int d^5x\; \bigg[\frac{1}{2!}\; \varepsilon^{0ijklm}\; (\partial_i K_{jk})\; \bar {\cal K}_{lm}
- (\partial^0 K^{\nu\eta} + \partial^\nu K^{\eta 0} 
+ \partial^\eta K^{0 \nu})\bar K_{\nu\eta}  \]
\[\qquad +\; \bar K^{0i}\; (\partial_i B_1) 
+ \;(\partial^0 \bar C^{\nu\eta} + \partial^\nu \bar C^{\eta 0} 
+ \partial^\eta \bar C^{0 \nu})\;(\partial_\nu f_\eta - \partial_\eta f_\nu)  + B_2\; \dot B \]
\[-\; (\partial^0 \beta^i - \partial^i \beta^0 )\;(\partial_i B_2)
+ (\partial^0 f^i - \partial^i f^0)\; \bar f_i  + B \; \dot B_2 \qquad\qquad\;\]
\[-\; (\partial^0 C^{\nu\eta} + \partial^\nu C^{\eta 0} 
+ \partial^\eta C^{0 \nu})\;(\partial_\nu \bar F_\eta - \partial_\eta \bar F_\nu) 
 + B_1 \;\dot B_1 \quad\;\;\; \quad\]
\[+\; (\partial^0 \bar \beta^i - \partial^i \bar \beta^0 )\;(\partial_i B) 
-  (\partial^0 \bar F^i - \partial^i  \bar F^0)\,F_i \bigg ]. \;\;\hskip 2cm
\eqno{(C.1)}\]
The above expression requires some involved algebraic computations in the  proof of 
$s_b Q_{ab} = - i\; \{Q_{ab},\; Q_b\} = 0$. Therefore, 
for the sake of brevity as well as step-by-step computations,  we divide the r.h.s. of 
$(C.1)$ into four parts as given below
\[I_1 = \int d^5 x\; \bigg[\frac{1}{2!}\; \varepsilon^{0ijklm}\; 
(\partial_i K_{jk})\; \bar {\cal K}_{lm} - (\partial^0 K^{\nu\eta} + \partial^\nu K^{\eta 0} 
+ \partial^\eta K^{0 \nu})\bar K_{\nu\eta}\bigg], \qquad\]
\[I_2 = \int d^5 x\; \Big[\bar K^{0i}\; (\partial_i B_1) 
+ B_1 \;\dot B_1 \Big], \hskip 6cm \quad\]
\[I_3 = \int d^5 x \;\Big[(\partial^0 \bar \beta^i 
- \partial^i \bar \beta^0 )(\partial_i B) + B \; \dot B_2 + B_2\; \dot B
- (\partial^0 \beta^i - \partial^i \beta^0 )(\partial_i B_2) \Big], \qquad\;\]
\[I_4 = \int d^5 x\; \Big[(\partial^0 \bar C^{\nu\eta} + \partial^\nu \bar C^{\eta 0} 
+ \partial^\eta \bar C^{0 \nu})\;(\partial_\nu f_\eta 
- \partial_\eta f_\nu) 
+(\partial^0 f^i - \partial^i f^0)\; \bar f_i \quad\;\]
\[- (\partial^0 C^{\nu\eta} + \partial^\nu C^{\eta 0} 
+ \partial^\eta C^{0 \nu})\;(\partial_\nu \bar F_\eta 
- \partial_\eta \bar F_\nu) 
-  (\partial^0 \bar F^i - \partial^i  \bar F^0)\;F_i \Big].
\eqno{(C.2)}\]
Exploiting the CF-type condition $K_{\mu\nu} + \bar K_{\mu\nu} = \partial_\mu \phi^{(1)}_\nu 
- \partial_\nu \phi^{(1)}_\mu$ [cf. (47)], the second term of the integral $I_1$ can be re-expressed as
\[- \int d^5x \;\Big(\partial^0 K^{\nu\eta} + \partial^\nu K^{\eta 0} 
+ \partial^\eta K^{0 \nu} \Big) \bar K_{\nu\eta}\]
\[=  \int d^5x \;\Big(\partial^0 \bar K^{\nu\eta} 
+ \partial^\nu \bar K^{\eta 0} + \partial^\eta \bar K^{0 \nu} \Big) \bar K_{\nu\eta}.
\eqno{(C.3)}\]
The E-L equations of motion $\frac {1}{2} \; \varepsilon_{\mu\nu\eta\kappa\lambda\rho}$
$(\partial^\mu \bar {\cal K}^{\nu\eta}) =  \partial _\kappa  \bar K_{\lambda\rho} 
+ \partial _\lambda  \bar K_{\rho\kappa} + \partial _\rho  \bar K_{\kappa\lambda}$ [cf. (62)],
derived from the Lagrangian density ${\cal L}_{(\bar b)}$, imply  
\[
\partial^0 \bar K^{\nu\eta} + \partial^\nu  \bar K^{\eta 0}  + \partial^\eta   \bar K^{0 \nu} 
= - \frac {1}{2}  \varepsilon^{0\mu\lambda\rho\nu\eta} 
(\partial_\mu \bar {\cal K}_{\lambda\rho}).
\eqno{(C.4)}\]
Substituting $(C.4)$ into the expression $I_1$ of $(C.2)$ and performing partial integration, we obtain
the following 
\[
 I_1 =\; - \frac{1}{2!}\,\int d^5 x\; \partial_i\Big(\varepsilon^{0ijklm}\, 
\bar K_{jk}\, \bar {\cal K}_{lm} \Big) \qquad \qquad\quad\]
\[+\; \frac{1}{2!}\,\int d^5 x\; \varepsilon^{0ijklm} \, 
\Big[\partial_i (K_{jk} +  \bar K_{jk}) \Big]\, \bar {\cal K}_{lm}. \;\quad
\eqno{(C.5)}\]
Thus, we note that $I_1 = 0$ if we use (47) and if we assume that the 
physical fields are those that vanish at infinity (due to the Gauss  divergence theorem).

Let us now take the integral $I_2$. Performing the partial integration, we obtain
\[
I_2 = \int d^5 x\; \partial_i\, \Big( B_1 \,\bar K^{0i} \Big)  
- \int d^5 x\; \Big [B_1\,(\partial_i \bar K^{0i}) -  B_1 \,\dot B_1 \Big].
\eqno{(C.6)}\]
We note that the first term goes to zero (due to the validity of  Gauss's divergence theorem)
and the second term vanishes if we use the equation of motion 
$\partial_\mu \bar K^{\mu\nu} + \partial^\nu B_1 = 0$ from (62)
(which implies $\partial_i \bar K^{0i} = \dot B_1$).
If we carry out the partial integration and 
throw away the total space derivative terms, we obtain the suitable form of integral $I_3$ as  
\[I_3 = - \int d^5 x\; \Big[(\partial^0 \partial_i \bar \beta^i 
- \partial_i\partial^i  \bar \beta^0 )\, B 
- (\partial^0 \partial_i \beta^i - \partial_i \partial^i \beta^0 )\,B_2 
- B \, \dot B_2 - B_2\, \dot B \Big].
\eqno{(C.7)}\]
Using the appropriate equations of motion $B = - (\partial \cdot \beta),\; 
B_2 =  (\partial \cdot \bar \beta) $ from (62) [which imply $\partial_i \beta^i = - B - \partial_0 \beta^0, \; 
\partial_i \bar \beta^i = - \partial_0 \bar \beta^0 + B_2$], the above expression can be re-expressed as
\[I_3 = \int d^5 x\; \Big[B \; (\Box \bar \beta^0) - B_2 \;(\Box \beta^0) \Big].
\eqno{(C.8)}\]
It is clear from the above equation that the  integral $I_3$ vanishes ($I_3 = 0$) 
if we use the E-L equations of motion $\Box \beta_0 = 0, \; \Box \bar \beta_0 = 0$ [cf. (62)].

The integral $I_4$ can be expressed in the component form as
\[I_4 = - 2 \int d^5 x\; \Big[\partial^0 (\partial_i \bar C^{ij}) \; f_j 
+ \partial_i (\partial^i \bar C^{j0} - \partial^j \bar C^{i0})\;f_j 
- \partial_i (\partial^i C^{j0} - \partial^j  C^{i0})\; \bar F_j \]
\[+ \;\frac {1}{2}\;(\partial^0 \bar F^i - \partial^i  \bar F^0)\;F_i 
- \frac {1}{2}\;(\partial^0 f^i - \partial^i f^0)\; \bar f_i  
-  \partial^0 (\partial_i C^{ij}) \; \bar F_j \Big ],
\eqno{(C.9)}\]
where we have performed the partial integration in the first and third terms of the 
integral $I_4$ and have thrown away the total space derivative terms.
Using the following equations of motion $\partial_\mu C^{\mu\nu} - \partial^\nu C_1 + 2 F^\nu = 0, \; 
\partial_\mu \bar C^{\mu\nu}  - \partial^\nu \bar C_1 + 2 \bar f^\nu = 0$, the above integral can be
reduced to the following form:
\[I_4 = 2 \int d^5x \;\Big [(\Box \bar C^{0i}) f_i  
+ 2(\partial^0 \bar f^i - \partial^i \bar f^0)\, f_i 
- 2 (\partial^0 F^i - \partial^i F^0) \; \bar F_i \]
\[\qquad\qquad \;\;\;- \; \big(\Box C^{0i}\big)\, \bar F_i 
+ \frac {1}{2}\;(\partial^0 f^i - \partial^i f^0) \; \bar f_i 
- \frac {1}{2}\;(\partial^0 \bar F^i - \partial^i \bar F^0) \; F_i \Big ].
\eqno{(C.10)}\]
Further, exploiting the appropriate equations of motion 
$\Box C^{0i} = -  \frac {3}{2} (\partial^0 F^i - \partial^i F^0),$
$\Box \bar C^{0i} =  - \frac {3}{2} (\partial^0 \bar f^i - \partial^i \bar f^0)$ [cf. (62)]
and the CF-type conditions $f_\mu + F_\mu = \partial_\mu C_1,$
$\bar f_\mu + \bar F_\mu = \partial_\mu \bar C_1$ from (66), the above expression 
can be re-written as
\[
I_4 = \int d^5x \; \Big [(\partial^0 \bar f^i - \partial^i \bar f^0) \; (f_i  + F_i) 
- (\partial^0  F^i - \partial^i F^0) \; (\bar f_i +  \bar F_i) \Big ]\]
\[\equiv\; - \frac{2}{3} \int d^5x\; \Big [(\Box \bar C^{0i}) (\partial_i C_1) 
- (\Box C^{0i})(\partial_i \bar C_1)  \Big]. \qquad\quad \;
\eqno{(C.11)}\]
After performing the partial integration and, then, using the Euler-Lagrange equations of motion
$\partial_i C^{0i} = - \partial^0 C_1 + 2 F^0,\;
\partial_i \bar C^{0i} =  - \partial^0  \bar C_1 + 2 \bar f^0$ [cf. (62)], the
above integral can be proved to be zero. The following form of $I_4$ corroborates the above statement, namely;
\[I_4 =  - \frac {2}{3}  \int d^5x \; \Big [\partial^0 (\Box\bar C_1) \, C_1 - 2\, (\Box \bar f^0)\, C_1\]
\[- \;\partial^0 (\Box C_1) \, \bar C_1 + 2\, (\Box F^0) \,\bar C_1 \Big] = 0, \quad
\eqno{(C.12)}\]
where we have thrown away the total space derivative terms and we have used the equations of motion 
$\Box C_1 = 0, \;\Box \bar C_1 = 0, \;\Box F^0 = 0, \;\Box \bar f^0 = 0$ [cf. (62)].

From the above computations, it is clear that $s_b Q_{ab}$ $= - i\; \{Q_{ab},\; Q_b\} = 0$  shows that 
$Q_b$ and $Q_{ab}$ are absolutely  anticommuting in nature on the constrained hypersurface defined by the 
CF-type equations (47) and (66). In an exactly similar fashion, one can also  prove that  
$s_{ab} Q_b = - i\; \{Q_b,\; Q_{ab}\} = 0$ (which imply the anticommutativity of $Q_b$ and $Q_{ab}$).
We would like to comment that the complete  extended BRST algebra (102) 
can be derived by using the concept of principle of symmetry transformations as we 
have demonstrated in the proof of nilpotency  ($Q^2_{(a)b} = 0$) and anticommutativity property 
($\{Q_b, \; Q_{ab}\} = 0$) (cf.  Appendices {\bf B} and {\bf C}). In this method of proof, the algebra is 
straightforward.   
\\

\begin{center}
\bf{\Large Appendix D: Arbitrary $p$-form theory}
\end{center}
Here we discuss the discrete symmetry transformations for the Abelian 
$p$-form gauge theory in $D = 2p$ dimensions 
of spacetime. It is clear that the kinetic term of this gauge theory will be 
constructed from the ($p+1$)-form $F^{(p+1)}$ as given below
\[F^{(p+1)} = d\;A^{(p)} 
= \left [ \frac{dx^{\mu_1} \wedge dx^{\mu_2}...\wedge dx^{\mu_{(p+1)}}}{(p+1)!}\right]
F_{\mu_1\mu_2...\mu_{(p+1)}},
\eqno{(D.1)}\]
where $F_{\mu_1\mu_2\mu_3...\mu_{(p+1)}}$ is the totally antisymmetric curvature tensor that 
is expressed in terms of the antisymmetrized  version of derivatives on the $p$-form potential 
$A_{\mu_1\mu_2\mu_3...\mu_p}$. The gauge-fixing term in the $D = 2p$
dimensional spacetime for the $p$-form gauge potential is a $(p-1)$-form antisymmetric tensor 
(defined in terms of the co-exterior derivative $\delta$) as 
\[\delta\; A^{(p)} = - *\;d\;* A^{(p)} 
= \bigg [ \frac {dx^{\mu_2} 
\wedge dx^{\mu_3}...\wedge dx^{\mu_p}}{(p-1)!}\bigg] 
\Big(\partial^{\mu_1}A_{\mu_1\mu_2\mu_3...\mu_p }\Big),
\eqno{(D.2)}\]
where ($\partial^{\mu_1}A_{\mu_1\mu_2\mu_3...\mu_p }$) is totally antisymmetric in all the indices 
from $\mu_2$ to $\mu_p$.

The gauge-fixed Lagrangian density for the $D = 2p$ dimensional Abelian $p$-form gauge 
theory can be written (in the Feynman gauge) as (see, e.g., [38])
\[
{\cal L} = \frac{1}{2\,(p+1)!}\;\Big(F_{\mu_1\mu_2...\mu_{(p+1)}}\Big)
\Big(F^{\mu_1\mu_2...\mu_{(p+1)}}\Big) \;\;\quad\]
\[+\;\frac{1}{2\,(p-1)!}\Big(\partial^{\mu_1}A_{\mu_1\mu_2...\mu_p}\Big) 
\Big(\partial_{\nu_1}A^{\nu_1\mu_2...\mu_p} \Big). 
\eqno{(D.3)}\]
For the 6D Abelian 3-form theory [cf. (32)], we note that an overall factor 
of (1/2) has been taken out, for the sake of simplicity. In fact, for $p = 3$, we can obtain
Lagrangian density (32) from the general Lagrangian density $(D.3)$ modulo the above factor of (1/2).
The above Lagrangian density would respect (dual-)gauge symmetry transformations analogous to (2), (7)
and (34). The key reason behind the existence of the dual-gauge symmetry transformations is the 
following {\it self-duality} condition
\[
*\;A^{(p)} = \frac{(- 1)^p}{p!} \;\varepsilon_{\mu_1\mu_2...\mu_{p}\mu_{(p+1)}...\mu_{2p}}\;
A^{\mu_{(p+1)} \mu_{(p+2)}...\mu_{2p}},
\eqno{(D.4)}\]
where the Hodge duality ($*$) operation is defined on a $2p$-dimensional flat spacetime manifold, on which, 
a $2p$-dimensional Levi-Civita tensor 
$\varepsilon_{\mu_1\mu_2\mu_3...\mu_{p}\mu_{p+1}...\mu_{2p}}$ can exist. The 
following discrete symmetry transformations on the $p$-form gauge potential, namely;
\[
A_{\mu_1\mu_2...\mu_{p}}\rightarrow \pm\; \frac{i}{p!}\; \varepsilon_{\mu_1\mu_2...\mu_{p}
\mu_{(p+1)}...\mu_{2p}}\;
 A^{\mu_{(p+1)}\mu_{(p+2)}...\mu_{2p}},
\eqno{(D.5)}\]
would turn out to be the {\it symmetry} transformations for the gauge-fixed Lagrangian density $(D.3)$
as we have seen the existence of  analogous symmetries in the cases of 2D Abelian 1-form, 4D Abelian 
2-form and 6D Abelian 3-form gauge theories.
In fact, under the discrete transformations $(D.5)$, the kinetic and gauge-fixing 
terms would exchange with each-other [in the gauge-fixed Lagrangian density $(D.3)$ in the Feynman gauge
for the Abelian $p$-form theory under consideration].

The ``classical'' dual-gauge symmetries can be generalized to the nilpotent
``quantum'' (anti-)dual-BRST symmetries in the same way as we have done for the 2D 
Abelian 1-form, 4D Abelian 2-form and 6D Abelian 3-form gauge theories. We
very briefly outline here the derivation of, first of all, the proper (anti-)BRST 
symmetry transformations by exploiting the geometrical superfield approach [31,32]. 
Primarily, we invoke here the following horizontality
condition $(\tilde F^{(p+1)} = F^{(p+1)})$. In other words, we demand 
\[
\tilde d \;\tilde A^{(p)} \;=\; d\;A^{(p)},
\eqno{(D.6)}\]
where $d = dx^\mu \partial_\mu \; (\mu = 0, 1, 2,...,2p-1)$ is the ordinary exterior 
derivative defined on the ordinary
$D = 2p$ dimensional spacetime and $A^{(p)}$ is the ordinary $p$-form connection. On the l.h.s. of $(D.6)$,
we have the super exterior derivative $\tilde d$ (with $\tilde d^2 = 0$) and super 
$p$-form connection $\tilde A^{(p)}$ which are the generalizations of their ordinary counterparts (in 
$2p$-dimensions of spacetime) onto a
($2p, 2$)-dimensional supermanifold. These generalizations can be succinctly expressed, 
in the mathematical form, as
\[
d \longrightarrow \tilde d = dZ^M \partial_M\;\equiv\;  dx^\mu\;\partial_\mu 
+ d\theta \;\partial_{\theta} + d\bar\theta\; \partial_{\bar\theta},\]
\[A ^{(p)} \longrightarrow  \tilde A^{(p)}
 = \left(\frac {dZ^{M_1}\wedge dZ^{M_2}... \wedge dZ^{M_p}}{p!}\right) \,
\tilde A_{M_1M_2...M_p} (x, \theta, \bar\theta),
\eqno{(D.7)}\]
where $Z^M = (x^\mu, \theta, \bar\theta)$ and $\partial_M 
= (\partial_\mu, \partial_{\theta}, \partial_{\bar\theta})$ 
are the superspace variables and corresponding derivatives that characterize the
($2p, 2$)-dimensional supermanifold on which the Abelian $p$-form gauge theory has been generalized. Here
the  bosonic variable $x^\mu (\mu = 0, 1, 2,...,2p - 1)$ are the ordinary spacetime 
variable and ($\theta, \bar\theta$) are a pair of Grassmannian variables 
(with $\theta^2 = \bar\theta^2 = 0, \; \theta\bar\theta + \bar\theta\theta = 0$). Furthermore,
the multiplet superfields  $\tilde A_{M_1M_2...M_{p-1}M_{p}} (x, \theta, \bar\theta)$ 
will have (anti)symmetric components 
(i.e. $\tilde A_{\mu_1\mu_2...\mu_{p}}, \; \tilde A_{\mu_1\mu_2,\theta,\bar\theta...\mu_{p}}$, etc.). These 
superfields $\tilde A_{M_1M_2...M_{p}} (x, \theta, \bar \theta)$ can be expanded along the 
Grassmannian  directions of the ($2p, 2$)-dimensional supermanifold (see, e.g., [21]) 
in terms of the basic, auxiliary and secondary fields of the ordinary 
$2p$-dimensional (anti-)BRST invariant field theory.

At this juncture, the horizontality condition $(D.6)$ plays a decisive role. 
This condition is basically the covariant reduction of the supercurvature ($p + 1$)-form 
$\tilde F^{(p + 1)}$ to the ordinary curvature ($p + 1$)-form 
$F^{(p + 1)}$.  In fact, first of all,
we substitute the values of $\tilde d$ and $\tilde A^{(p)}$ from (D.7) into the l.h.s. of $(D.6)$.
The covariant reduction of the supercurvature ($p + 1$)-form $\tilde F^{(p + 1)}$ (defined on the 
($2p, 2$)-dimensional supermanifold) to the ordinary curvature ($p + 1$)-form $F^{(p + 1)}$  (defined on 
$2p$-dimensional supermanifold), leads to the derivation of off-shell
nilpotent ($s_{(a)b}^2 = 0$) and absolutely anticommuting ($s_b\, s_{ab} + s_{ab} \,s_b = 0$) (anti-)BRST
transformations (see, e.g., [21] for Abelian 2-form and 3-form gauge theories). From these proper 
(anti-)BRST transformations, one can always obtain the (anti-)BRST invariant coupled Lagrangian densities
(see, e.g., Appendix {\bf A} for 3-form gauge theory) that incorporate the gauge-fixing and Faddeev-Popov ghost terms.
Furthermore, these Lagrangian densities would always respect the (anti-)BRST symmetry transformations derived 
from the horizontality condition (within the framework of superfield formalism [31,32].

Finally, the discrete symmetry transformations for the gauge field [cf. $(D.5)$] can be generalized 
to incorporate such discrete symmetry transformations on the (anti-)ghost and auxiliary fields 
of the theory. The full set of discrete symmetry transformations would  define the analogue of the Hodge 
duality ($*$) operation of differential geometry. As a consequence, we can define the analogue 
of relation (100) that would enable us to deduce the (anti-)dual-BRST symmetry transformations 
$s_{(a)d}$. The latter symmetries also turn out to be off-shell nilpotent ($s_{(a)d}^2 = 0$) and absolutely 
anticommuting ($s_d s_{ad} + s_{ab} s_b = 0$) in nature. Having obtained these basic 
nilpotent ($s_{(a)b}^2 = 0,\; s_{(a)d}^2 = 0$) (anti-)BRST and (anti-)co-BRST symmetry
transformations $s_{(a)b}$ and $s_{(a)d}$,  we can derive bosonic as well as ghost-scale 
symmetries of the theory and show that the Abelian $p$-form gauge theory, in $D = 2p$ dimensions 
of spacetime, provides a model for the Hodge theory where we obtain the physical realizations of all 
the cohomological quantities of differential geometry [17--20].
\\

\begin{center}
\bf{\Large Appendix E: Extra bosonic symmetries}
\end{center}
Here we show that the Lagrangian densities (53) and (54), in addition to the  six continuous symmetries 
$(s_{(a)b},$ $s_{(a)d}, s_\omega, s_g)$,
respect two {\it new bosonic} symmetries but these new bosonic symmetries [cf. (90)] can not be 
regarded as the analogue of Laplacian operator of differential geometry. 
 These new bosonic symmetries are defined as:
$s_{\omega}^{(1)} = \{s_b, \; s_{ad}\}, \; s_{\omega}^{(2)} = \{s_d, \; s_{ab}\}$.
It is interesting to point out that these bosonic symmetries $s_{\omega}^{(1)}$ and $s_{\omega}^{(2)}$ 
do not exist (i.e. $\{s_d, s_{ab}\} = \{s_b, s_{ad}\} = 0$) for the 2D Abelian 1-form [10--12] and 
4D Abelian 2-form gauge theories [13--16]. Under the following bosonic transformations 
$s_{\omega}^{(1)}$ and $s_{\omega}^{(2)}$:
\[
s_{\omega}^{(1)} \phi_\mu^{(1)} = - \;\partial_\mu B, 
\qquad s_{\omega}^{(1)} \phi_\mu^{(2)} = - \partial_\mu B,\qquad
s_{\omega}^{(1)} \bar \beta_\mu = \partial_\mu (B_1 - B_3), \]
\[s_{\omega}^{(1)}\big[A_{\mu\nu\eta},\; K_{\mu\nu}, \;
\bar K_{\mu\nu},\; {\cal K}_{\mu\nu},\;
\bar {\cal K}_{\mu\nu},\;
C_{\mu\nu},\; \bar C_{\mu\nu},\; F_\mu,\; \bar F_\mu, \;f_\mu, \;\bar f_\mu,\; \beta_\mu,\]
\[\qquad B,\; B_1,\; B_2,\; B_3,\; C_1,\; \bar C_1,\; C_2,\; \bar C_2\big] = 0,
\eqno{(E.1)}\]
\[
 s_{\omega}^{(2)} \phi_\mu^{(1)} = \partial_\mu B_2, 
\quad s_{\omega}^{(2)} \phi_\mu^{(2)} = -\; \partial_\mu B_2,\qquad
s_{\omega}^{(2)} \beta_\mu = \partial_\mu (B_1 + B_3), \]
\[s_{\omega}^{(2)}\big[A_{\mu\nu\eta},\; K_{\mu\nu}, \;
\bar K_{\mu\nu}, \;{\cal K}_{\mu\nu},\; \bar {\cal K}_{\mu\nu},\;
C_{\mu\nu}, \bar C_{\mu\nu}, F_\mu, \bar F_\mu, f_\mu,\;
\bar f_\mu,\; \beta_\mu,\]
\[\qquad B,\; B_1,\; B_2,\; B_3,\; C_1,\; \bar C_1,\;
C_2, \;\bar C_2\big] = 0,
\eqno{(E.2)}\]
the Lagrangian densities (53) and (54) transform as follows:
\[s_{\omega}^{(1)} {\cal L}_{(b)} \,=\, s_{\omega}^{(1)} {\cal L}_{(\bar b)} 
\,=\, - \partial_\mu \Big[(B_1 - B_3) (\partial^\mu B) 
- B\; \partial^\mu (B_1 - B_3) \Big],
\eqno{(E.3)}\]
\[s_{\omega}^{(2)} {\cal L}_{(b)}\, =\, s_{\omega}^{(2)} {\cal L}_{(\bar b)} 
\,=\, \partial_\mu \Big[(B_1 + B_3) (\partial^\mu B_2) 
- B_2\; \partial^\mu (B_1 + B_3) \Big].
\eqno{(E.4)}\]
Furthermore, it can be checked that the above bosonic symmetry transformations
$s_\omega^{(1)}$ and $s_\omega^{(2)}$ lead to the following conserved currents
\[J^{\mu \;(1)}_{(\omega, b)} \,=\, ({\cal K}^{\mu\nu} - K^{\mu\nu}) \;(\partial_\nu B)
- (\partial^\mu \beta^\nu - \partial^\nu \beta^\mu)\; \partial_\nu (B_1 - B_3),
\eqno{(E.5)}\]
\[J^{\mu \;(1)}_{(\omega, \bar b)} \,=\, (\bar {\cal K}^{\mu\nu} - \bar K^{\mu\nu}) \;(\partial_\nu B)
- (\partial^\mu \beta^\nu - \partial^\nu \beta^\mu)\; \partial_\nu (B_1 - B_3),
\eqno{(E.6)}\]
\[J^{\mu \;(2)}_{(\omega,  b)} \,=\, ({\cal K}^{\mu\nu} + K^{\mu\nu}) \;(\partial_\nu B_2)
- (\partial^\mu \bar \beta^\nu - \partial^\nu \bar \beta^\mu)\; \partial_\nu (B_1 + B_3),
\eqno{(E.7)}\]
\[J^{\mu \;(2)}_{(\omega, \bar b)} \,=\, (\bar {\cal K}^{\mu\nu} + \bar K^{\mu\nu}) \;(\partial_\nu B_2)
- (\partial^\mu \bar \beta^\nu - \partial^\nu \bar \beta^\mu)\; \partial_\nu (B_1 + B_3),
\eqno{(E.8)}\]
where the conservation law can be proven by exploiting the Euler-Lagrange equations of motion (61) and (62).
The above conserved currents lead to the following expressions for the conserved charges
($Q^{(r)}_{(\omega, s)} = \int d^5x \; J^{0 \;(r)}_{(\omega, s)}$ where  $r= 1, 2$ and $s = b, \bar b$)
\[
Q_{(\omega, b)}^{(1)} = \int d^5x \;\Big[({\cal K}^{0i} - K^{0i}) \;(\partial_i B)
- (\partial^0 \beta^i - \partial^i \beta^0)\; \partial_i (B_1 - B_3) \Big],
\eqno{(E.9)}\]
\[Q_{(\omega, \bar b)}^{(1)} = \int d^5x\; \Big[(\bar {\cal K}^{0i} - \bar K^{0i}) \;(\partial_i B)
- (\partial^0 \beta^i - \partial^i \beta^0)\; \partial_i (B_1 - B_3) \Big], 
\eqno{(E.10)}\]
\[Q_{(\omega, b)}^{(2)} = \int d^5x \;\Big[({\cal K}^{0i} + K^{0i}) \;(\partial_i B_2)
- (\partial^0 \bar \beta^i - \partial^i \bar \beta^0)\; \partial_i (B_1 + B_3) \Big],
\eqno{(E.11)}\]
\[Q_{(\omega, \bar b)}^{(2)} = \int d^5x\; \Big[(\bar {\cal K}^{0i} + \bar K^{0i}) \;(\partial_i B_2)
- (\partial^0 \bar \beta^i - \partial^i \bar \beta^0)\; \partial_i (B_1 + B_3) \Big].
\eqno{(E.12)}\]
Even though, the above bosonic charges remain invariant under the (anti-)BRST as well as 
(anti-)dual BRST symmetry transformations [cf. (57), (55), (72) and  (70)],
under the ghost-scale symmetry transformations $s_g$ [cf. (95)], we obtain the following 
\[s_g Q_{(\omega, b)}^{(1)} = - i \,\big[Q^{(1)}_{(\omega, b)}, \; Q_g\big] 
=  + 2 \;Q_{(\omega, b)}^{(1)}, \]
\[ s_g Q_{(\omega, \bar b)}^{(1)} =  - i\, \big[Q^{(1)}_{(\omega, \bar b)}, \; Q_g\big] 
= + 2\; Q_{(\omega, \bar b)}^{(1)},
\eqno{(E.13)}\]

\[s_g Q_{(\omega, b)}^{(2)} = - i \,\big[Q^{(2)}_{(\omega, b)}, \; Q_g\big] 
=  - 2 \;Q_{(\omega, b)}^{(2)}, \]
\[s_g Q_{(\omega, \bar b)}^{(2)} =  - i \,\big[Q^{(2)}_{(\omega, \bar b)}, \; Q_g\big] 
= - 2\; Q_{(\omega, \bar b)}^{(2)}.
\eqno{(E.14)}\]
Thus, the ghost numbers of 
$\Bigl (Q_{(\omega, b)}^{(1)}, \; Q_{(\omega, \bar b)}^{(1)} \Bigr )$ are $(+2)$ and the ghost
numbers of $\Bigl (Q_{(\omega, b)}^{(2)}, \; Q_{(\omega, \bar b)}^{(2)} \Bigr )$ 
are $(-2)$ [cf. $(E.13)$ and $(E.14)$] and, as a consequence, they {\it do not} commute with the ghost charge 
of the present theory. Moreover, at the level of symmetry transformations, the bosonic symmetries 
 $s_{\omega}^{(1)}$ and $s_{\omega}^{(2)}$ commute with $s_b, s_{ab}, s_d, s_{ad}$ but they 
{\it do not} commute with the ghost-scale transformation $s_g$ for the generic fields
$\Psi_1$ and $\Psi_2$, namely; 
\[\big[s_{\omega}^{(1)}, \; s_g\big]\Psi_1 \ne 0,  
\qquad \Psi_1 = \phi_\mu^{(1)}, \phi_\mu^{(2)},  \bar \beta_\mu, 
\eqno{(E.15)}\]
\[
\big[s_{\omega}^{(2)}, \; s_g\big] \Psi_2 \ne 0, 
\qquad \Psi_2 = \phi_\mu^{(1)}, \phi_\mu^{(2)}, \beta_\mu,
\eqno{(E.16)}\]
However,  the bosonic symmetry $s_\omega$ [cf. (81)] commutes with all the 
symmetry transformations $s_b, s_{ab}, s_d, s_{ad}, s_g$.
Thus, due to the above mentioned reasons, these bosonic symmetries $s_{\omega}^{(1)}$ and $s_{\omega}^{(2)}$ 
(and the corresponding generators $Q_\omega^{(1)}$ and $Q_\omega^{(2)}$) are {\it not} the Casimir operators for the 
algebras (98) and (102). It should be noted that, for the sake of brevity, we have taken
$Q_\omega^{(1)} \equiv \left(Q_{(\omega, b)}^{(1)}, \;Q_{(\omega, \bar b)}^{(1)} \right)$ and
$Q_\omega^{(2)} \equiv \left (Q_{(\omega, b)}^{(2)},\; Q_{(\omega, \bar b)}^{(2)} \right)$ 
for our further discussions.

The clinching evidence that $Q_\omega^{(1)}$ and $Q_\omega^{(2)}$ are {\it not} the physical realization of the 
Laplacian operator of differential geometry emerges out from the ghost number considerations.
From the algebra in $(E.13)$ and $(E.14)$, it is obvious that 
\[
 i\, Q_g\, Q_\omega^{(1)} \,|\chi\rangle_n = (n + 2)\; Q_\omega^{(1)} \,|\chi\rangle_n,
\eqno{(E.17)}\]
\[
i\, Q_g \,Q_\omega^{(2)} \,|\chi\rangle_n = (n - 2)\; Q_\omega^{(2)} \,|\chi\rangle_n. 
\eqno{(E.18)}\]
Thus, we establish  that $Q_\omega^{(1)}$ and $Q_\omega^{(2)}$ are {\it not} the analogue of Laplacian operator
as is evident from the comparison between [$(E.17), (E.18)$] and (101).

\end{document}